\documentclass[10pt,twocolumn,final,3p]{elsarticle}

\usepackage{amsmath}
\usepackage{amssymb}
\usepackage{graphicx}
\usepackage{subfigure} 
\usepackage[utf8]{inputenc}
\usepackage[]{hyperref} %
\usepackage{xspace} 

\newcommand{\etal}{\textit{et al}.\@\xspace}
\newcommand{\ie}{\textit{i.e.}\@\xspace}

\newcommand{\pwscf}{\textsc{Pwscf}\@\xspace}
\newcommand{\quantumespresso}{\textsc{Quantum Espresso}\@\xspace}

\newcommand{\abinitio}{\textit{ab initio}\@\xspace}
\newcommand{\Abinitio}{\textit{Ab initio}\@\xspace}

\journal{Acta Materialia}

\begin{document}

\begin{frontmatter}

\title{Hydrogen and vacancy clustering in zirconium}

\author{Céline Varvenne\fnref{1}}
\fntext[1]{Present address: Institute of Mechanical Engineering, 
École Polytechnique Fédérale de Lausanne, 
Lausanne CH-1015, Switzerland}

\author{Olivier Mackain}

\author{Laurent Proville}

\author{Emmanuel Clouet\corref{1}}
\ead{emmanuel.clouet@cea.fr}
\cortext[1]{Corresponding author}

\address{CEA, DEN, Service de Recherches de Métallurgie Physique, F-91191 Gif-sur-Yvette, France}

\begin{abstract}
	The effect of solute hydrogen on the stability of vacancy clusters in hexagonal closed packed zirconium 
	is investigated with an \abinitio approach, including contributions of H vibrations.
	Atomistic simulations within the density functional theory evidence
	a strong binding of H to small vacancy clusters. 
	The hydrogen effect on large vacancy loops is modeled through its interaction with the stacking faults.
	A thermodynamic modeling of H segregation on the various faults, relying on \abinitio binding energies,
	shows that these faults are enriched in H, leading to a decrease of the stacking fault energies.
	This is consistent with the trapping of H by vacancy loops observed experimentally.
	The stronger trapping, and thus the stronger stabilization, is obtained for vacancy loops
	lying in the basal planes, \ie the loops responsible for the breakaway growth 
	observed under high irradiation dose.

\end{abstract}


\end{frontmatter}

\section{Introduction}

Macroscopic properties of  metals and their alloys are known to be strongly affected by hydrogen, which is a common impurity found in structural materials. 
Among hydrogen effects, one can mention phase transformations leading to hydrides \cite{Sandrock1999}, interactions with structural defects modifying the plastic activity of materials \cite{Taketomi2008}, the interplay with point defects provoking swelling \cite{Garner2006}, or the so-called H embrittlement, via different possible mechanisms \cite{Oriani1977,Birnbaum1994,Song2013}.

Hexagonal closed packed (hcp) zirconium alloys, that are used as a cladding material in nuclear industry, suffer an in-service hydrogen pickup due to the oxidation of the rods by water. 
Under irradiation, the apparent solubility limit of hydrogen in zirconium increases \cite{Lewis1984,McMinn2000,Vizcaino2002,Vizcaino2007}. 
This has been associated with a trapping of hydrogen by the defects created by irradiation, in particular the vacancy clusters \cite{Lewis1984}.
This hydrogen in solid solution influences not only the mechanical properties of hcp Zr alloys \cite{Domain2004,Udagawa2010}, 
but also their macroscopic structure: it impacts the stress free dimensional change experienced by these alloys under irradiation \cite{McGrath2011}.  
Indeed, under irradiation, a Zr single crystal undergoes an elongation along the $\left\langle a \right\rangle$ axis 
and a shortening along the $\left\langle c\right\rangle$ axis, without significant volume change \cite{Carpenter1988}.
The growth strain remains small at low irradiation dose, whereas a breakaway growth is observed at higher fluence \cite{Carpenter1988,Fidleris1988}. 
This breakaway growth at high doses has been correlated to the appearance of faulted vacancy dislocation loops 
lying in the basal planes.  These vacancy clusters are called $\langle c \rangle$ loops 
because of the $\langle c \rangle$ component of their Burgers vectors.
At low irradiation dose, where the growth strain remains moderate, 
only perfect dislocation loops, either of interstitial or vacancy types,
with $\left\langle a\right\rangle = 1/3\left\langle 11\bar{2}0\right\rangle$ Burgers vector
and with an habit plane close to the prismatic planes of the hcp structure,
are observed.  

Experiments have shown that an increase of the H content leads to an increase of irradiation growth in Zr alloys under neutron irradiation \cite{McGrath2011}.
This is consistent with the TEM observations of Tournadre \etal \cite{Tournadre2013,Tournadre2014} 
who demonstrated that the amount of $\left\langle c\right\rangle$ loops formed under proton irradiation
is notably higher when pre-hydriding the zirconium alloys.
This suggests that hydrogen promotes the stability of these $\langle c \rangle$ loops compared to other vacancy clusters, 
an hypothesis also supported by the calorimetry and X-ray diffraction experiments of Vizcaíno \etal \cite{Vizcaino2002,Vizcaino2007}.
These experiments showed that a higher annealing temperature is needed
in zirconium alloys exposed to a high irradiation dose, hence when $\langle c \rangle$ vacancy loops are present, 
for the H solubility limit to reach its unirradiated value.

Understanding the potential trapping of hydrogen by irradiation defects as well as the hydrogen influence on the stability of the different
vacancy clusters is therefore of prime importance for modeling then the kinetic evolution of Zr alloys under irradiation
and the associated macroscopic behavior in presence of hydrogen.  
In this paper, we study the elementary interactions between H and vacancy clusters, 
going from small clusters containing only a few vacancies to large dislocation loops. 
Atomistic simulations are a well-suited tool to address this question, since they give information at defect sizes not accessible by other techniques.
In particular, \abinitio methods, having a high level of accuracy and transferability, will be used all along this work,
as they appear unavoidable to study structural defects in hcp Zr.
A proper account of angular contributions is indeed necessary to correctly describe the stacking fault energies and the interactions between vacancies 
\cite{Legrand1984,Clouet2012,Varvenne2014}. 
Previous \abinitio studies of hydrogen interaction with defects in zirconium mainly focused on plasticity 
and embrittlement \cite{Domain2004,Udagawa2010} or on single point-defect properties \cite{Christensen2015a,Christensen2015b}.
The work presented in this paper extends thus these studies by modeling H interaction with vacancy clusters of various sizes.

The paper is organized as follows. We first investigate the interactions of hydrogen with small vacancy clusters and with the different stacking faults of interest.
We then model the evolution of the stacking fault energies with the H bulk concentration. 
Finally, including this variation into analytical laws describing the formation energies of extended vacancy clusters,
we discuss the influence of H on the relative stability of vacancy dislocation loops in hcp Zr.

\section{Modeling method}

\subsection{\Abinitio simulations}

In this work, all the \abinitio calculations  are based on the Density Functional Theory (DFT), as implemented in the \pwscf code of the \quantumespresso package \cite{Giannozzi2009}. 
Calculations  are performed in the Generalized Gradient Approximation with the exchange-correlation functional of Perdew, Burke and Ernzerhof \cite{Perdew1996}. Valence electrons are described with plane waves, using a cutoff of $28$~Ry.
The pseudo-potential approach is used to describe the electron-ion interaction.
For Zr and H, ultrasoft pseudo-potentials of Vanderbilt type have been chosen, including 4s and 4p electrons as semicore in the case of Zr.
The electronic density of state is broadened with the Methfessel-Paxton function, with a broadening of $0.3$~eV. The integration is performed on a regular grid of $14 \times 14 \times 8$ k-points for the primitive cell and an equivalent density of k-points for larger supercells. 

Atomic relaxations are performed at constant volume, using a conjugate gradient algorithm.
For calculations involving point defects, we use supercells of $5 \times 5 \times 4$ repeated unit cells (200 atoms) with fully periodic boundary conditions.
The elastic correction of Ref.~\cite{Varvenne2013} is applied, so as to remove the spurious interaction energy between the defect and its periodic images. \ref{ap:CVg_SC_size} 
demonstrates that the obtained corrected energies are well converged with respect to the supercell size.

\subsection{Vibrations}

To consider the contribution of vibrations in the computed energies,
we use the harmonic approximation \cite{Ashcroft1976}, 
which validity is assessed in \ref{sec:appendix_harmonic} for both a single H atom
and an H interacting with a vacancy in an hcp Zr matrix. 
Within this approximation, the vibrational free energy is 
\begin{equation}
	F^{\rm vib}(T) = kT \sum_i{ \ln{\left[ 2 \sinh{\left( \frac{\hbar \omega_i}{kT} \right)} \right]}},
	\label{eq:Fvib}
\end{equation}
with the sum running over all vibration modes of pulsation $\omega_i$. 
We further make the assumption that the vibration modes of the Zr atoms are not affected 
by the presence of hydrogen and consider only the three vibration modes of the H atom.
This is justified by the small mass and the small radius of hydrogen 
compared to zirconium.

To obtain the H vibration frequencies for a given atomic configuration, 
we rigidly displace the H atom from its equilibrium position in six different directions,
using four intermediate positions for each direction with a maximal displacement of 0.04\,\AA.
We then make a least-square fitting of  the calculated energy variations to obtain the $3\times3$ Hessian matrix, 
from which the eigenfrequencies are obtained by diagonalization.

\subsection{Validation of the modeling approach}

Our choice of cutoffs, k-mesh, GGA functional for the exchange correlation and pseudo-potential for zirconium have already been validated both on the hcp bulk and on vacancy cluster properties in previous studies \cite{Clouet2012, Varvenne2014, Varvenne2013}. 
For H, the validation of the chosen pseudo-potential consists first in comparing with experimental data the intrinsic properties of the H$_2$ molecule, such as the equilibrium distance $d_{\rm H_2}$, the vibrational frequency $\nu_{\rm H_2}$ and the dissociation energy $E^{\rm diss}_{\rm H_2}$  (see Table \ref{tab:Zr_H_prop}). The dissociation energy includes both the binding and vibrational $\frac{1}{2}h\nu_{\rm H_2}$ contributions. The vibrational frequency $ \nu_{\rm H_2}$ is calculated according to the harmonic oscillator approximation: 
\begin{equation}
\nu_{\rm H_2}= \frac{1}{2\pi}\sqrt{\frac{2k}{\mu}}
\end{equation}
where $k$ is the bond stiffness obtained by \abinitio calculations and $\mu=m_{\rm H}/2$ is the reduced mass of the H$_2$ molecule. 
The obtained quantities are all in good agreement with the experimental ones, within less than $5\%$. This validates our choice of pseudo-potential for the calculations involving hydrogen. 
 
\begin{table*}[!tb]
\caption{Properties of the H$_2$ molecule (equilibrium distance $d_{\rm H_2}$, 
vibration frequency $\nu_{\rm H_2}$ and dissociation energy $E^{\rm diss}_{\rm H_2}$ )
and solution energies $E^{\rm sol}_{\rm H}$ of H in the hcp Zr matrix.
Values in parenthesis include zero point vibration energy of H in the Zr matrix.} 
\label{tab:Zr_H_prop}
\begin{center}
\begin{tabular}{lccccc}
\hline
& \multicolumn{4}{c}{\Abinitio} & Exp. \cite{Fukai1993} \\
&   This work & Udagawa \cite{Udagawa2010} & Domain \cite{Domain2002} & Lumley \cite{Lumley2014} \\ 
  \hline
  $d_{\rm H_2}$ (\AA) & $0.75$   & $0.75$ & $0.76$  	& $0.75$ &  $0.74$ \\ %
  $\nu_{\rm H_2}$ (THz) &  $130.1$  &  $-$ & $134.4$ 	& $129.8$ & $130.8$ \\ %
  $E^{\rm diss}_{\rm H_2}$ (eV)  & $4.25$ & $4.50$ & $4.27$ 	& $4.53$ & $4.48$  \\ %
  $E^{\rm sol}_{\rm H}$ (eV) T site & $-0.609$ ($-0.387$)  &  $-0.52$ & $-0.604$  	& $-0.60$ &  $-0.66$ \\ 
  $E^{\rm sol}_{\rm H}$ (eV) O site & $-0.549$  ($-0.414$)&  $-0.44$ & $-0.532$  	& $-0.56$ &  - \\  
\hline
\end{tabular}
\end{center}
\end{table*}

As we are interested in the interplay between hydrogen solute and vacancies in an hcp zirconium matrix, we also check the insertion of the H atom in substitutional position and in different interstitial sites: the tetrahedral (T) and the octahedral (O) sites. 
Our DFT calculations show that the substitutional position for the H atom, corresponding actually to an H atom inside a vacancy,
is metastable and has a higher energy than the interstitial positions. More details are given in the following section of this paper. 
For the interstitial sites, we calculate the solution energy, defined as follows: 
\begin{equation}
	E^{\rm sol}_{\rm H} = E_{\rm ZrH} - E_{\rm Zr} - \frac{1}{2}E_{\rm H_2}
	\label{eq:Esol}
\end{equation}
where $E_{\rm Zr}$ refers to the energy of the Zr bulk supercell, $E_{\rm H_2}$ to the energy of a H$_2$ molecule (including the vibrational contribution)  and $E_{\rm ZrH}$ to the energy of the supercell containing an H atom located in the different possible sites.   
The calculated solution energy, without the contribution of H vibration in the Zr matrix,
is lower for H in the T site than for H in the O site,
with an energy difference $\Delta E^{\rm O/T} = 60$\,meV, in good agreement
with other \abinitio calculations although variations are observed between different studies
(72\,meV in \cite{Domain2002}, 80\,meV in \cite{Udagawa2010},
86\,meV in \cite{Burr2013a}, 40\,meV in \cite{Lumley2014}, 
 and 61\,meV in \cite{Christensen2015b}). 
Neutron diffraction experiments \cite{Narang1977} have shown that the H atoms occupy the T sites in hcp Zr.
The correct interstitial site is thus predicted by \abinitio calculations
and our value of solution energy $E^{\rm sol}_{\rm H}$ is very close to the experimental one \cite{Fukai1993} (see Table~\ref{tab:Zr_H_prop}). 
But this is not true anymore when including the H vibration energy in the calculation. 
With this contribution, the energy difference between the T and O insertion sites becomes 
$\Delta E^{\rm O/T} = -26$\,meV at 0\,K and $\Delta E^{\rm O/T} = -47.$\,meV at 600\,K, 
thus corresponding to the O site being more stable than the T site, 
with an increasing stability with the temperature. 
As pointed by Christensen \etal \cite{Christensen2015b} and  shown in \ref{sec:appendix_harmonic_bulk}, the energy difference between these two interstitial sites 
is too small to be able to conclude on the preferential insertion site from \abinitio calculations 
without a much more elaborated treatment of vibrations. 
Consequently, both T and O insertions sites are considered in the following. 
Although it may not be the most stable when vibrations of H are considered, we take the T insertion site 
as the H reference state in the hcp Zr matrix for the various calculations of binding energies.

We also verify that the insertion of H$_2$ in Zr is not favorable: it is unstable inside a T site, and metastable inside an O site,
but with a negative binding energy corresponding to a strong repulsion ($E^{\rm b}_{\rm H_2}=-0.70$~eV).  
This shows that hydrogen at the molecular state does not need to be considered at low H concentrations.

\section{Hydrogen and small vacancy clusters}

We first examine the interaction between H and small vacancy clusters, denoted as V$_n$, with $n$ the number of vacancies. 
We define the binding energy between H and the vacancy cluster as:
\begin{equation}
\label{Eb_HVn}
E^{\rm b}_{\rm HV_n} = E_{\rm ZrV_n} + E_{\rm ZrH} - E_{\rm ZrHV_n} - E_{\rm Zr} 
\end{equation}
where $ E_{\rm ZrV_n}$, $E_{\rm ZrH}$, $E_{\rm ZrHV_n}$ and $E_{\rm Zr}$ are the energies of the same simulation cell containing respectively the vacancy cluster, an H atom in T site, the complex HV$_n$ and no defect.
A positive value of $E^{\rm b}_{\rm HV_n}$ indicates an attractive interaction between the H atom and the vacancy cluster. 
For a given cluster V$_n$, different possibilities exist to insert the H atom. It can be located either in a substitutional or in an interstitial site (T and O, here), and many of these sites can be found close to the vacancy cluster. All of them should be considered,  
but the amount of different configurations to explore strongly increases with the cluster size, making an exhaustive study out of reach. 
An approach based on the understanding of the elementary interactions between H and vacancies therefore appears necessary, so as to reduce the configuration space and to focus on important clusters. 

\subsection{H-V interaction}

We start by studying the interaction between H and a single vacancy. 
We first check the direct insertion of H into a vacancy. It is metastable, but with a strongly repulsive interaction: we obtain $E^{\rm b}_{\rm HV}=-1.16$~eV for this position, a value close to the results of Domain \etal \cite{Domain2002}. So we do not consider anymore this substitutional position in the following.

\begin{figure}[!bthp]
\begin{center}
	\begin{footnotesize}
	\begin{tabular}{cc}
	\includegraphics[scale=0.2]{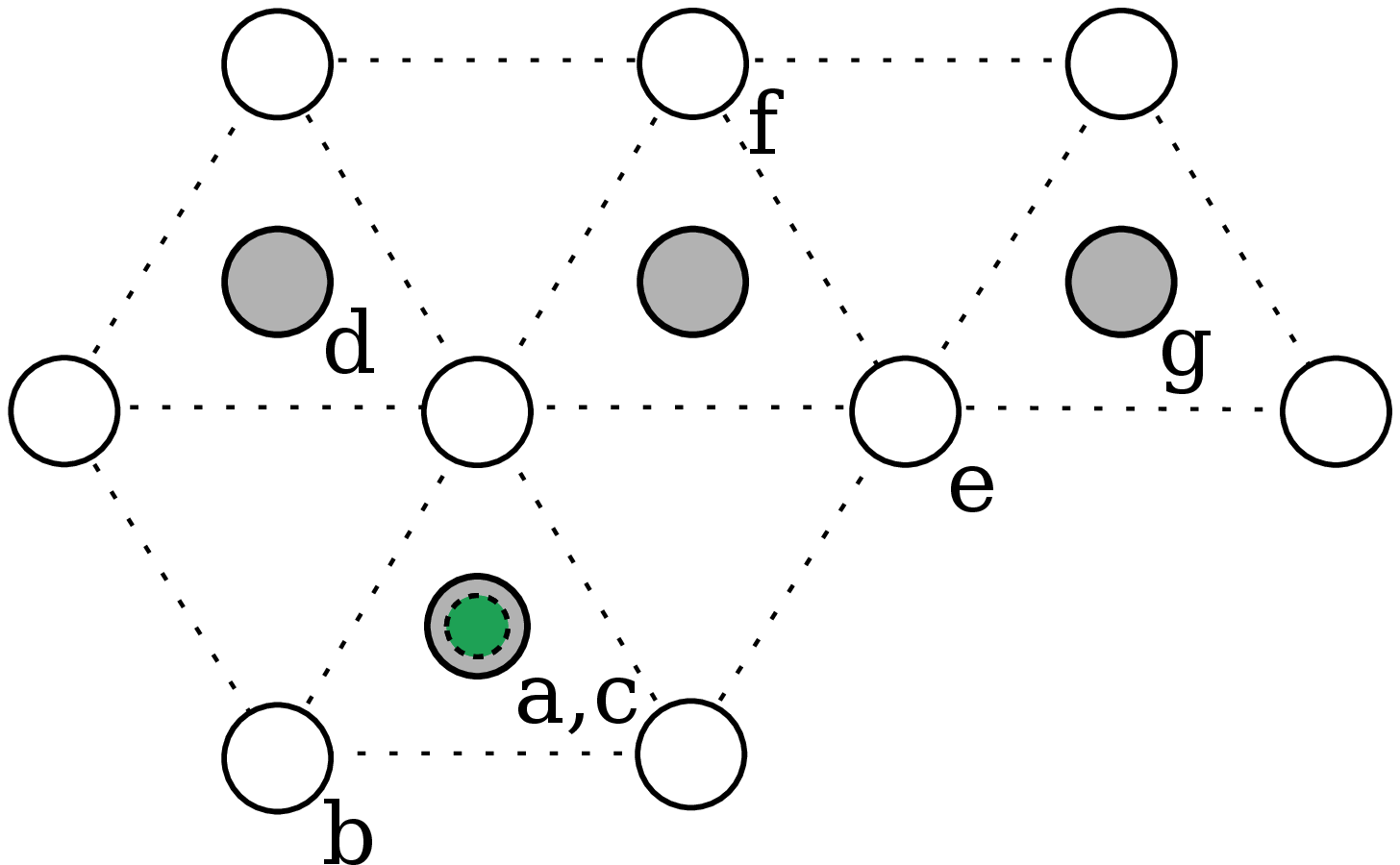} &
	\includegraphics[scale=0.2]{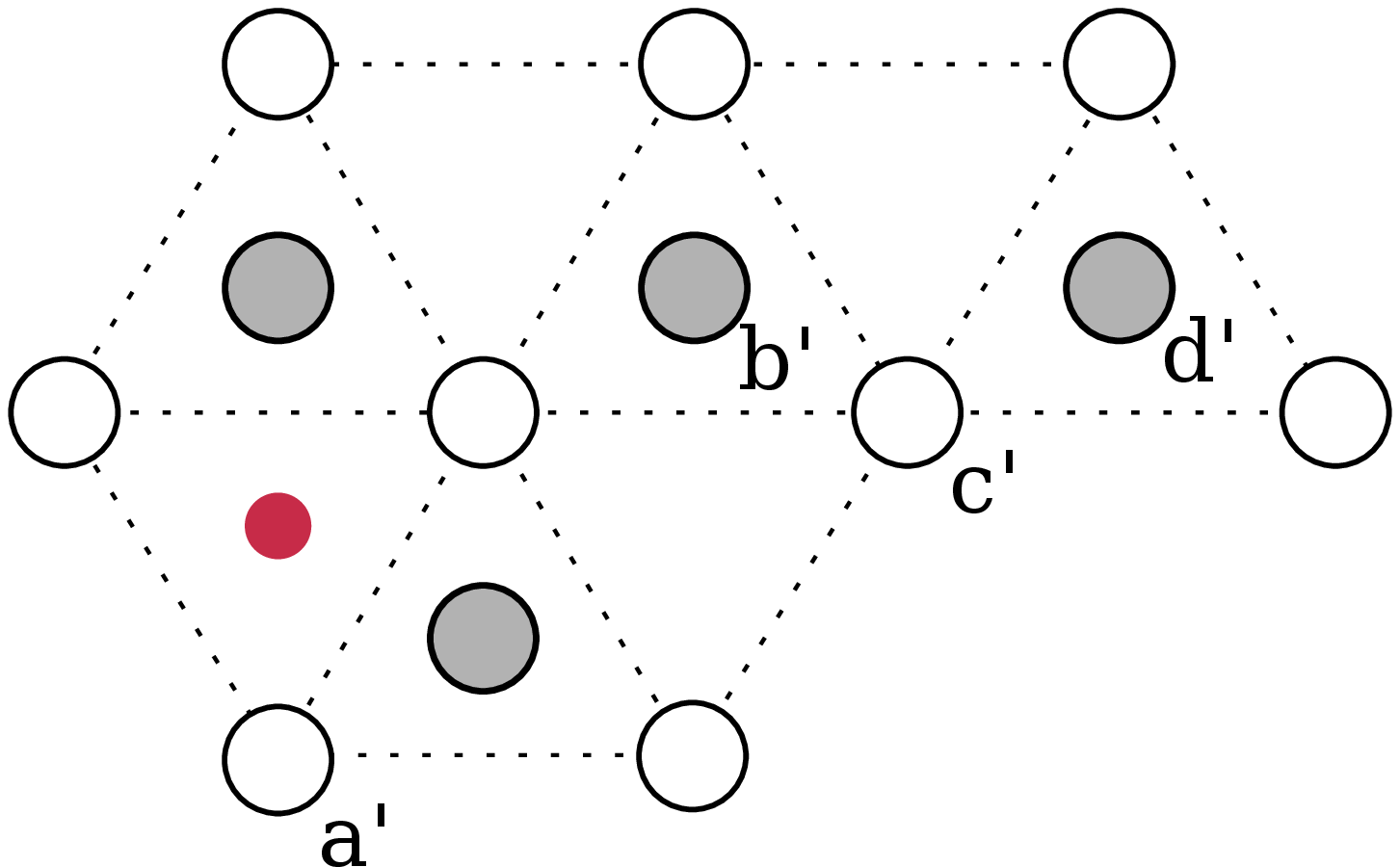} \\
	T site neighboring shells &
	O site neighboring shells \\
	\\
	\includegraphics[scale=0.2]{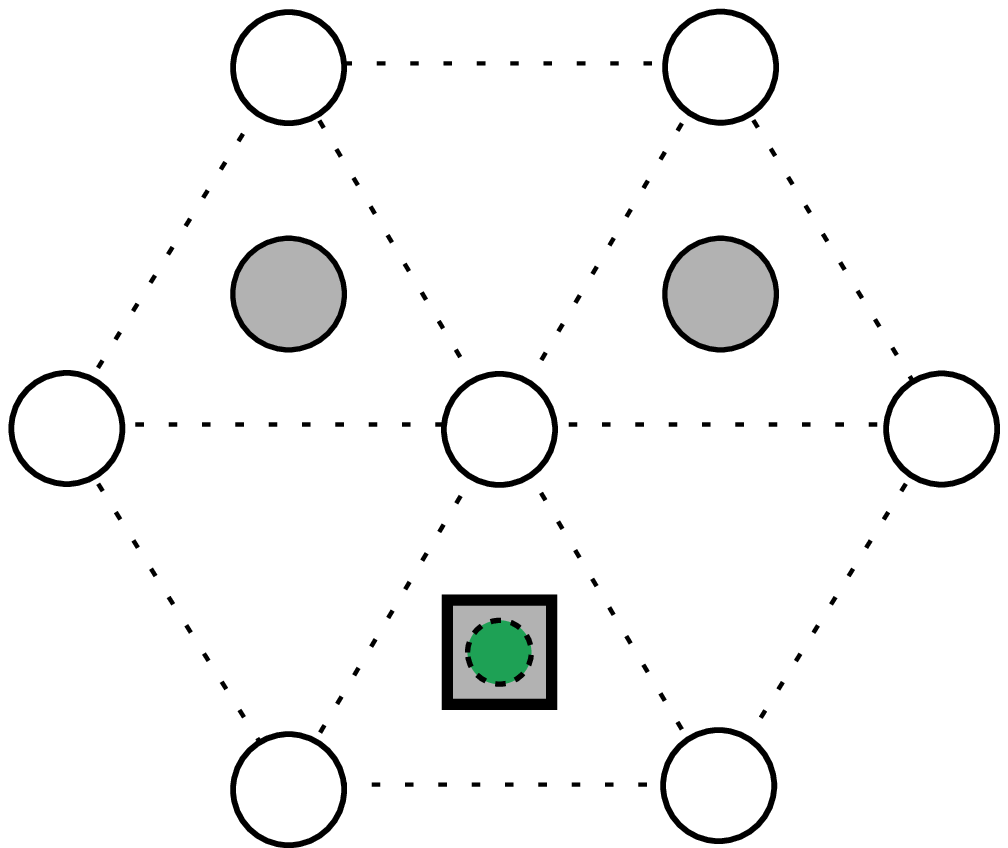} &
	\includegraphics[scale=0.2]{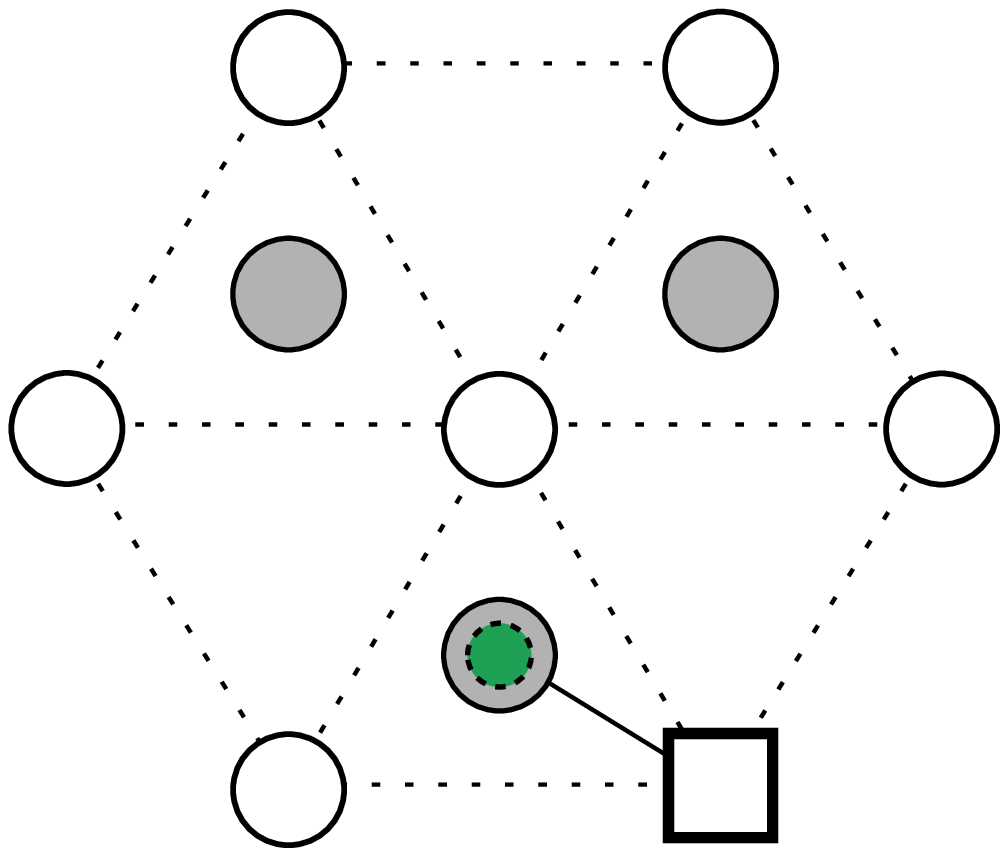} \\
	Configuration a &
	Configuration b \\
	\\
	\includegraphics[scale=0.2]{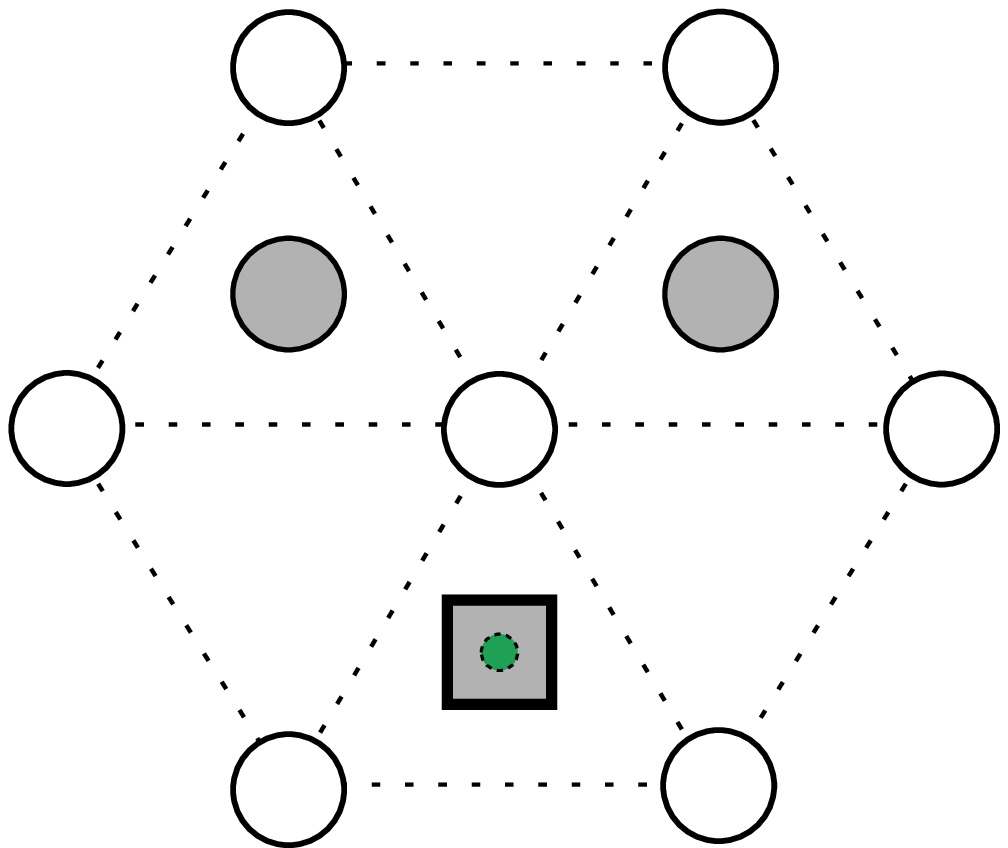} &
	\includegraphics[scale=0.2]{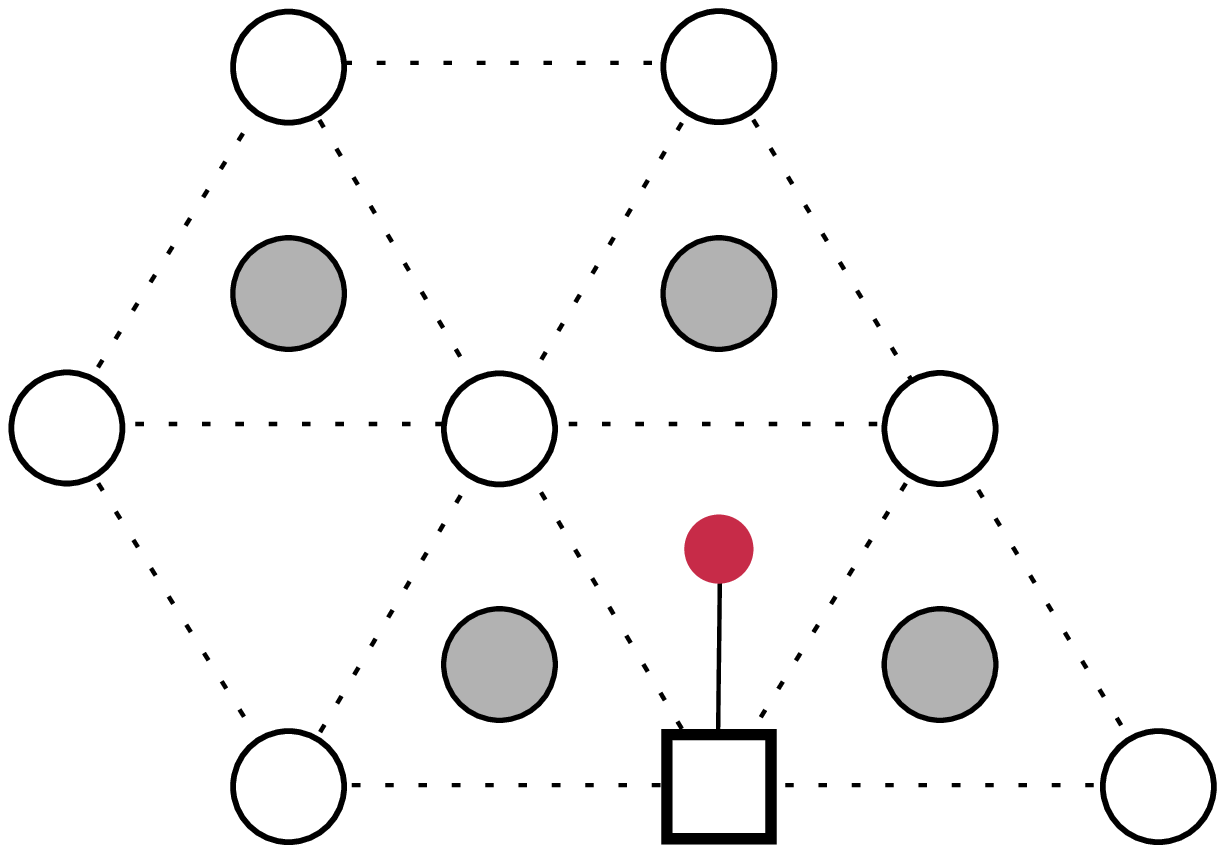}       \\
	Configuration c &
	Configuration a'
	\end{tabular}
	\end{footnotesize}
\end{center}
\caption{Projection in the basal plane of the different H-V pairs investigated. 
The successive neighbor shells of an H atom lying either in a T or O interstitial site
are respectively denoted a to g and a' to d'.
The white spheres represent the Zr atoms at $z=0$, the grey spheres the Zr atoms at $z=c/2$ 
and the squares symbolize the vacancy. The purple spheres represent the H atom in an O site at $z=c/4$,
the green spheres the H atom in a T site at $z=c/8$ (at $z=-c/8$ for the small symbol used in configuration c).}
\label{fig:conf_HV_1nn}
\end{figure}

\begin{figure}[!bth]
	\subfigure[without H vibration]{\includegraphics[scale=0.63]{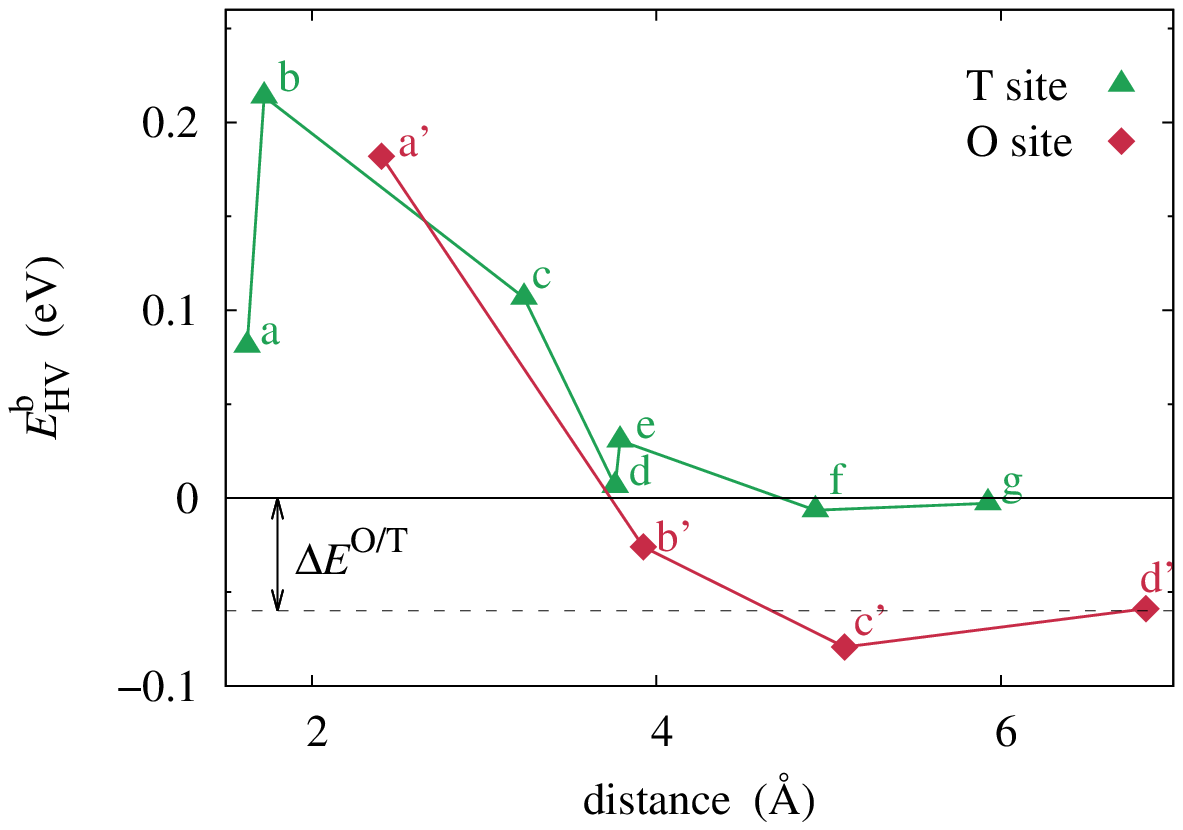}}
	\subfigure[with H vibration]{\includegraphics[scale=0.63]{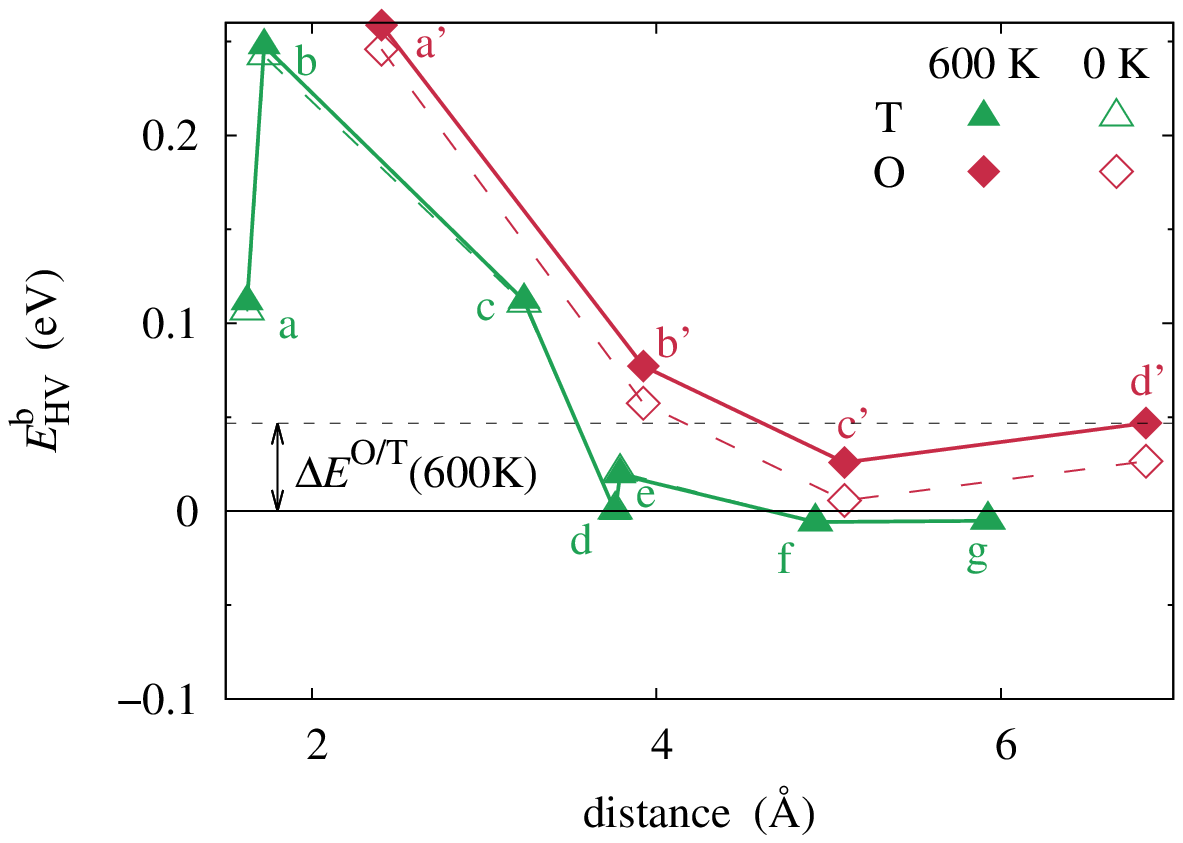}}
\caption{Binding energy between H and a vacancy versus their separation distance,
(a) without and (b) with H vibration free energy. 
Both O and T insertion sites are considered for H.
Label sites a to g and a' to d' are detailed in Fig.~\ref{fig:conf_HV_1nn}.}
\label{fig:El_HV_dist}
\end{figure}

We then examine both types of interstitial sites, T and O, for the insertion of hydrogen in the vicinity of the vacancy. 
Different configurations are investigated, where the hydrogen atom is placed in the successive neighboring shells of the vacancy.
The configurations corresponding to the different nearest neighbor positions are displayed in Fig.~\ref{fig:conf_HV_1nn}. 
The resulting binding  energies between H and V are provided as a function of their separation distance in Fig.~\ref{fig:El_HV_dist}.
The maximal binding energy is 0.21\,eV  without H vibration and 0.24\,eV with the inclusion of zero point energy,
in good agreement with the experimental value $0.28\pm0.06$\,eV obtained by Lewis \cite{Lewis1984} 
from a kinetic modeling of his recovery experiments performed after deuterium implantation in irradiated zirconium.

We first analyse the case where H is in a T site. The two non-equivalent first neighbor positions lead to different binding energies: the value for the configuration b is twice the one of the configuration a. 
Accounting for the directional aspect of the H-V interaction therefore appears to be important. 
The $c/a$ ratio of the hcp Zr structure is $1.601$ in our \abinitio approach \cite{Clouet2012}:
this is a lower value than the ideal one ($\sqrt{8/3}$). Imposing an ideal $c/a$ ratio, we calculate again these two binding energies. We obtain in that case $E^{\rm b}_{HV}=0.04$~eV for the configuration a and $E^{\rm b}_{HV}=0.20$~eV for the configuration b. The two interactions remain different even when applying an ideal $c/a$ ratio. The in-plane interaction (configuration b) is rather unaffected by the $c/a$ ratio, while the out-of-plane interaction of configuration a is very sensitive to its value. It goes from attractive to zero when increasing $c/a$: consequently, we expect this binding energy between H and V to be strongly affected when applying a strain. 
Configuration c shows an attractive binding of slightly larger magnitude than the one of configuration a. 
For larger distances (configurations d to g), the binding energy rapidly goes to zero, showing that long range interactions do not exist in this case. 

We finally look at the insertion of H into the O site. Its binding energy with the vacancy is attractive only for the nearest neighbor configuration, denoted a' here. Its magnitude is close to the magnitude of the most attractive configuration for H in T site (configuration b). This is surprising, as in pure zirconium, H is located in the tetrahedral sites: the presence of the vacancy hence increases the stability of the octahedral site. 
This is an additional reason to always consider this insertion site in the rest of our study.
For the configurations b' to d', the binding energy decreases to reach the energy difference between O and T sites in bulk Zr, thus indicating that H does not interact anymore with the vacancy. 

The main change caused by the vibrational free energy is a constant shift of the binding energies 
for the configurations with an H atom in an O site (Fig. \ref{fig:El_HV_dist}b). 
This merely corresponds to the increase of stability observed for the O site in bulk Zr. 
Apart from this constant shift, H vibrations slightly modify the interaction between the H and the vacancy 
when they are first nearest neighbours: it increases the binding when the H atom is in a T site (configurations a and b)
and decreases it for the O site (configuration a'). 
When the H atom and the vacancy are separated by a larger distance, the H vibrations does not change their binding.

This analysis of the H-V interaction suggests that only the very short ranged interactions need to be considered for the stability of the complexes between hydrogen and vacancies. This point is discussed in the following. 

\subsection{Position of H close to a vacancy cluster}

\begin{figure}[!b]
	\subfigure[without H vibration]{\includegraphics[scale=0.63]{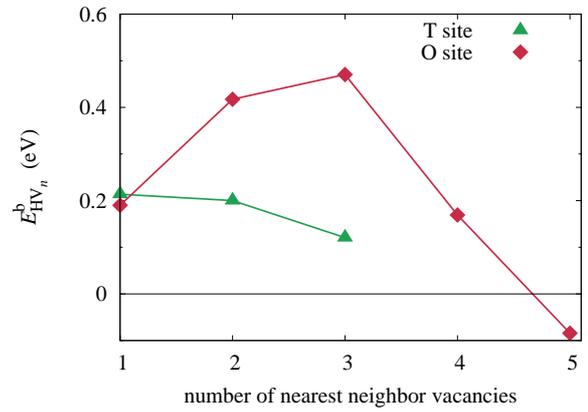}}
	\subfigure[with H vibration]{\includegraphics[scale=0.63]{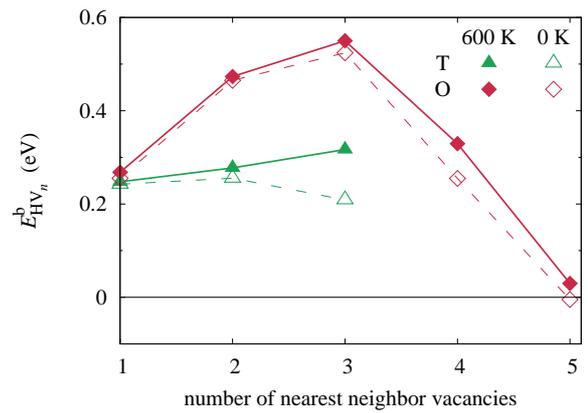}}
	\caption{Binding energies between H and an increasing number of vacancies in nearest neighbor position, 
		(a) without and (b) with H vibration free energy.
		The H atom is located either in a T or O site.}
	\label{fig:El_HVn_additiv}
\end{figure}

Vacancies are clustering under irradiation in zirconium alloys, and we have shown that H and V interact at very short distances. In order to understand where the H atom is located with respect to a preexisting vacancy cluster, we consider now the accumulation of vacancies around a single H atom.  
To this aim, we calculate the binding energies, as defined in Eq.~\ref{Eb_HVn}, of H with clusters of $n$ vacancies, where $n$ vacancies surround the H atom in nearest neighbor positions, \ie vacancy positions corresponding to configurations a, b and a' for the H-V pair (Fig.~\ref{fig:conf_HV_1nn}). Again, both T and O interstitial sites are considered. 
For each number $n$ of vacancies, the different possible configurations are tested (see Tab.~\ref{tab:config_HVn}), and the most stable one is retained.
The contribution of H vibrations to the binding energy is considered only for these most stable clusters.
The resulting binding energies versus the number of vacancies surrounding the H atom are displayed in Fig.~\ref{fig:El_HVn_additiv}. 
H vibrations generally increase the stability of the HV$_n$ clusters.

\begin{table}[!bth]
	\caption{Different possible configurations of the H-V$_n$ complexes 
		involving $n$ nearest neighbor bonds between H and V.
		The H atom is located either in T or O site.
		Values in parenthesis include zero point vibration energy of H.}
	\label{tab:config_HVn}
	\begin{center}
	\begin{tabular}{l c c c}
	\hline
	T site \\
	\hline
	$n=2$ & \includegraphics[scale=0.12]{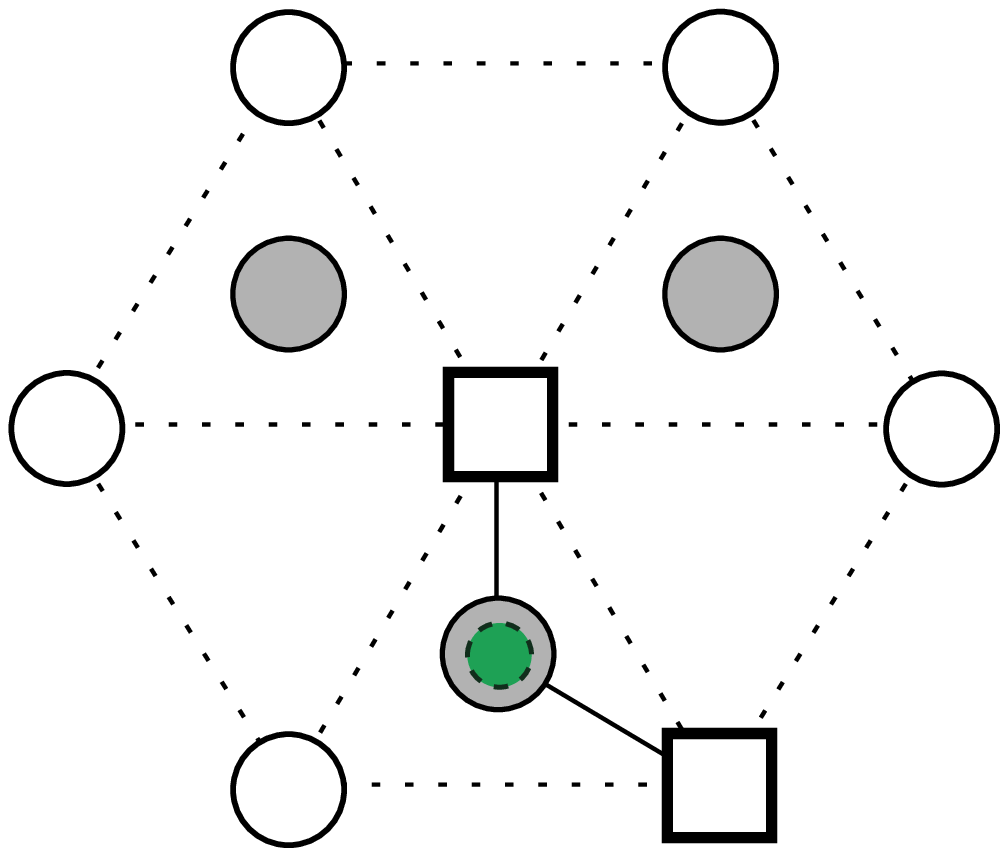} & \includegraphics[scale=0.12]{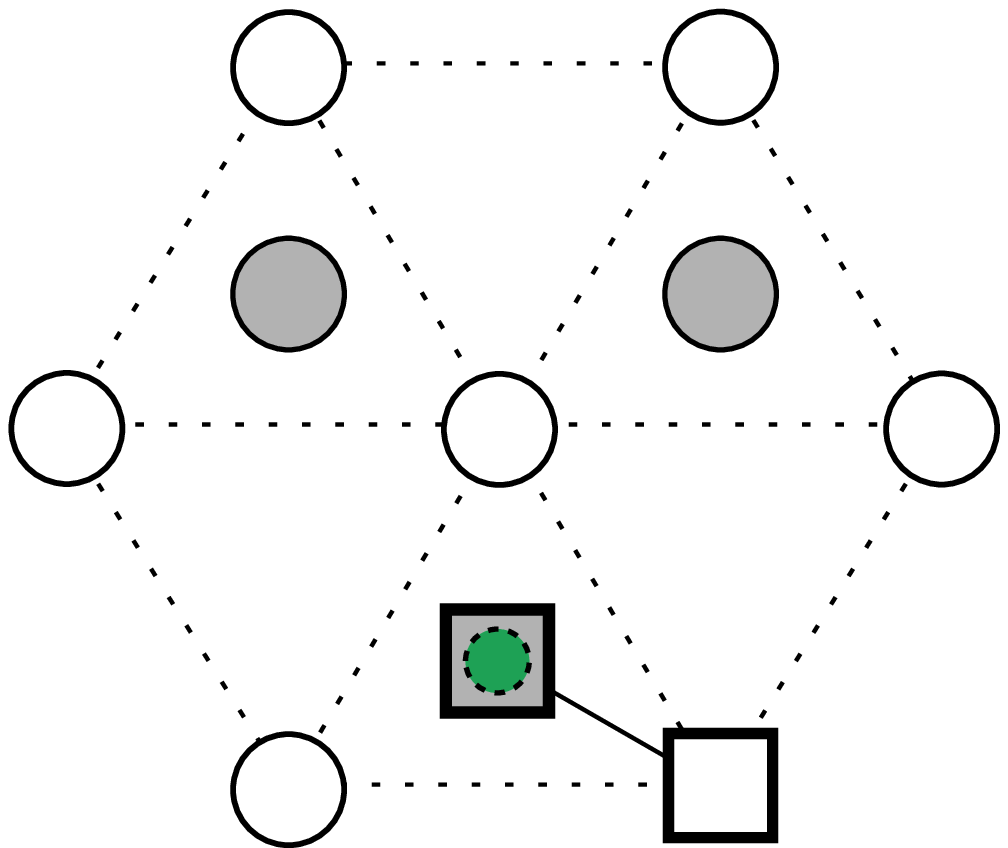} \\
	& 0.20\,eV & unstable \\ 
	& (0.25\,eV) \\ 
	$n=3$ & \includegraphics[scale=0.12]{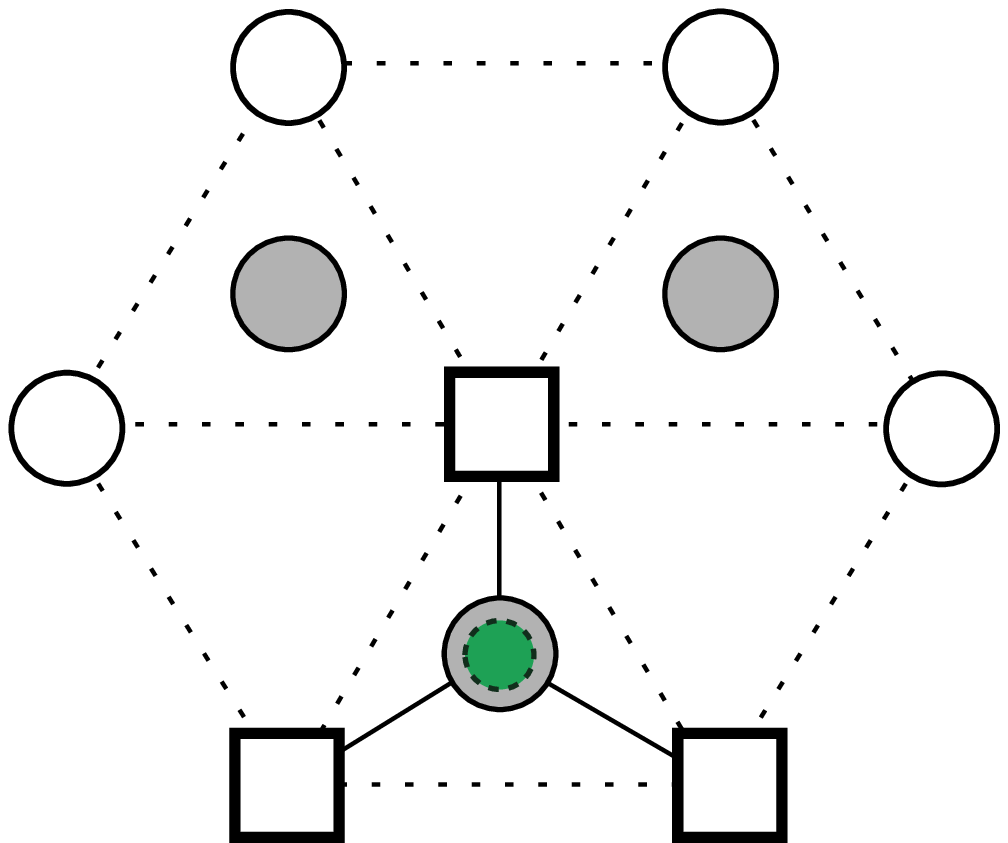} & \includegraphics[scale=0.12]{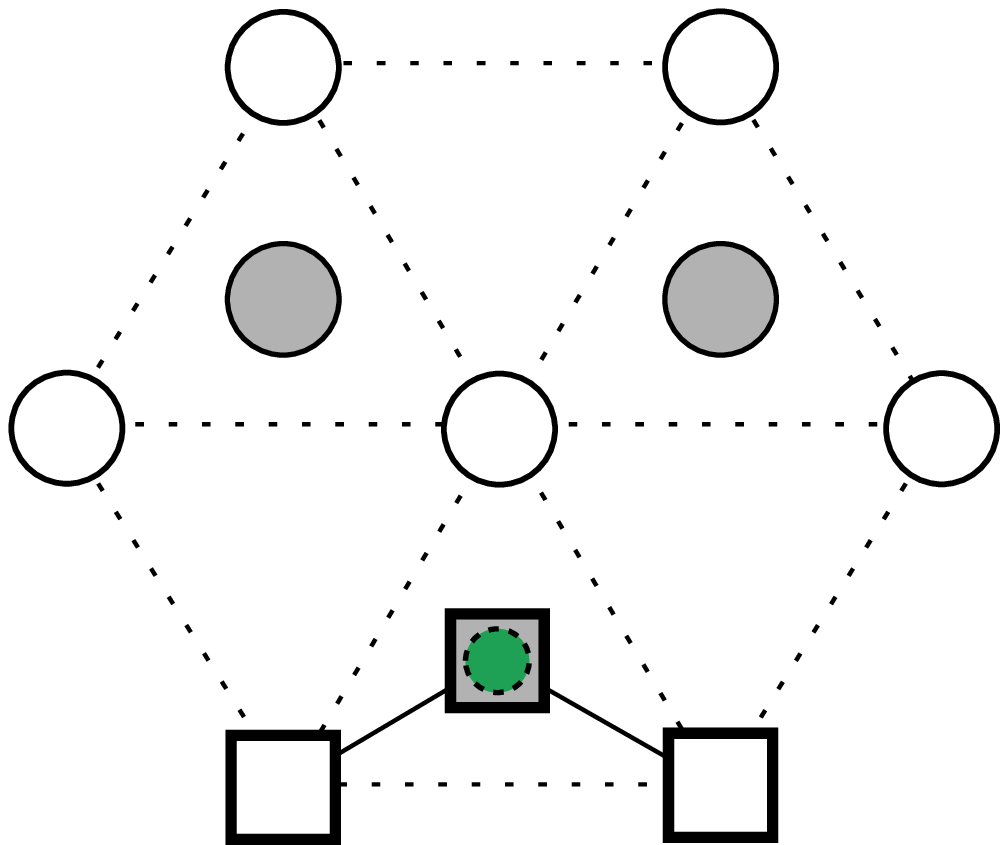} \\
	& 0.12\,eV & unstable \\
	& (0.21\,eV) \\ 
	$n=4$ & \includegraphics[scale=0.12]{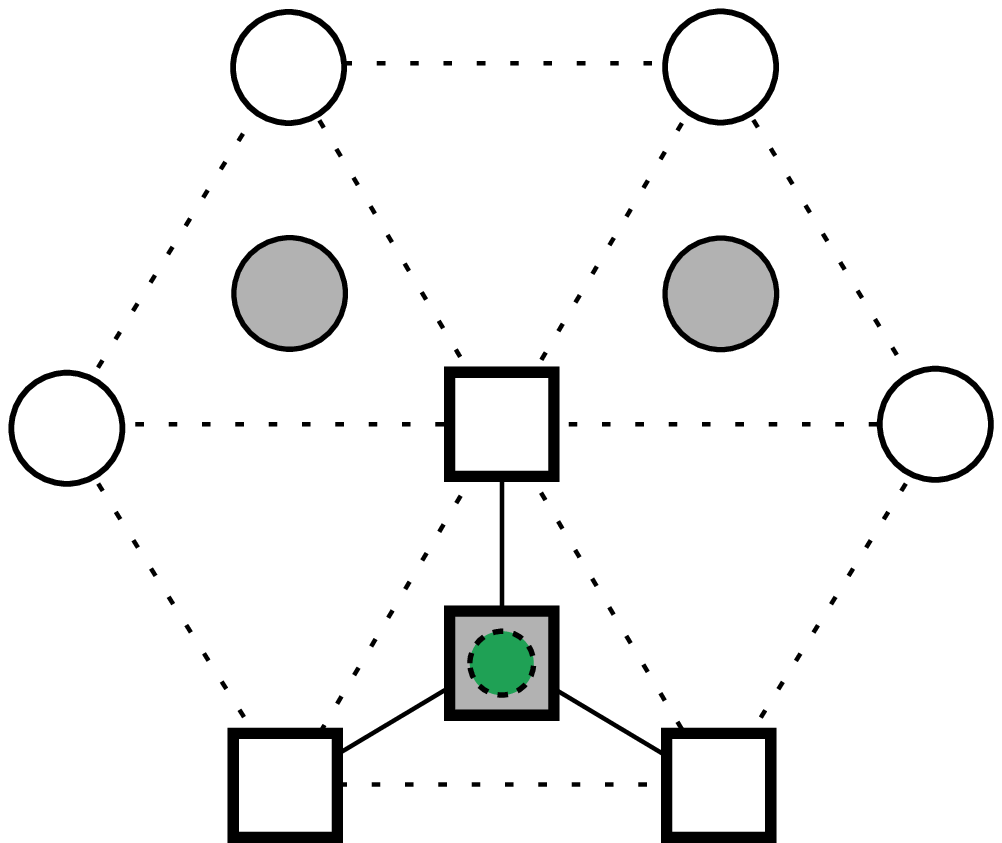} \\
	 & unstable \\
	\hline
	O site \\
	\hline
	$n=2$ & \includegraphics[scale=0.12]{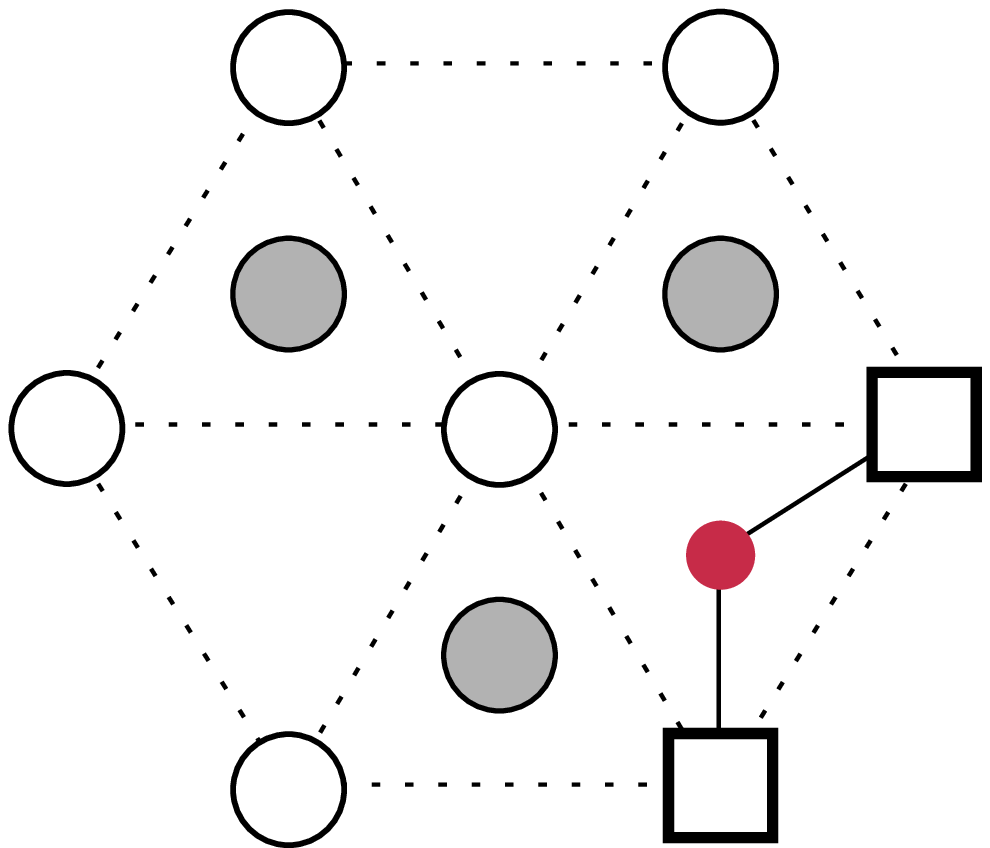} & \includegraphics[scale=0.12]{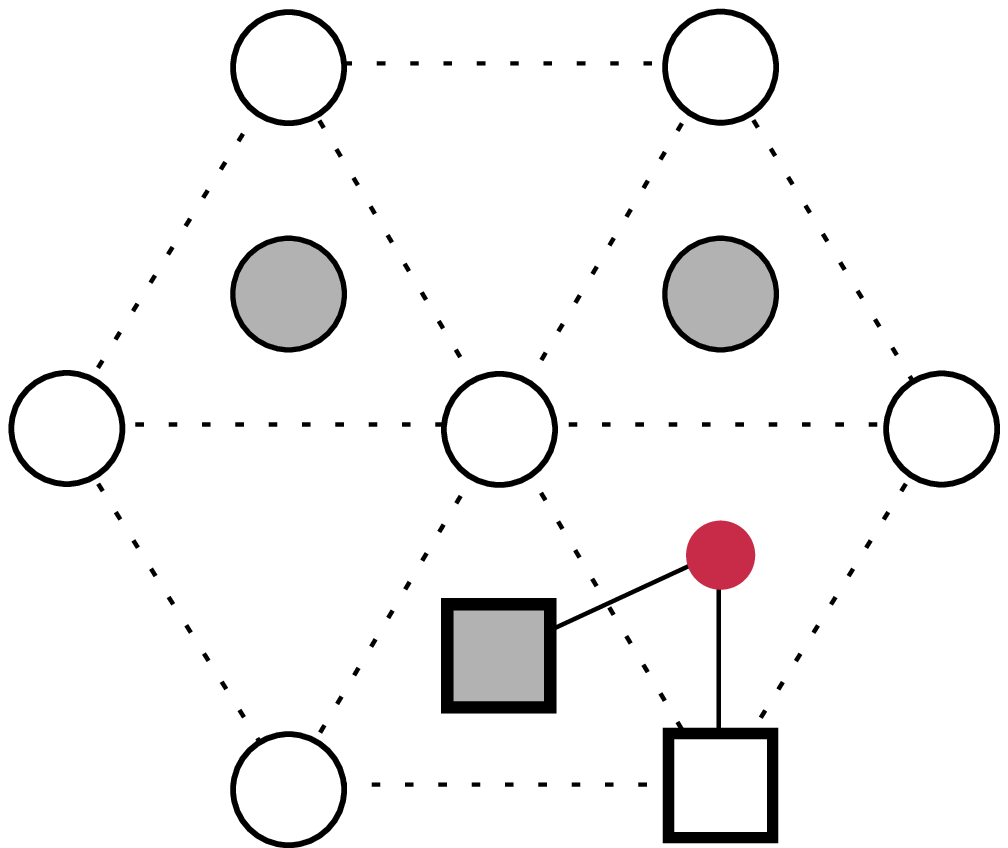} & \includegraphics[scale=0.12]{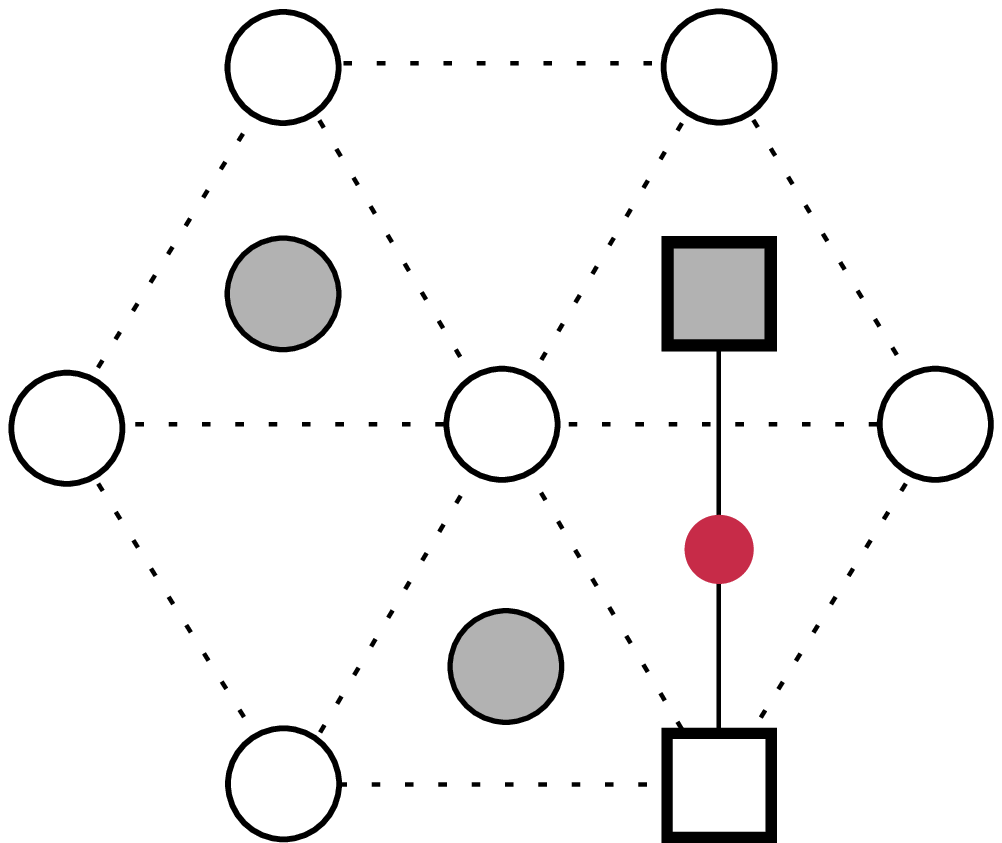} \\
	& 0.42\,eV & 0.36\,eV & 0.37\,eV \\
	& (0.46\,eV) \\ 
	$n=3$ & \includegraphics[scale=0.12]{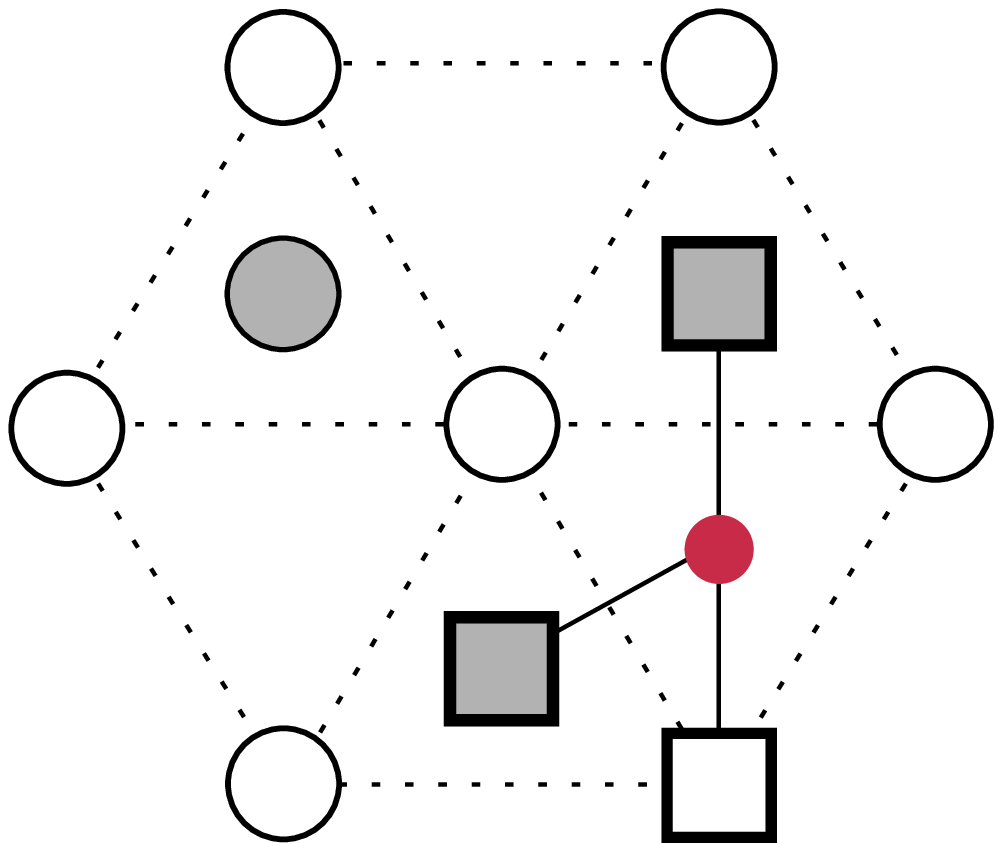} & \includegraphics[scale=0.12]{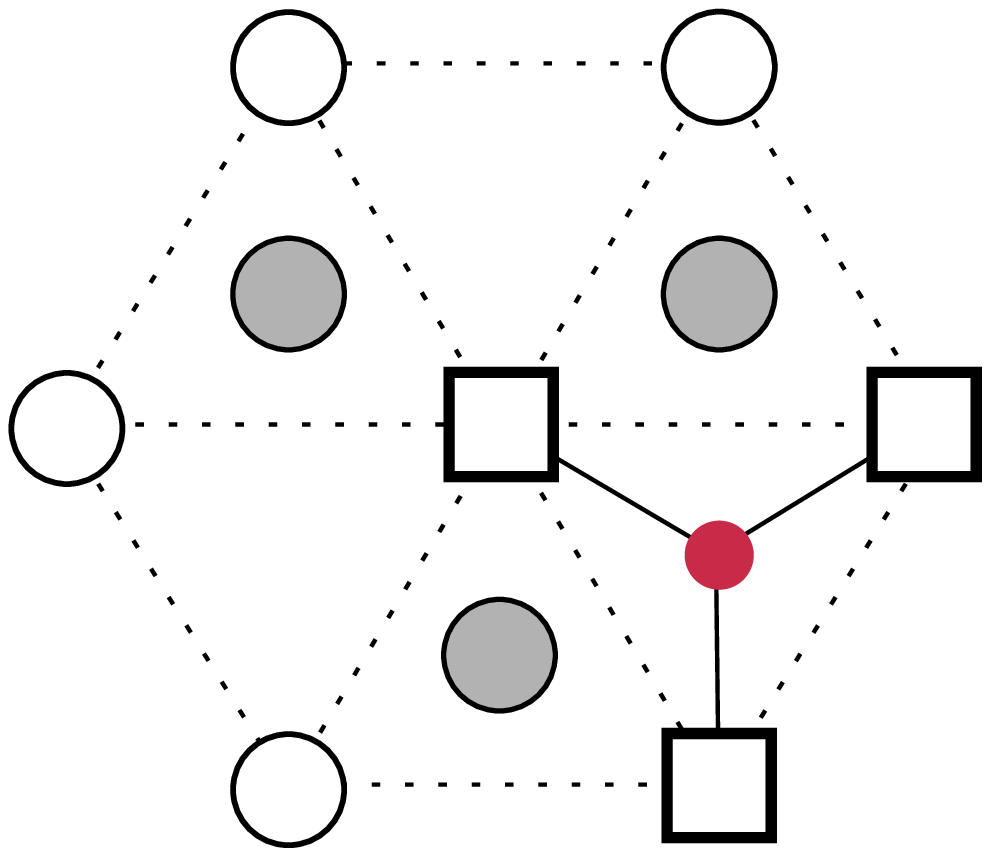} & \includegraphics[scale=0.12]{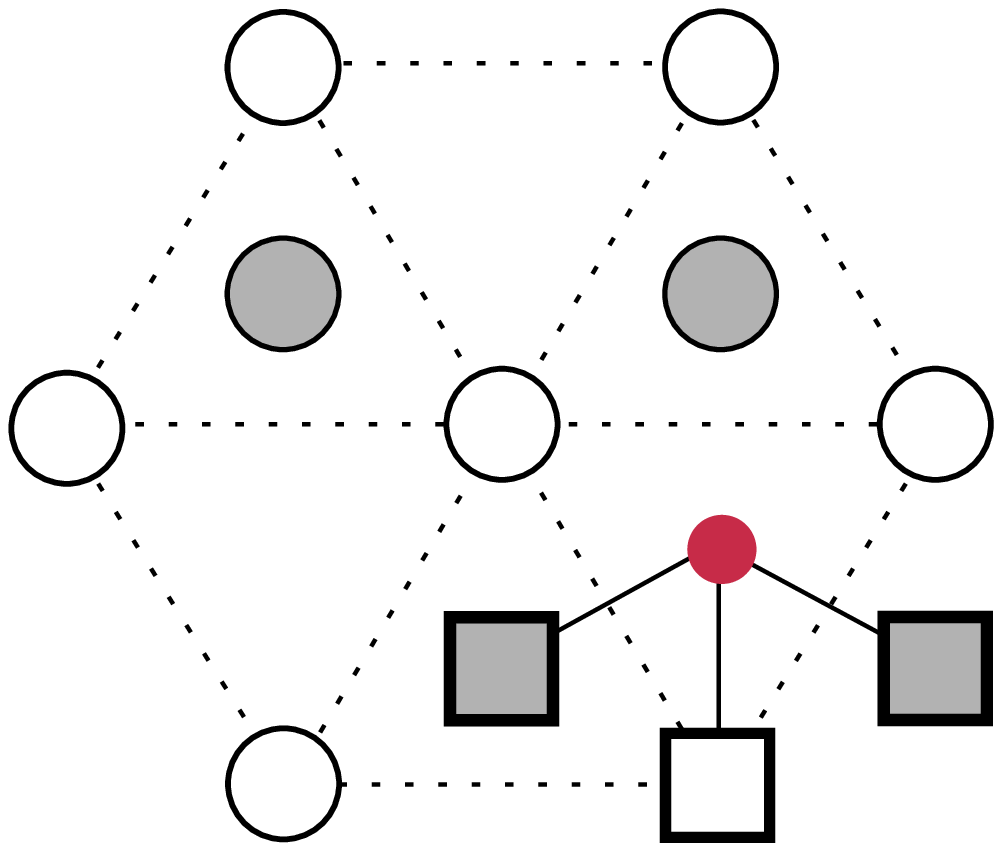} \\
	 & 0.47\,eV & 0.39\,eV & 0.44\,eV \\
	& (0.52\,eV) \\ 
	$n=4$ & \includegraphics[scale=0.12]{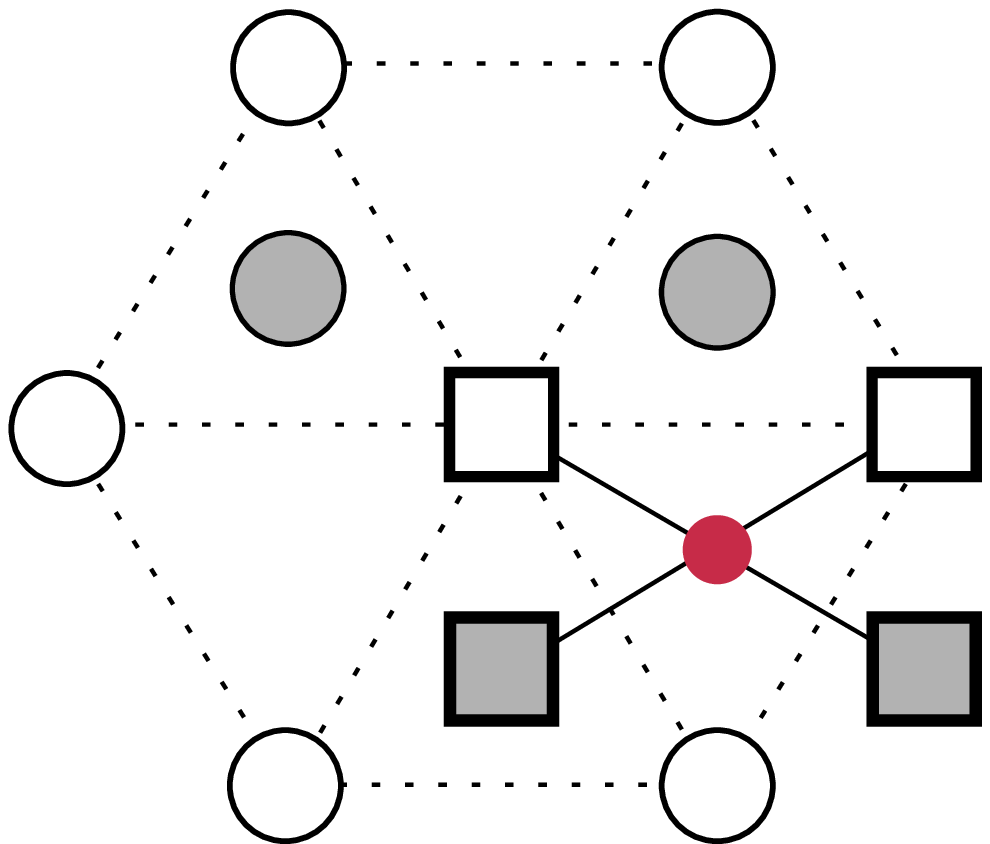} & \includegraphics[scale=0.12]{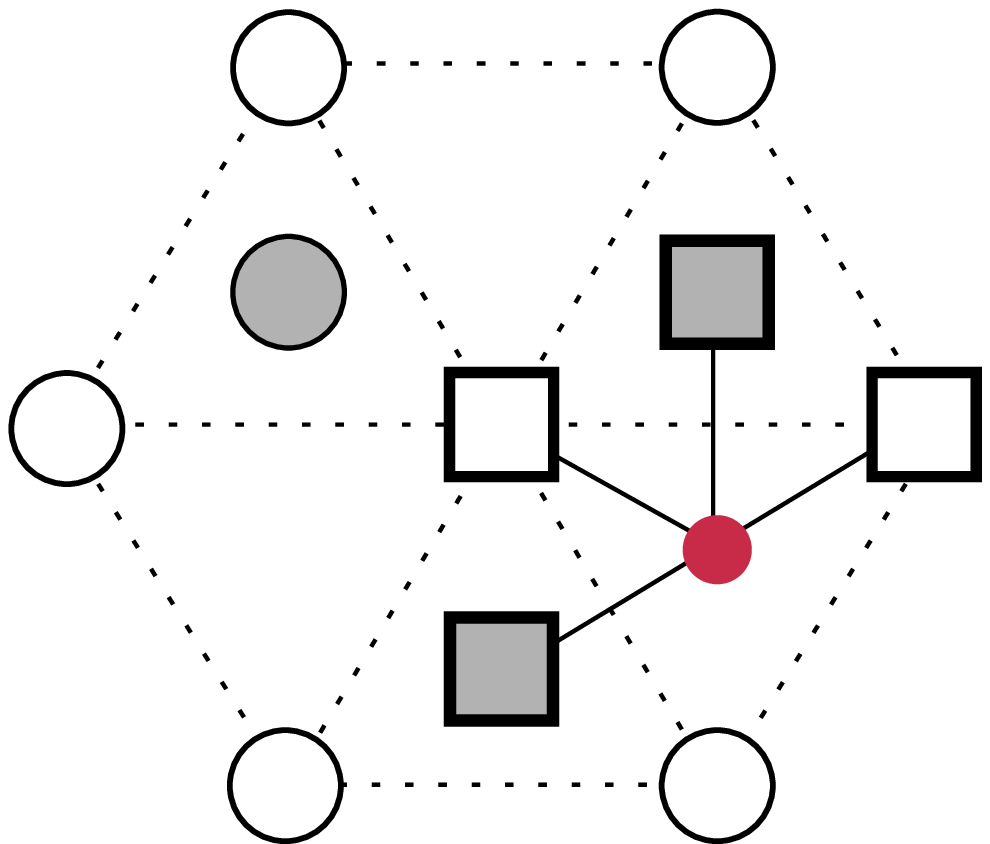} \\
	 & 0.17\,eV & 0.15\,eV \\
	& (0.25\,eV) \\ 
	$n=5$ & \includegraphics[scale=0.12]{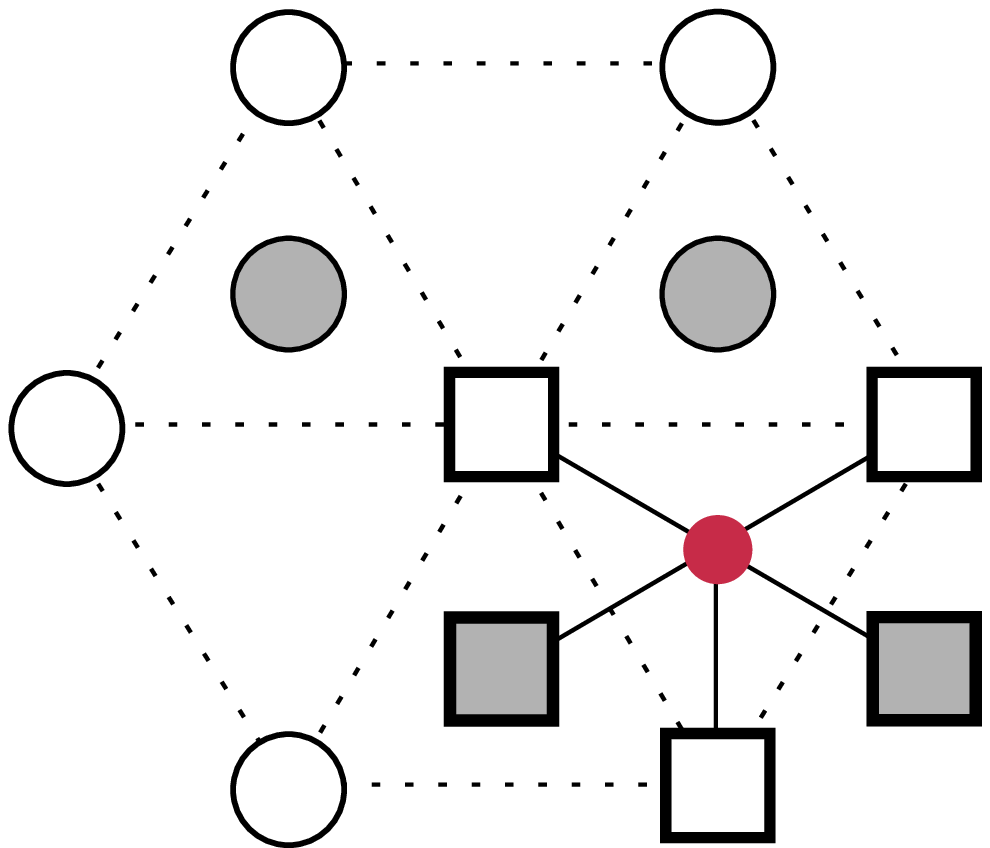} &  &  \\
	 & $-0.08$\,eV \\
	& ($-0.005$\,eV) \\ 
	$n=6$ & \includegraphics[scale=0.12]{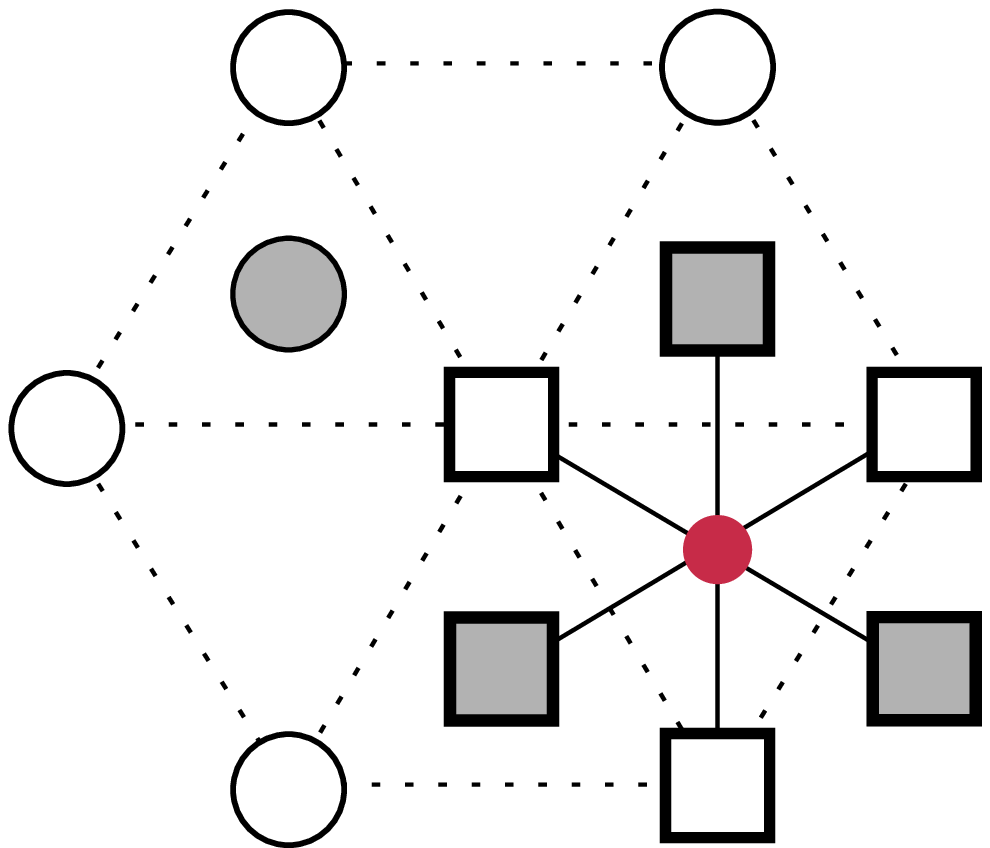} & &  \\
	& unstable \\
	\hline
	\end{tabular}
\end{center}
\end{table}

For H in T site, the binding energy decreases with the number $n$ of nearest neighbor vacancies, 
except at high enough temperature when the H vibrations are considered.
The nearest neighbor interactions are non additive and the highest binding energy is obtained for the HV complex. In addition, some configurations, involving $2$, $3$ and $4$ nearest neighbor vacancies, are unstable. This suggests a more favorable insertion of the H atom at the boundary of a vacancy cluster, rather than inside the cluster. 

The situation is quite different when H is inserted into the O site. 
Up to $3$ vacancies, the binding energy is partially additive: 
adding a new nearest neighbor vacancy to the HV$_n$ cluster stabilizes it a bit more.
For $n \geq 3$, the binding energy rapidly decreases, becoming repulsive for $n = 5$. 
For $n=6$, the cluster corresponds to an H atom inside a small cavity. 
As a consequence, the HV$_6$ cluster is unstable, with negative frequencies for the H vibration.
Again, this  suggests that the insertion inside a vacancy cluster is not favorable. 
Interestingly, the configurations where H occupies an O site with $2$ or $3$ first nearest neighbor vacancies lead to higher binding energies than any configuration where H is in a T site.
This was not expected since H is located in the T site in pure bulk hcp zirconium. 
A similar behavior for H close to vacancy clusters has already been observed recently in bcc iron \cite{Hayward2013}.

\subsection{Interaction with small vacancy clusters}

In light of the two preceding sections, a very simple approach can be proposed so as to reduce the configuration space to explore 
for finding the most stable insertion sites of H close to a vacancy cluster.  
We have identified three preferential insertion sites for H : T site with one vacancy in nearest neighbor position, and O site with $2$ or $3$ nearest neighbor vacancies. Only those sites will be considered in the rest of our study. 
The magnitude of the H-V interaction rapidly decreases with the distance: we therefore suppose that only the nearest neighbor interactions are  important, and that the longer ranged terms ($d_{\rm HV}>3.5$~\AA) do not influence the stability of H close to a vacancy cluster.  
Consequently, we assume that the binding energy between H and the vacancy cluster mainly depends on the nature and the number of nearest neighbor interactions between H and the vacancies. Only the configurations that maximize the binding energy according to this assumption are considered and relaxed with \abinitio calculations. 
We first check the validity of this assumption, and we investigate then the interaction of H with vacancy clusters of different types.
As H vibrations do not strongly modify the H interaction with vacancy clusters,
we do not include them in our calculation of binding energies in this section.

\subsubsection{Validation of the approach}

We test the validity of this approach on two different clusters containing four vacancies: one sitting in the basal plane and the other one sitting in the prismatic $(10\bar{1}0)$ plane. These two clusters are the most stable plane clusters of $4$ vacancies in hcp Zr \cite{Varvenne2014}. 
We compute the binding energy of these clusters with H for different positions of the H atom involving $p$ first nearest neighbor interactions ($1\le p \le 3$) and compare with the binding energy of H with a cluster containing only $p$ vacancies, and where the same $p$ interactions are involved.  
T and O sites are explored, and the calculated binding energies are displayed in Table~\ref{tab:valid_modele1nnV4_TO}.
Calculations for this validation test are performed in a smaller $4\times4\times3$ supercell 
containing 96 lattice sites.

\begin{table}[!bth]
\caption{Comparison of binding energy of H with $V_4$ clusters where $p$ nearest neighbor H-V interactions are present, 
with the binding energy of H with $V_p$ clusters where the same $p$ nearest neighbor H-V interactions are present.
Both T and O insertion sites are explored.}
\label{tab:valid_modele1nnV4_TO}
\begin{center}
\begin{tabular}{ l c c c c}
\hline
 T site & $p$ & 1 & 2 & 3 \\ 
\hline
&& \includegraphics[scale=0.12]{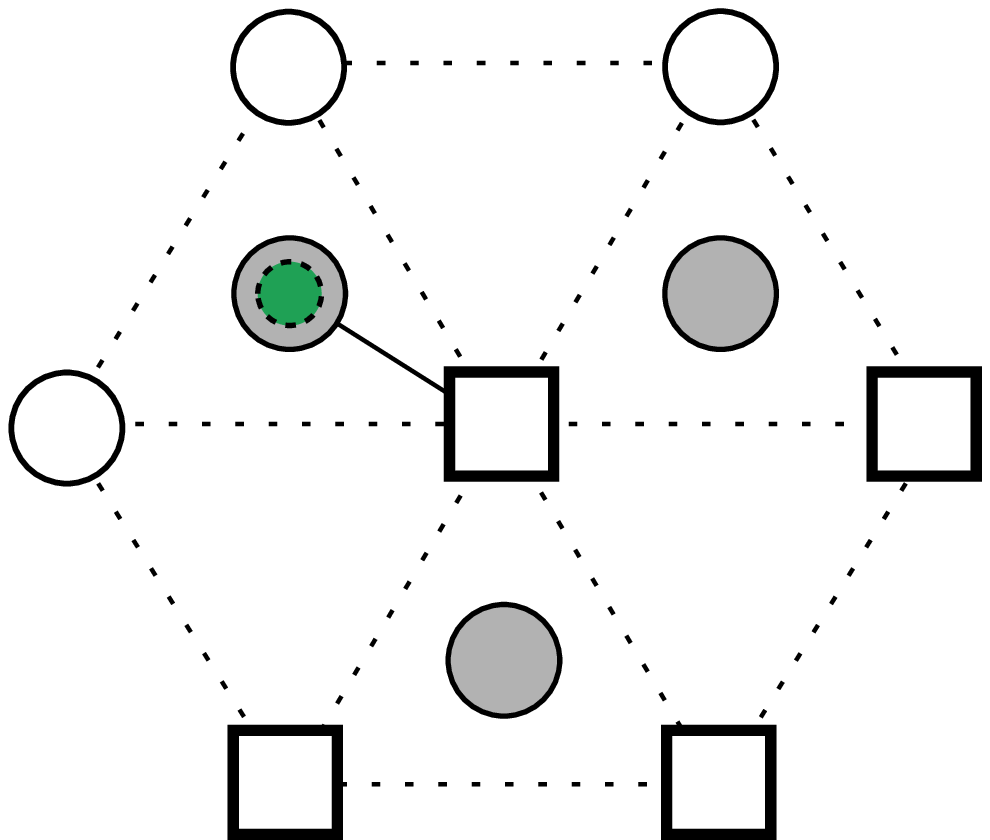} 
 & \includegraphics[scale=0.12]{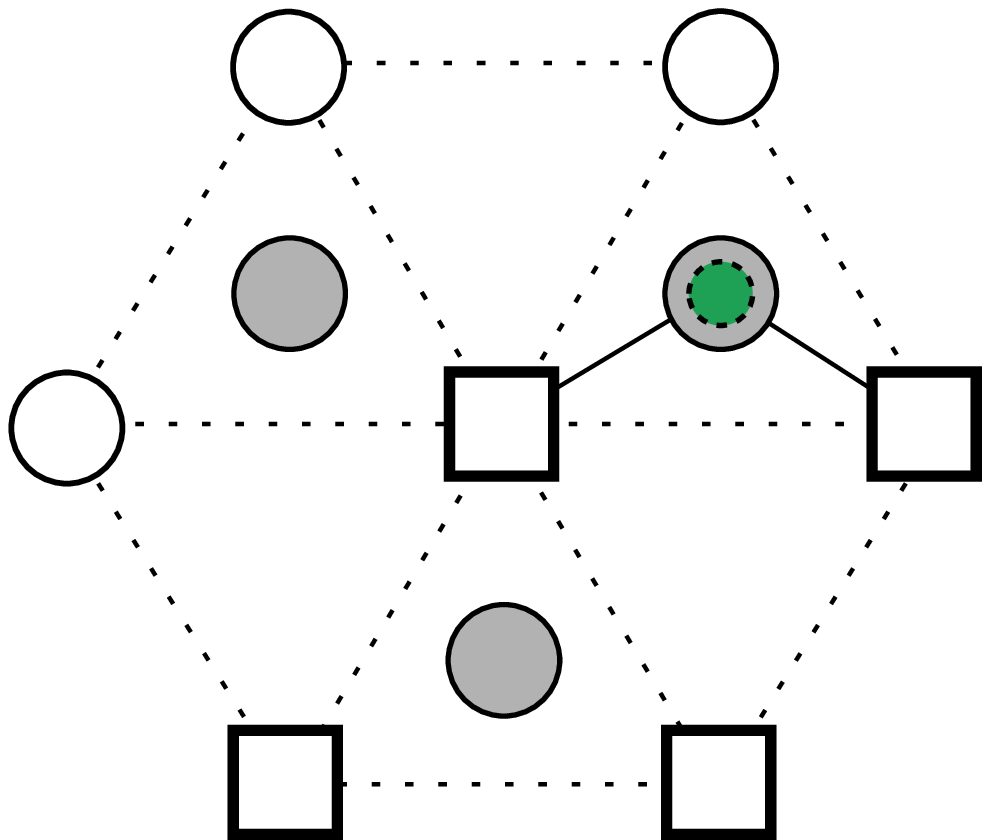}
 & \includegraphics[scale=0.12]{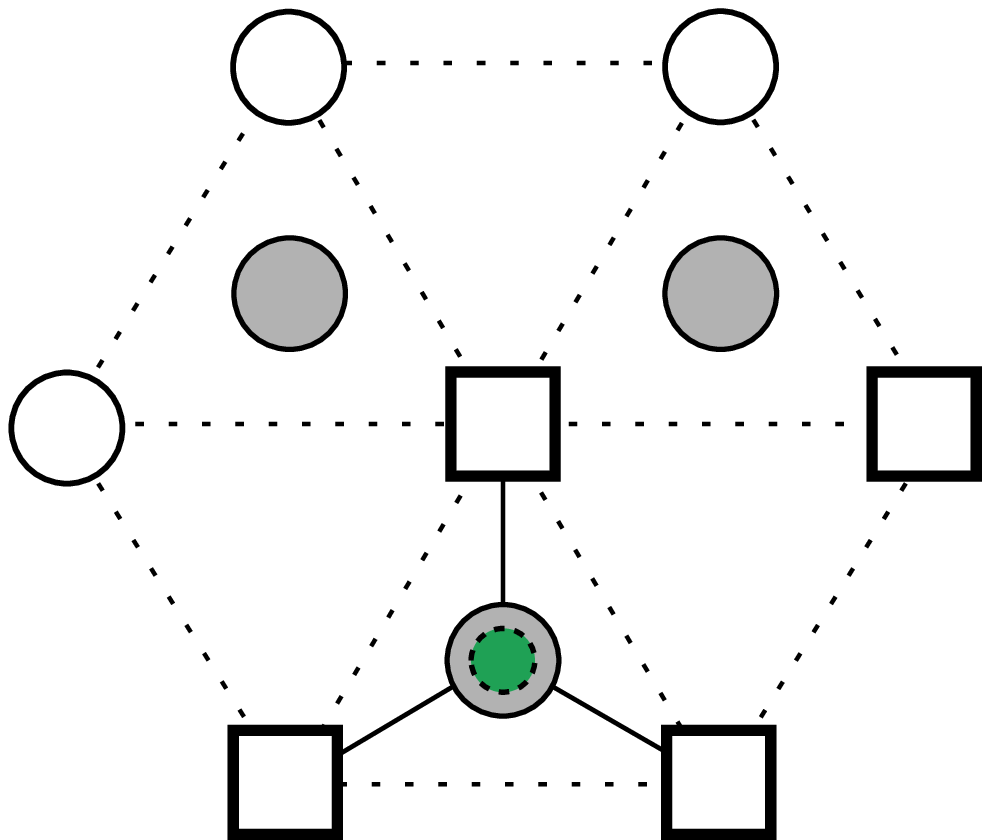} \\ 
Basal & & $0.24$~eV & $0.14$~eV & $0.06$~eV \\ 	
&& \includegraphics[scale=0.12]{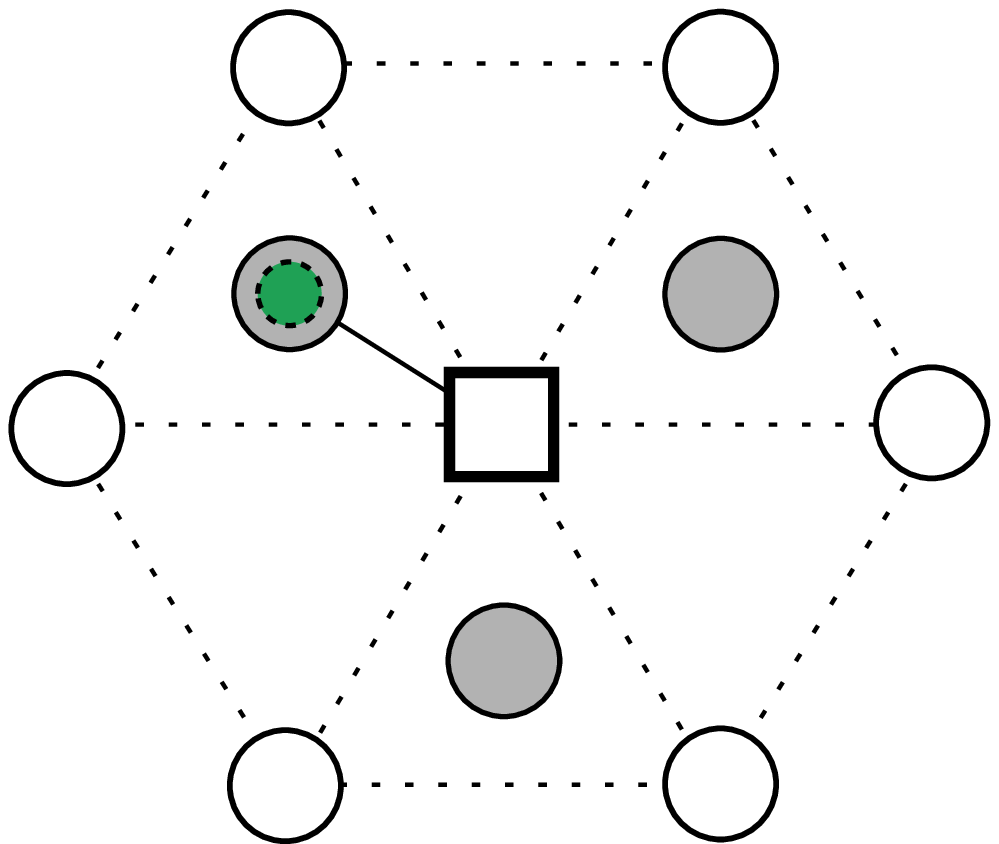}
 & \includegraphics[scale=0.12]{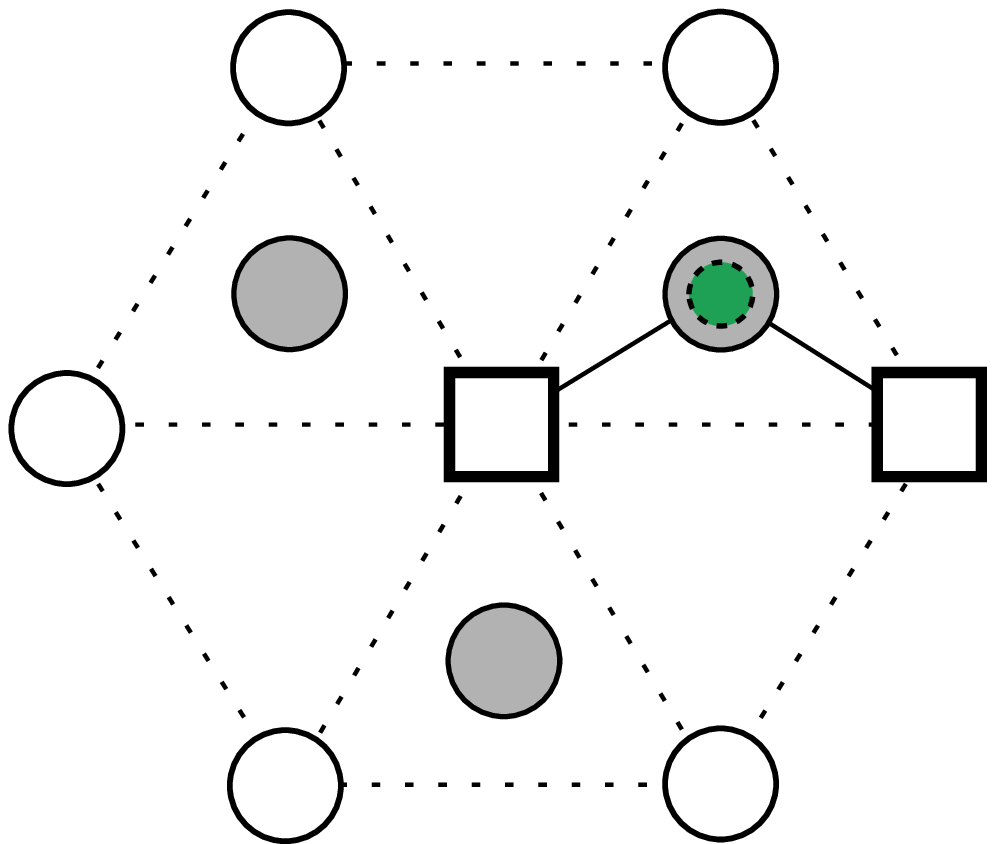}
 & \includegraphics[scale=0.12]{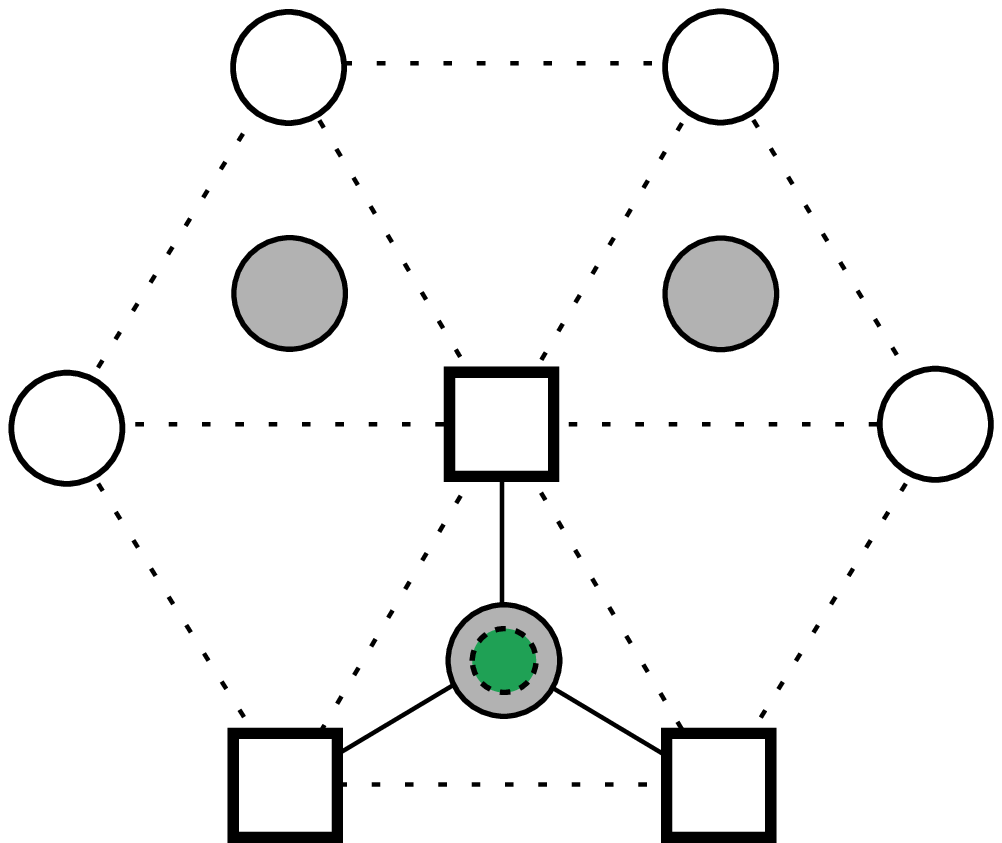} \\ 
 && $0.22$~eV & $0.20$~eV & $0.10$~eV \\ 	
\hline
&& \includegraphics[scale=0.12]{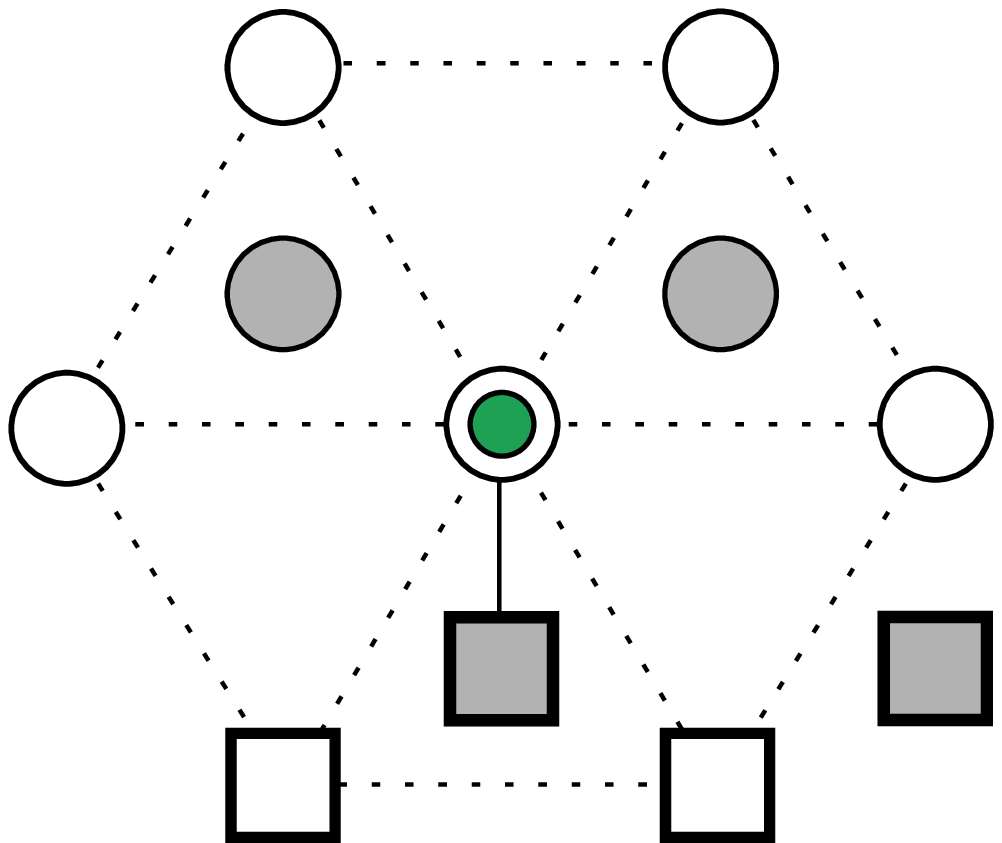} 
 & \includegraphics[scale=0.12]{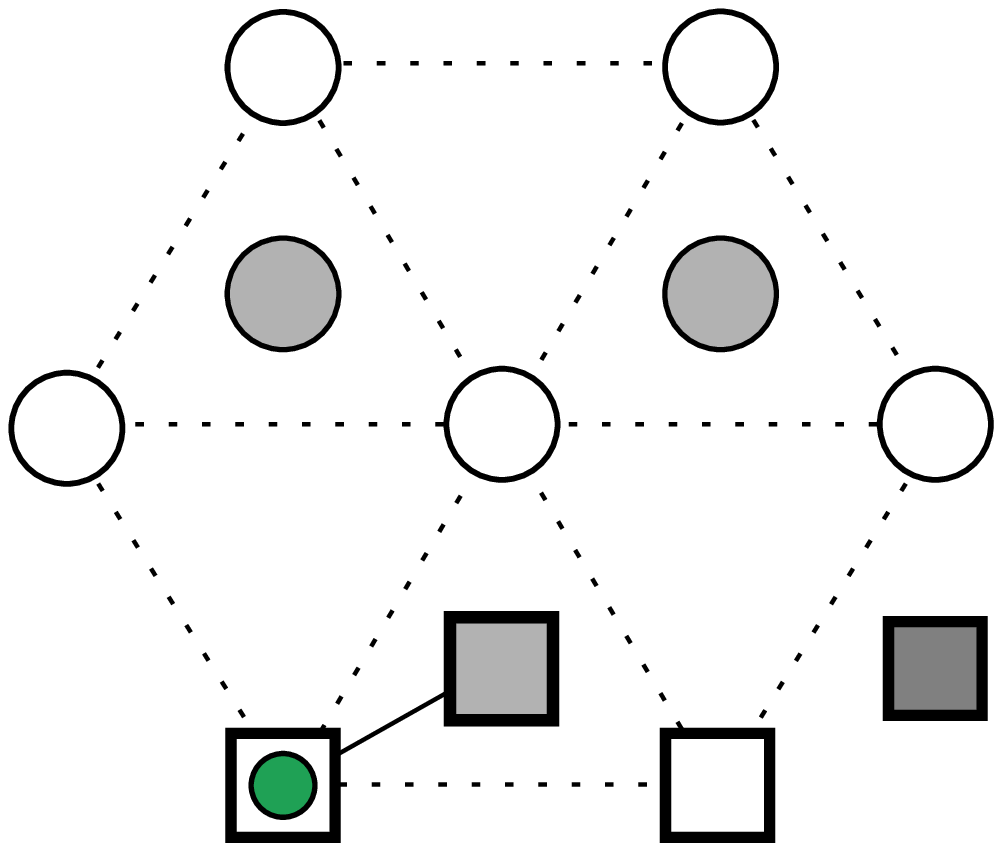}
 & \includegraphics[scale=0.12]{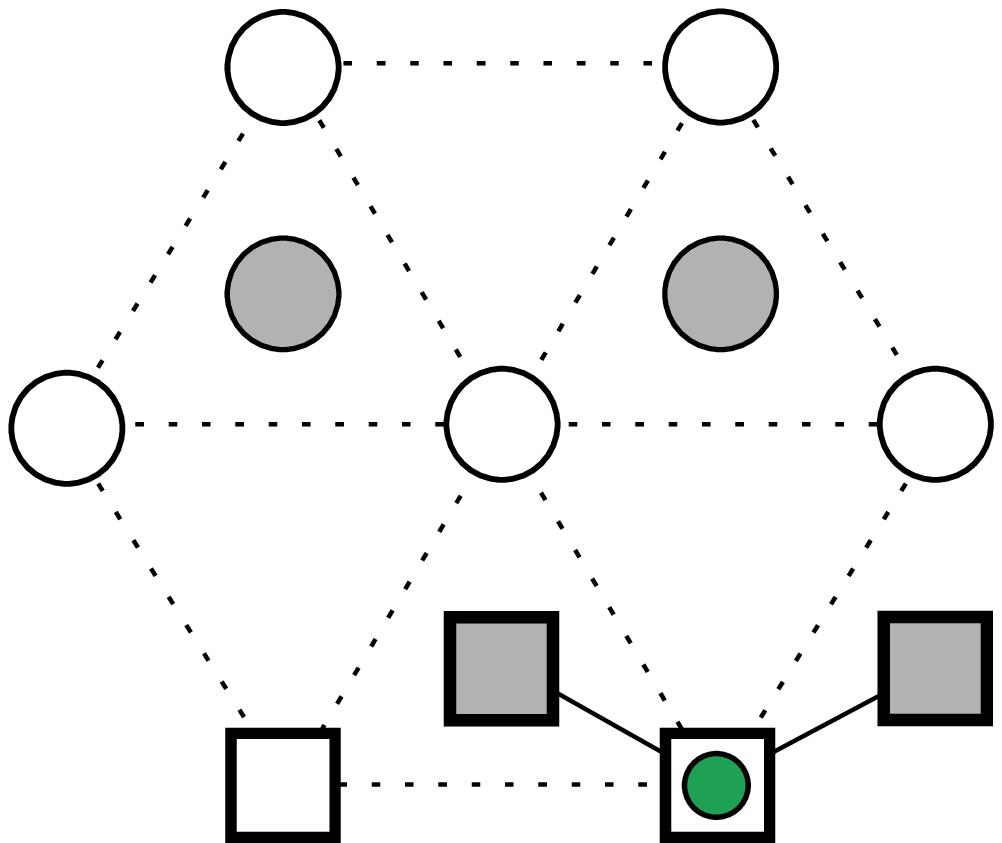} \\ 
Prismatic && $0.29$~eV & unstable & unstable \\ 	
&& \includegraphics[scale=0.12]{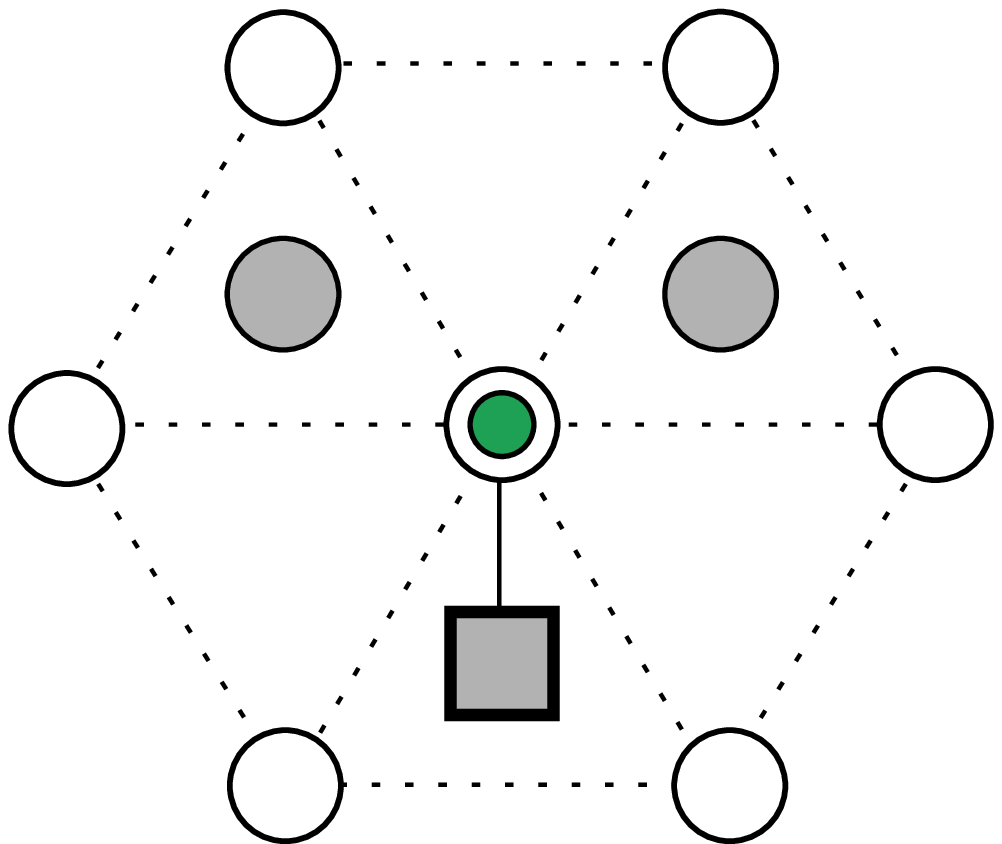}
 & \includegraphics[scale=0.12]{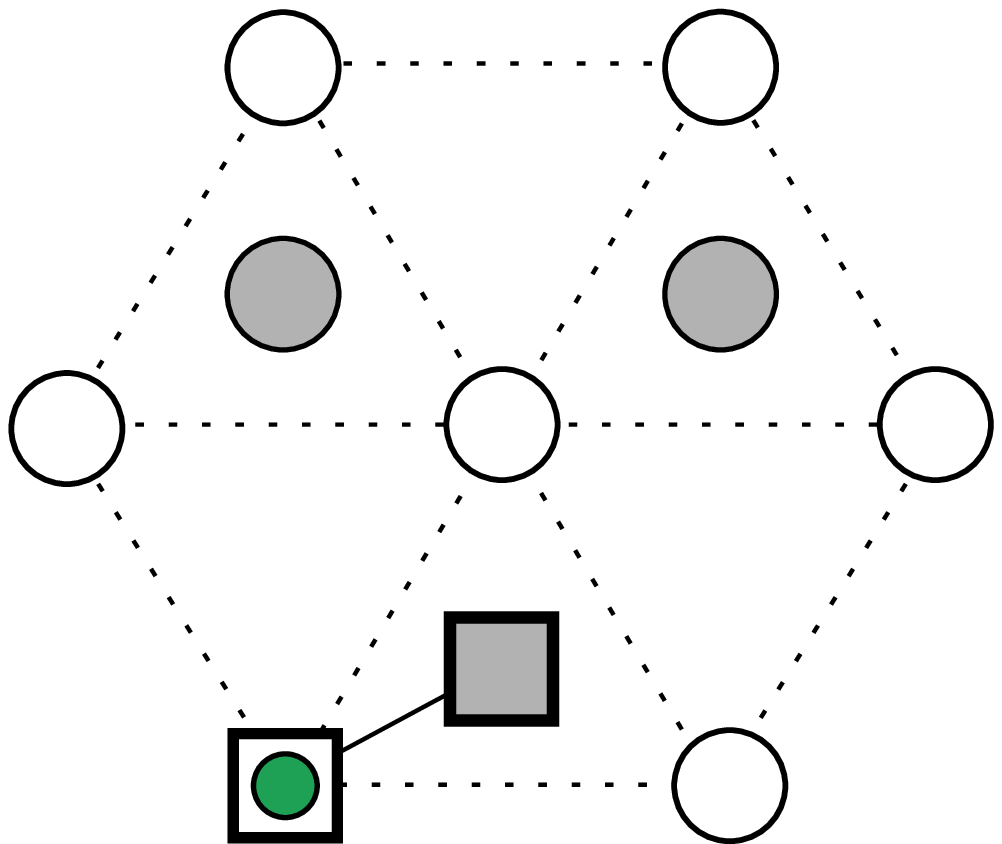}
 & \includegraphics[scale=0.12]{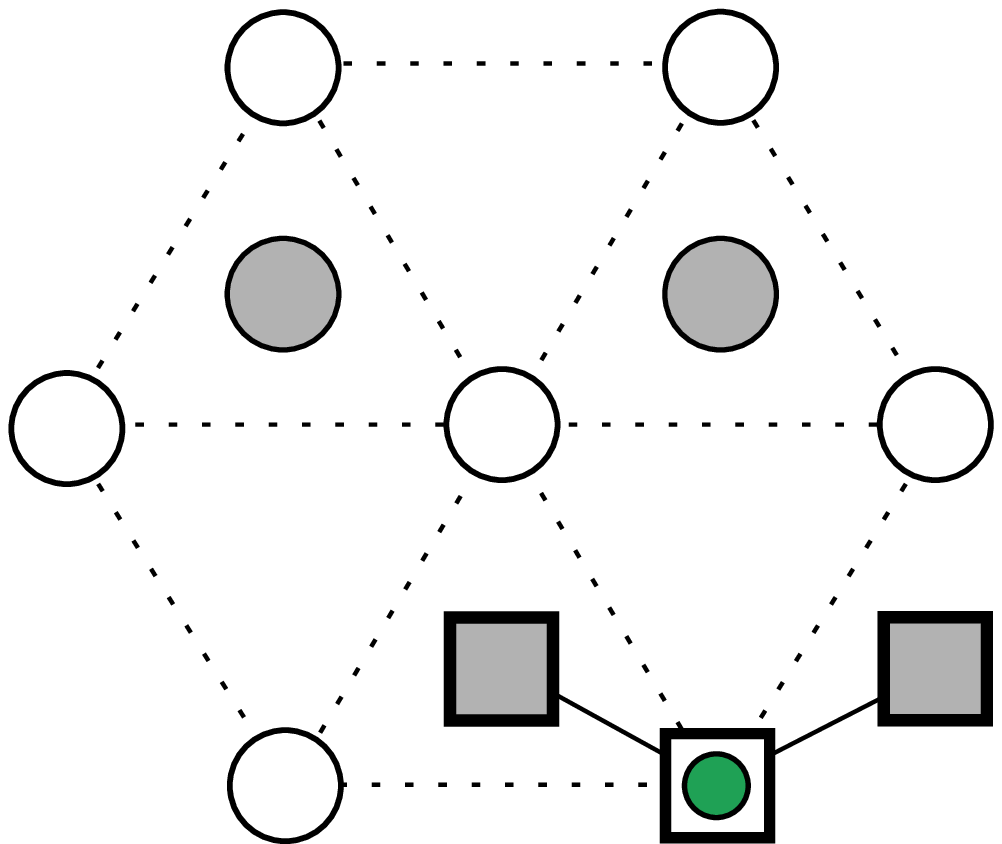} \\ 
 && $0.22$~eV & unstable & unstable \\ 	
\hline
O site & $p$ & 1  & 2  & 3  \\ 
\hline
&& \includegraphics[scale=0.12]{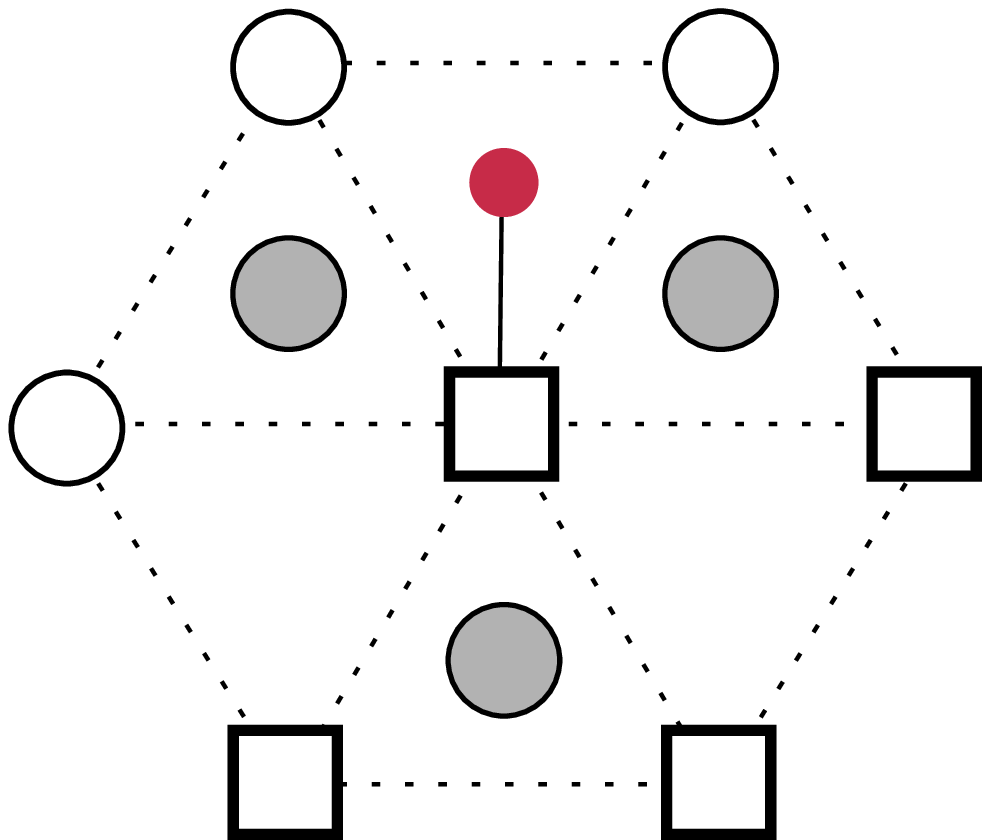} 
 & \includegraphics[scale=0.12]{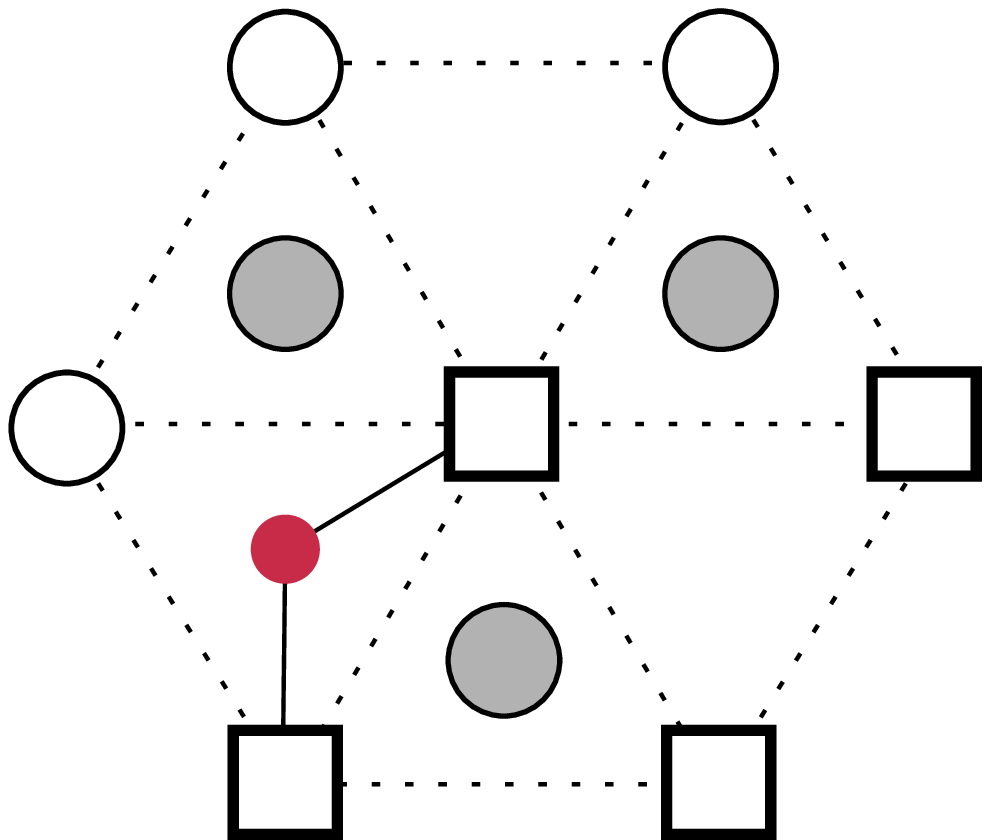}
 & \includegraphics[scale=0.12]{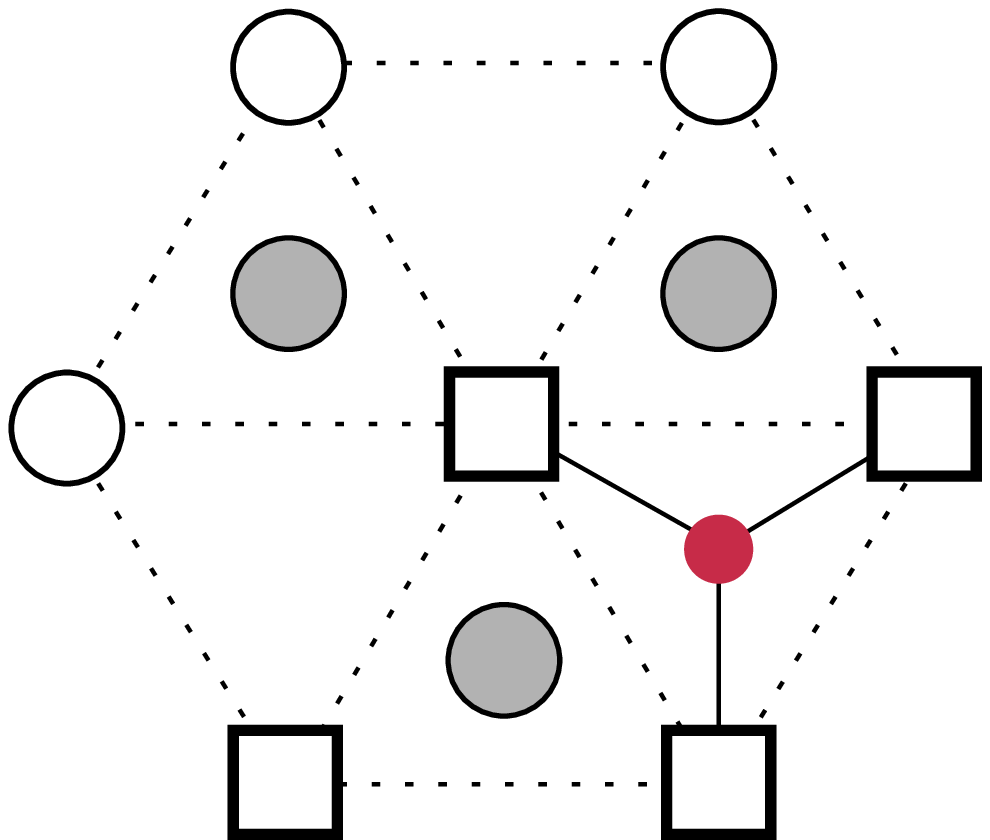} \\
Basal && $0.17$~eV & $0.41$~eV & $0.40$~eV \\ 	
&& \includegraphics[scale=0.12]{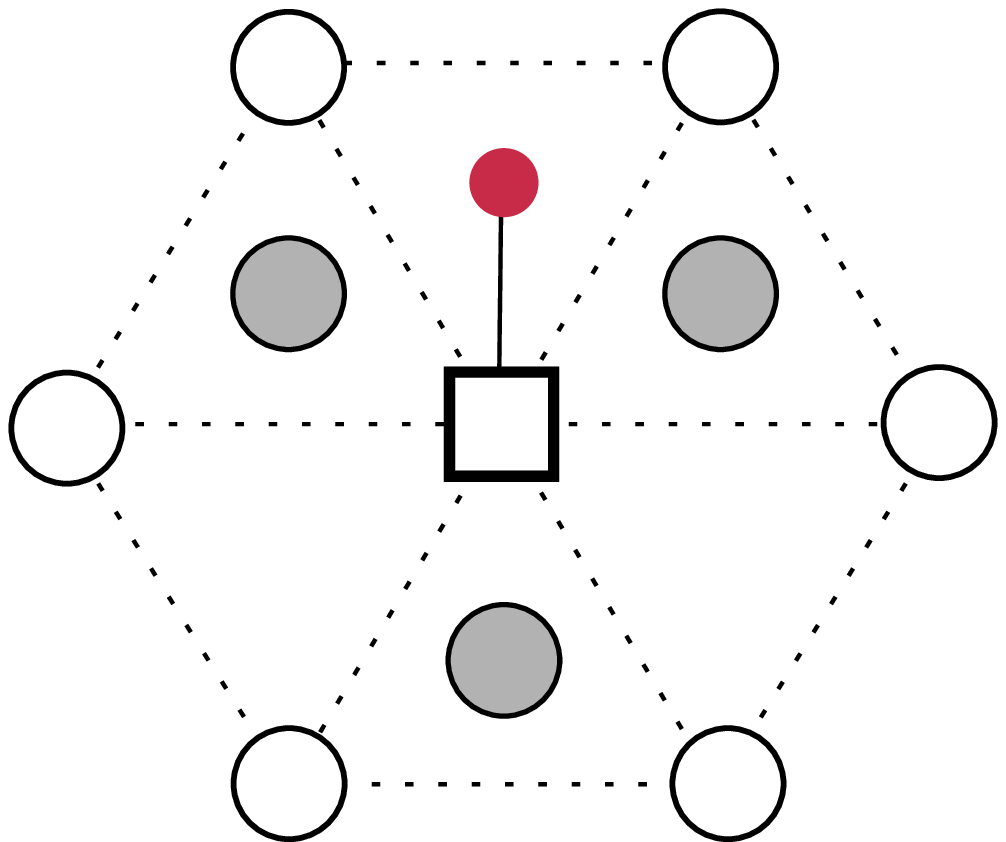}
 & \includegraphics[scale=0.12]{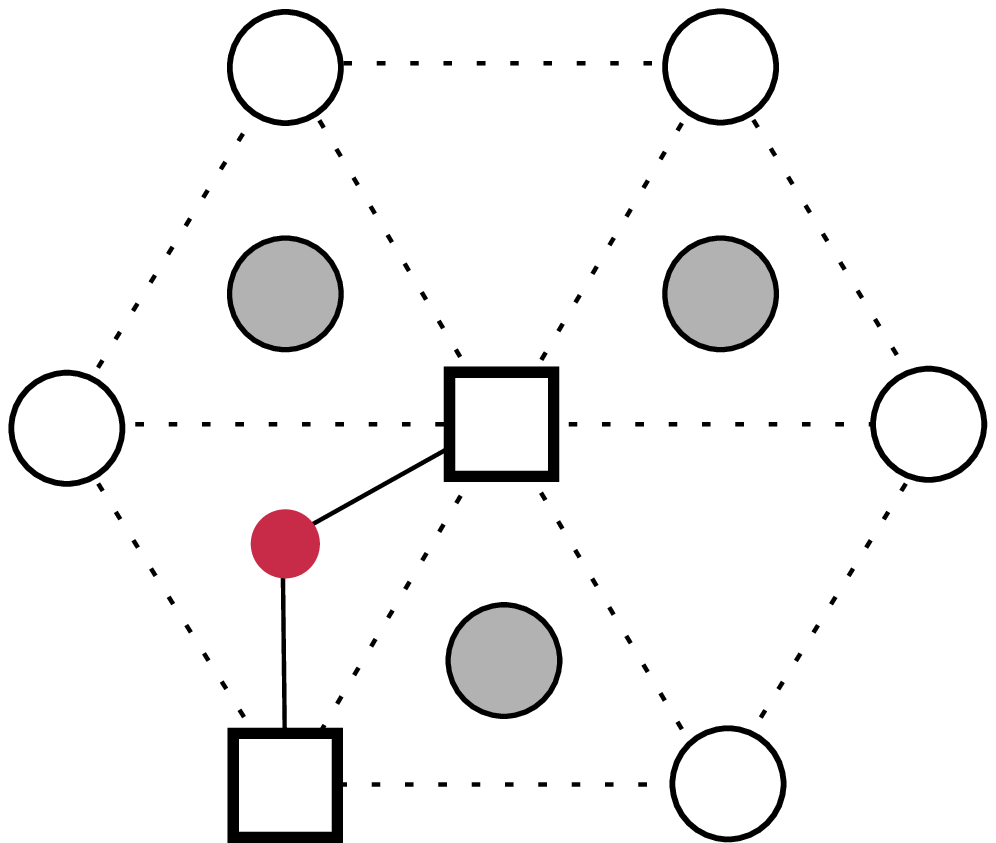}
 & \includegraphics[scale=0.12]{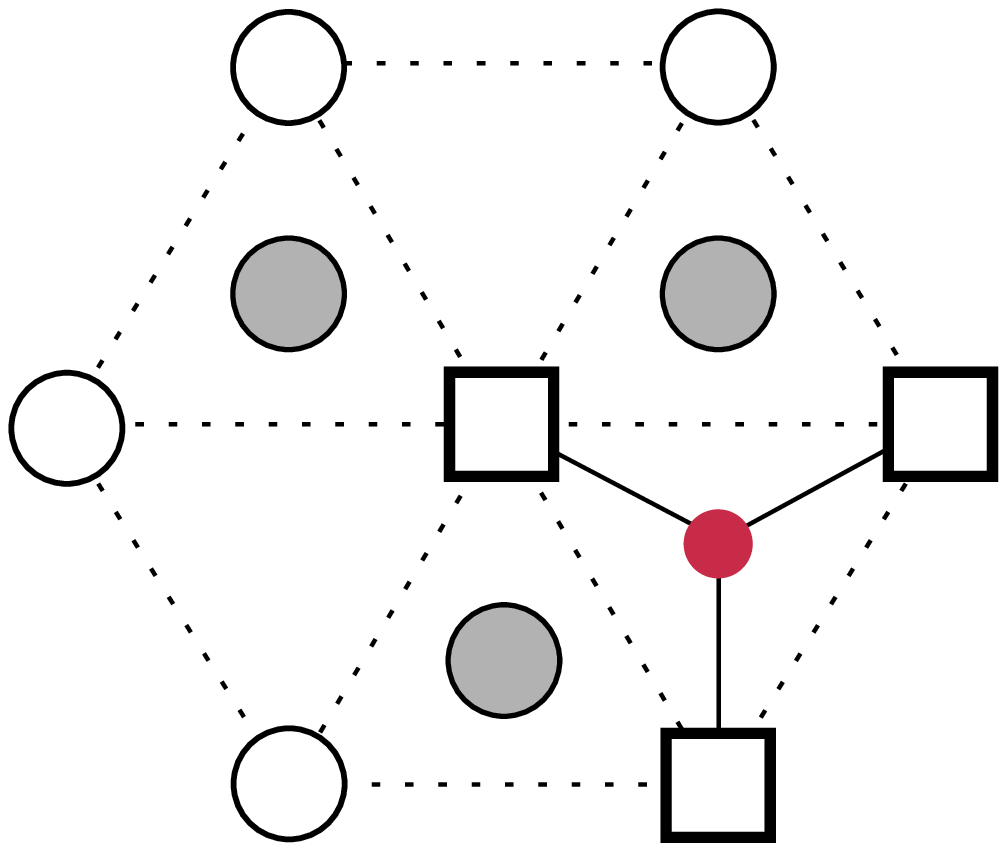} \\
 && $0.19$~eV & $0.42$~eV & $0.39$~eV \\ 	
\hline
&& \includegraphics[scale=0.12]{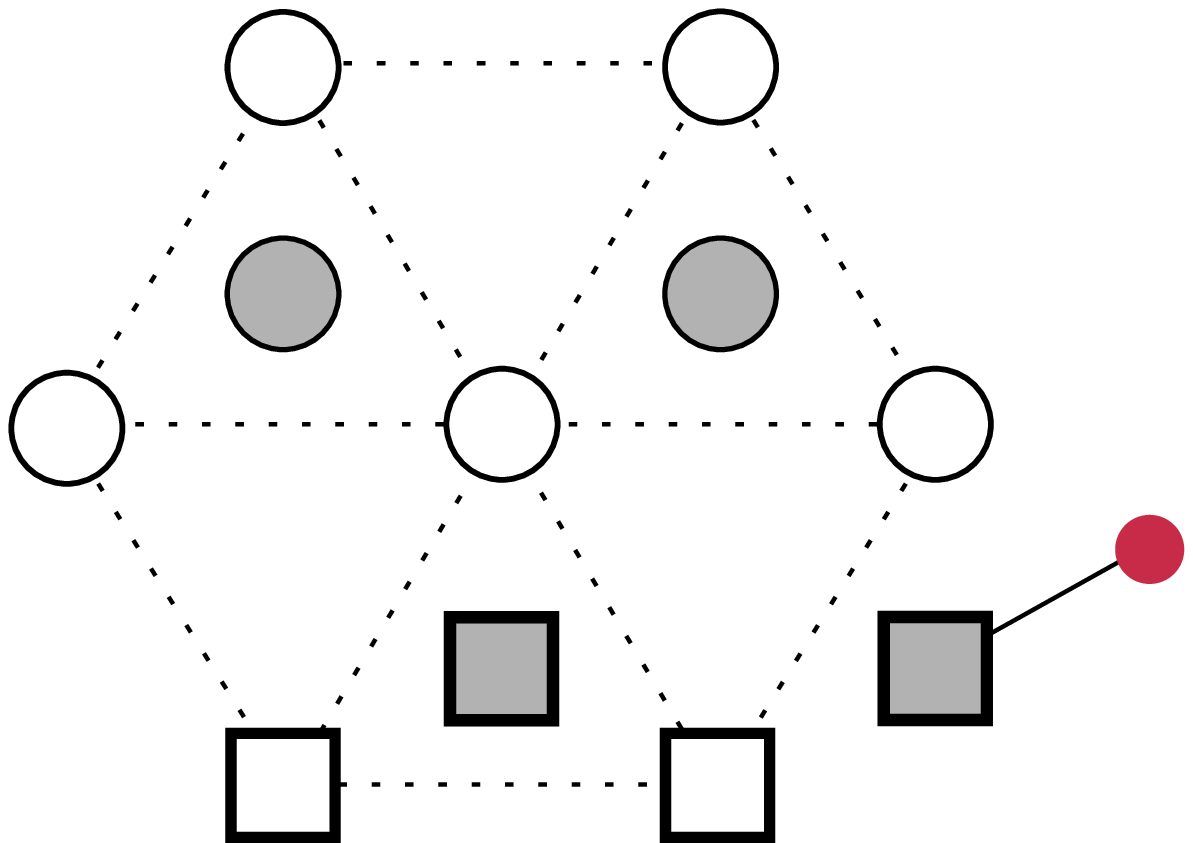} 
 & \includegraphics[scale=0.12]{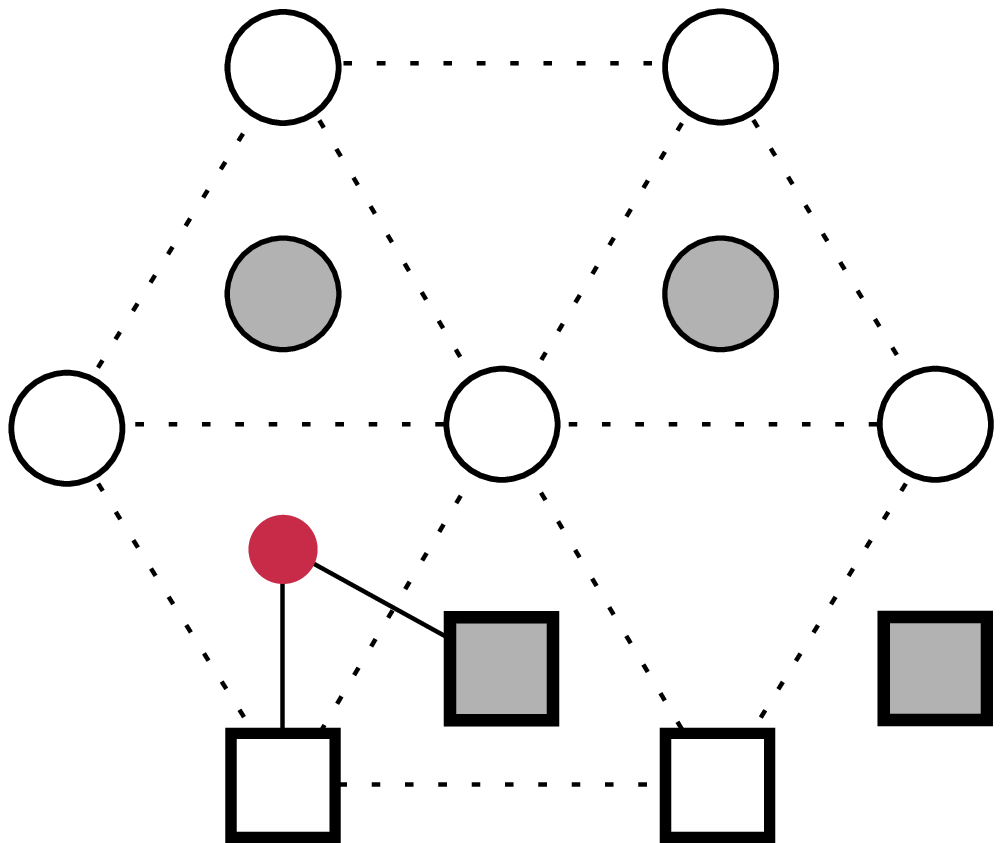}
 & \includegraphics[scale=0.12]{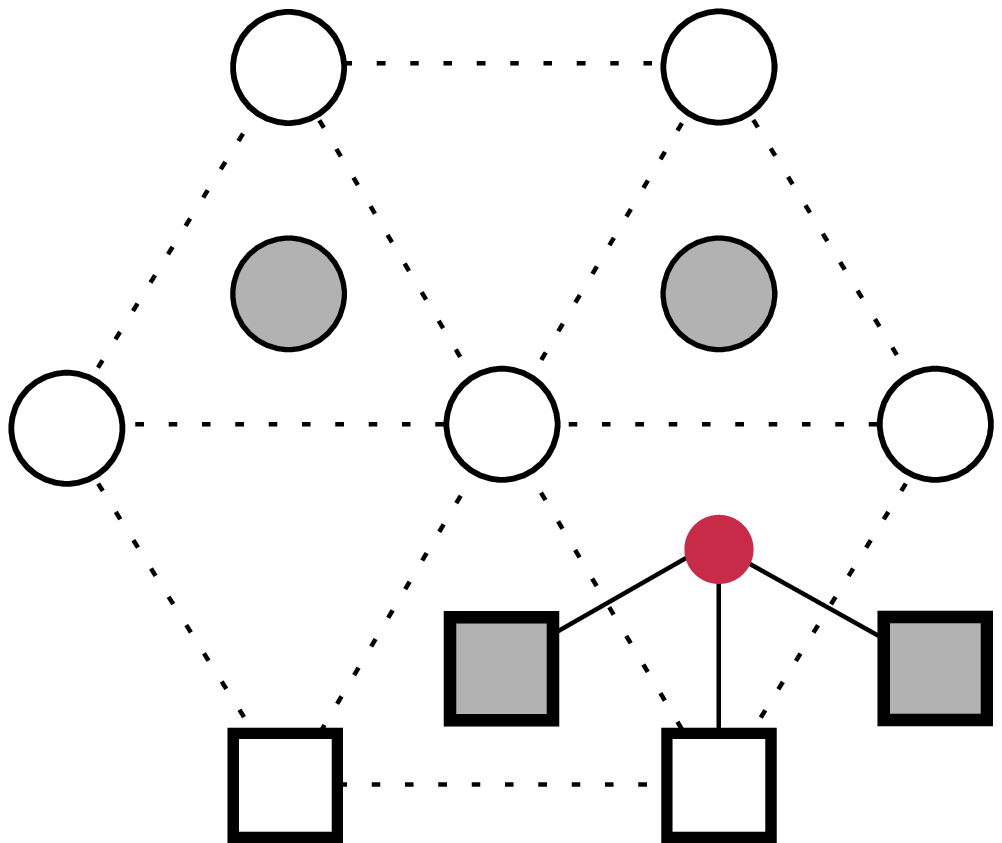} \\
Prismatic && $0.21$~eV & $0.37$~eV & $0.45$~eV \\ 	
&& \includegraphics[scale=0.12]{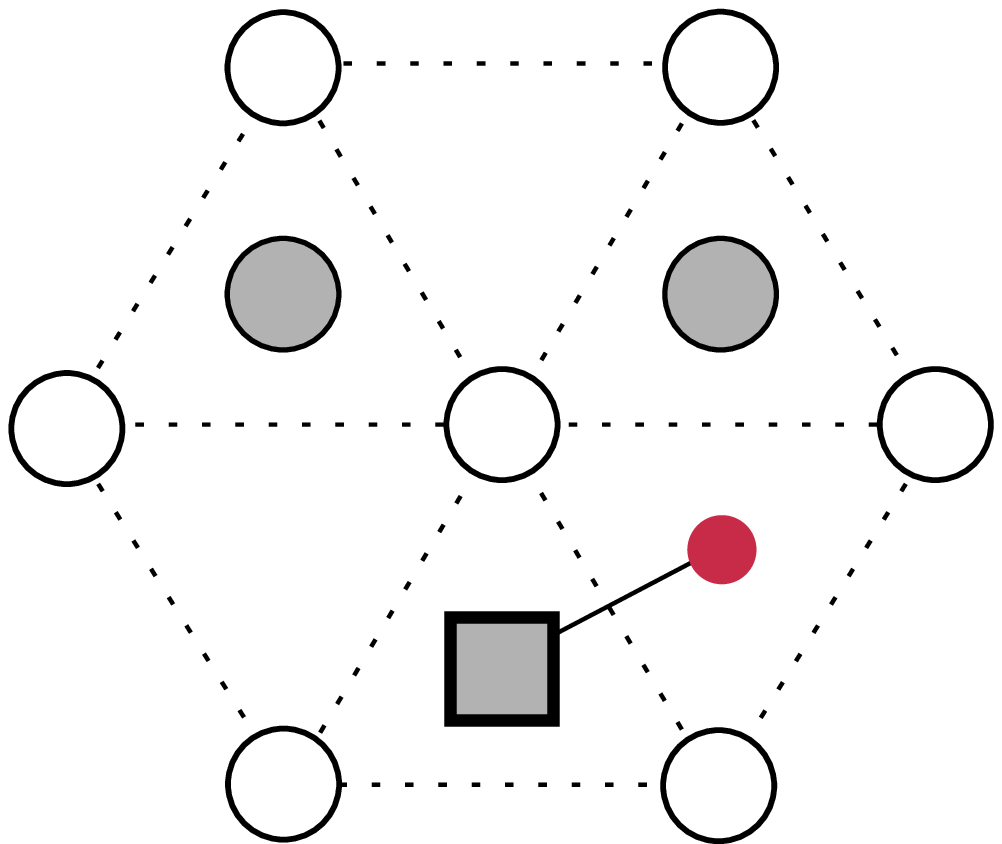}
 & \includegraphics[scale=0.12]{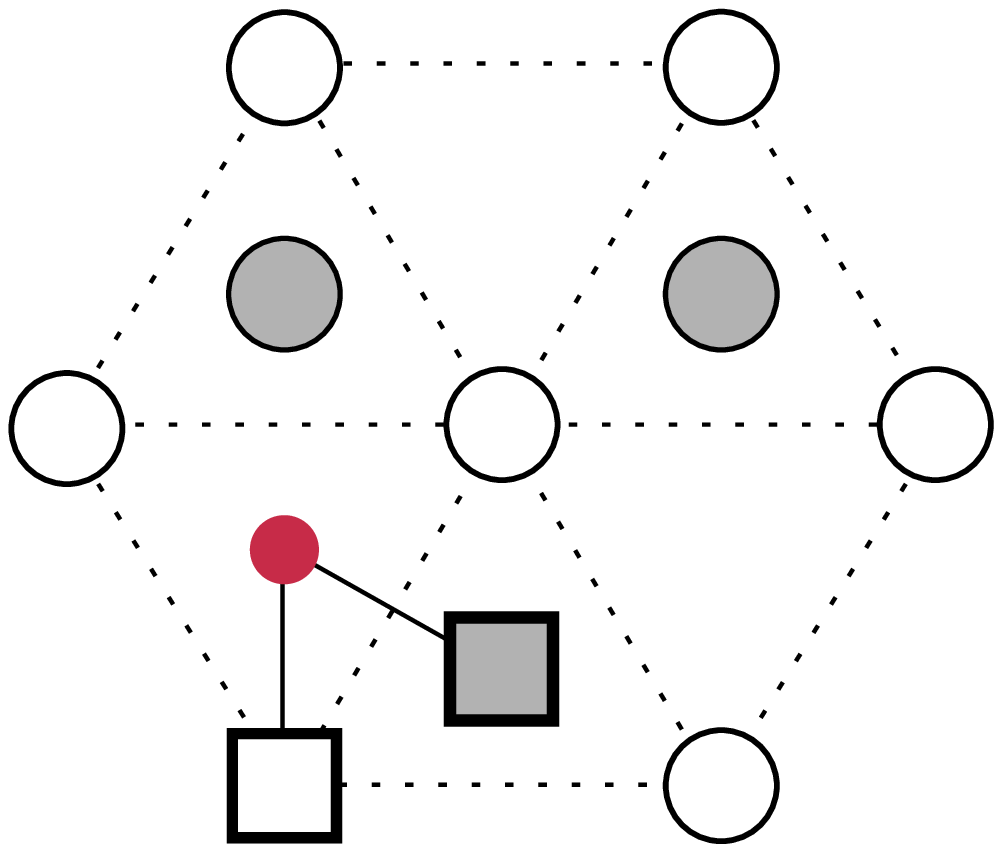}
 & \includegraphics[scale=0.12]{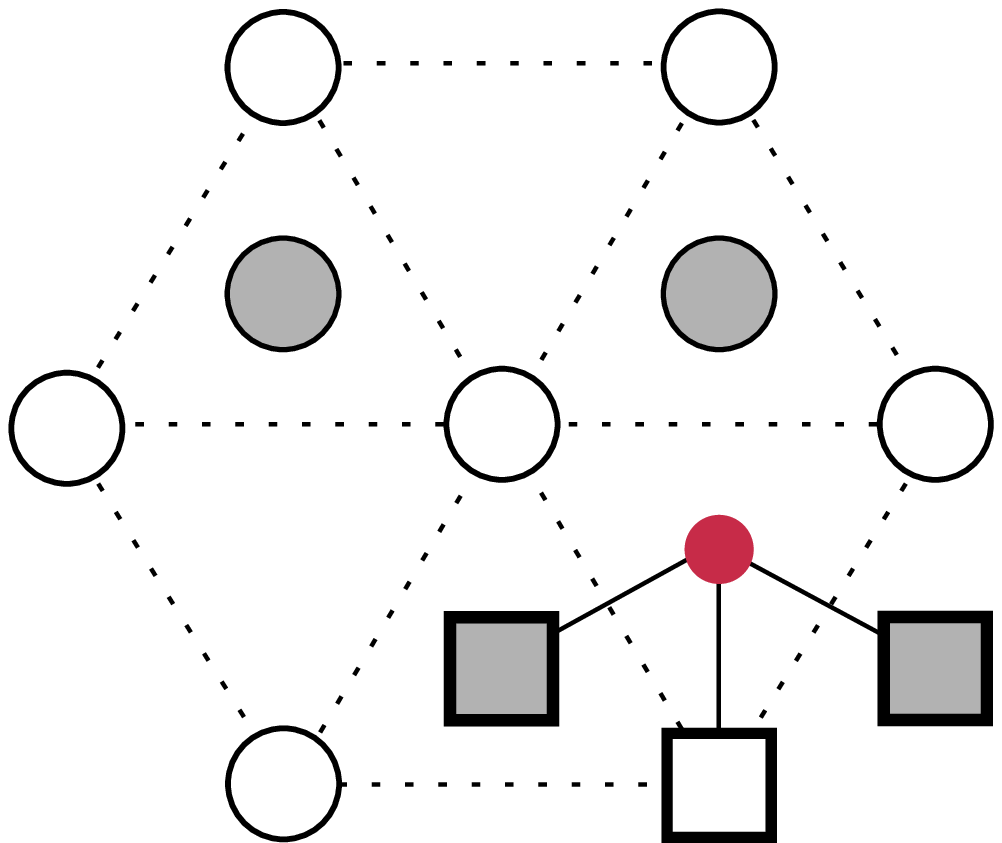} \\
 && $0.19$~eV & $0.37$~eV & $0.44$~eV \\ 		
\hline
\end{tabular}
\end{center}
\end{table}

When H is inserted into a T site, assuming that the number and the nature of nearest neighbor vacancies surrounding the H atom determines the stability of the cluster provides a rather good approximation.
First, the instabilities are well predicted by this simple approach (Tab.~\ref{tab:valid_modele1nnV4_TO}), and the final configurations have the same structure after atomic relaxations. 
Second, the differences between $E^{\rm b}_{\rm HV_4}$ and the corresponding $E^{\rm b}_{\rm HV_p}$
are smaller than $0.07$\,eV. 
For H in O site, the agreement between $E^{\rm b}_{\rm HV_4}$ and $E^{\rm b}_{\rm HV_p}$ is even better (Tab.~\ref{tab:valid_modele1nnV4_TO}). 
The obtained values for the clusters HV$_4$ and HV$_p$ with $p$ nearest neighbor interactions are very close, within less than $0.02$~eV in this case. 

This study confirms that vacancies farther than nearest neighbor positions have only a weak influence on the binding energy value. 
This latter is mainly fixed by the number and the position of vacancies that are nearest neighbors of the H atom.
Again, we note on the HV$_4$ clusters that the binding energies for H in O site with $p=2$ or $3$ are higher than the binding energies for H in T site.

\subsubsection{Influence of H on small clusters}

We finally examine the stability properties of the HV$_n$ clusters for $2 \le n \le 7$. 
We separate vacancy clusters into three different groups : 
\begin{itemize}
	\item basal clusters, where all vacancies are lying in the same basal plane. 
		These clusters can be seen as precursors of $\langle c \rangle$ loops.
	\item prismatic clusters, where all vacancies are lying in the same prismatic corrugated plane. 
		These clusters can be seen as precursors of $\langle a \rangle$ loops.
	\item 3D clusters (precursors of cavities). 
\end{itemize}

In the absence of hydrogen, we have shown in a previous study \cite{Varvenne2014} that the 3D small vacancy clusters are more stable than the plane vacancy clusters, until $7$ vacancies. 
As stated in our introduction, under irradiation, the amount of $\left\langle c\right\rangle$ vacancy loops increases when zirconium samples are pre-hydrided \cite{Tournadre2013,Tournadre2014}. 
This could suggest that H in solid solution affects the relative stability of the different types of vacancy clusters. 
Studying the binding properties of H with small V$_n$ clusters of different types should give some insight about the validity of such an assumption.  

\begin{figure}
\includegraphics[scale=0.63]{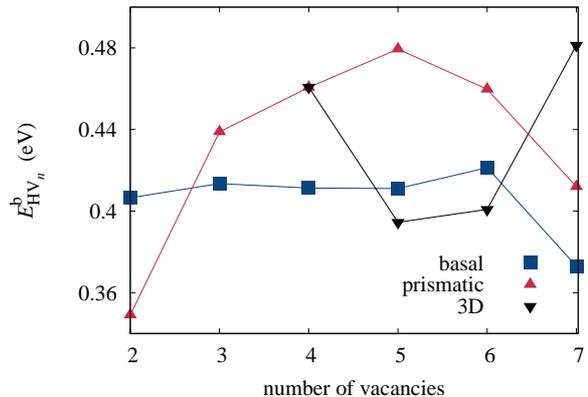}
\caption{Maximum binding energies of H with vacancy clusters containing between $2$ and $7$ vacancies. 
Three types of clusters are considered: basal, prismatic and 3D clusters.}
\label{fig:comp_amas_diff}
\end{figure}

We proceed as follows. The chosen vacancy clusters V$_n$, with $n=2$ to $7$, correspond to the most stable ones, previously found in pure Zr \cite{Varvenne2014}.  For each cluster, the previously identified insertion sites for H are tested. The most stable one is retained and we plot the binding energies versus the cluster size for each group of cluster (see Fig.~\ref{fig:comp_amas_diff}). 
The resulting binding energies are all positive, and are included in a range from $0.35$ to $0.48$~eV.

No clear preference is seen for any type of vacancy cluster. This is not surprising, since, as established above, H mainly interacts with its first nearest neighbor vacancies.
H therefore does not have any discriminating effect on the small vacancy clusters.  

\section{Hydrogen interaction with stacking faults}

In hcp Zr, vacancy clustering under irradiation leads to planar extended defects: $\left\langle a \right\rangle$ dislocation loops, lying in planes close to the prismatic planes, and $\left\langle c \right\rangle$ component loops, lying in the basal planes.
Cavities are hardly ever observed, in agreement with their lower stability \cite{Varvenne2014},
and will not be considered in the following.
The $\left\langle a \right\rangle$ loops are the most stable defects \cite{Varvenne2014}.
They are faulted at small sizes and perfect at large sizes. 
The $\left\langle c \right\rangle$ loops on the other hand are always faulted, 
with an extrinsic basal fault at small sizes and an intrinsic I$_1$ fault at large size.
These $\langle c \rangle$ loops are believed to play a crucial role in the breakaway growth
at high irradiation dose.

In order to determine the influence of H solutes on vacancy dislocation loops, we perform a careful study of their binding energies with the different stacking faults involved in the extended defects of interest. 
We assume that this interaction with the stacking fault is the main H contribution 
on the energetics of vacancy loops.

\subsection{Prismatic stacking fault}

When removing a vacancy platelet in a corrugated $\{10\bar{1}0\}$ plane, a prismatic stacking fault is formed \cite{Varvenne2014}. This stacking fault is associated with the formation of faulted $\left\langle a \right\rangle$ loops of Burgers vector $\vec{b}= 1/2\left\langle10\bar{1}0\right\rangle$. The unfaulting of the vacancy loop occurs by a $1/6\left\langle 1\bar{2}10\right\rangle$ shearing of the fault plane, leading to a perfect  $\left\langle a \right\rangle$ loop of Burgers vector $\vec{b}= 1/3\left\langle 2\bar{1}\bar{1}0\right\rangle$. We investigate here the interaction of an H atom with this intrinsic stacking fault. 

\begin{figure} 
\centering
\begin{tabular}{c}
\subfigure[H insertion sites]{\label{fig:Eb_prism_SF_1}\includegraphics[scale=0.12]{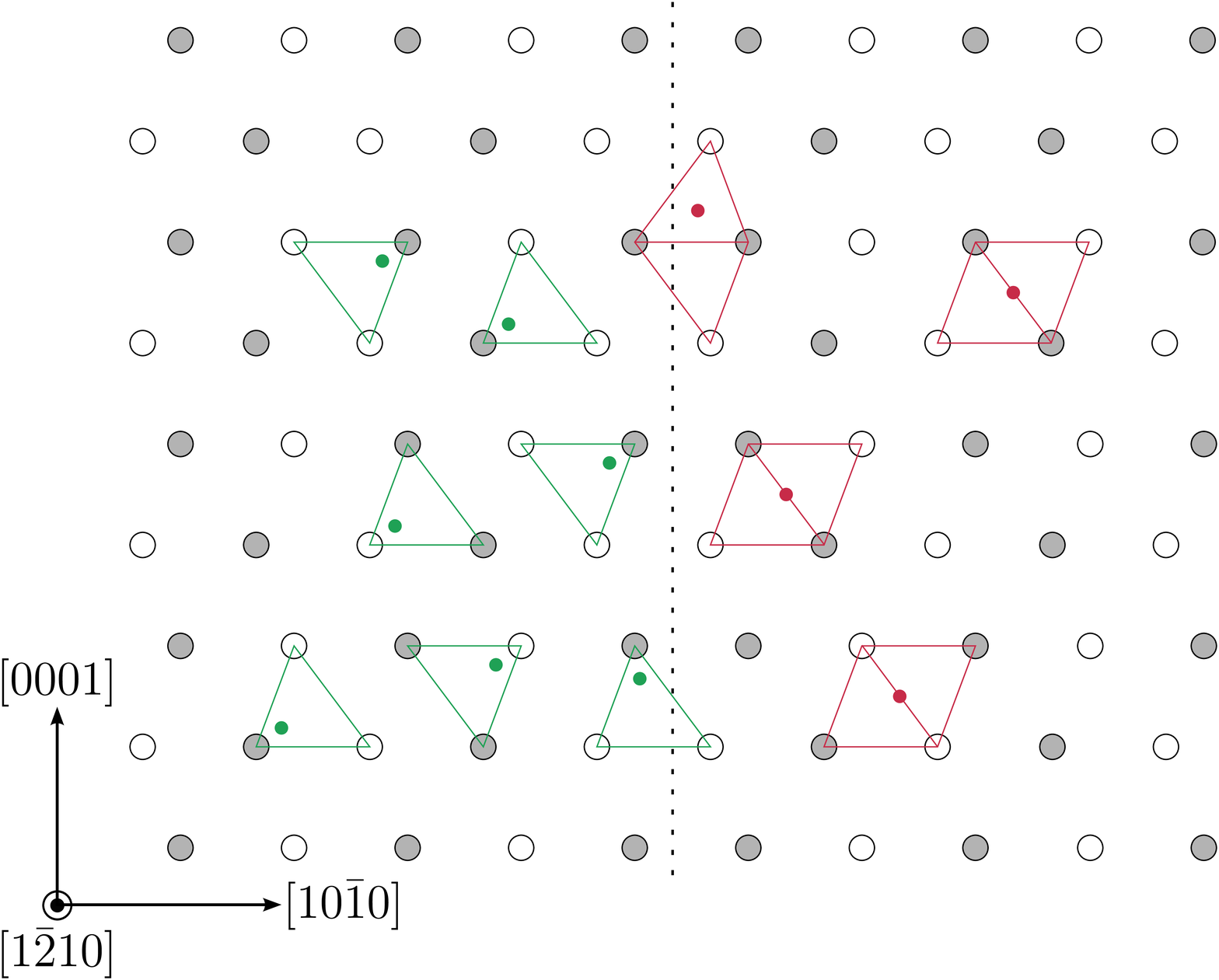}} \\
\subfigure[Binding energy without H vibration]{\label{fig:Eb_prism_SF_2}\includegraphics[scale=0.6]{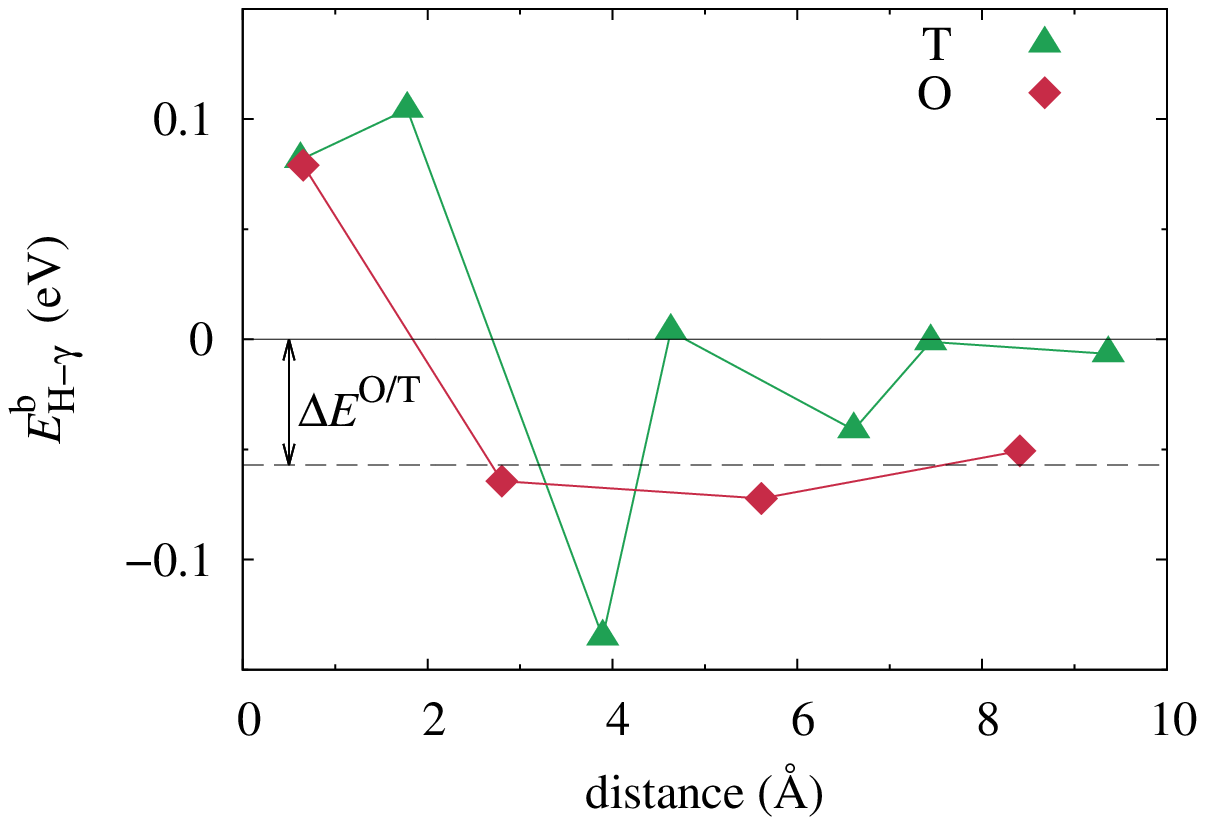}} \\
\subfigure[Binding energy with H vibration]{\label{fig:Eb_prism_SF_3}\includegraphics[scale=0.6]{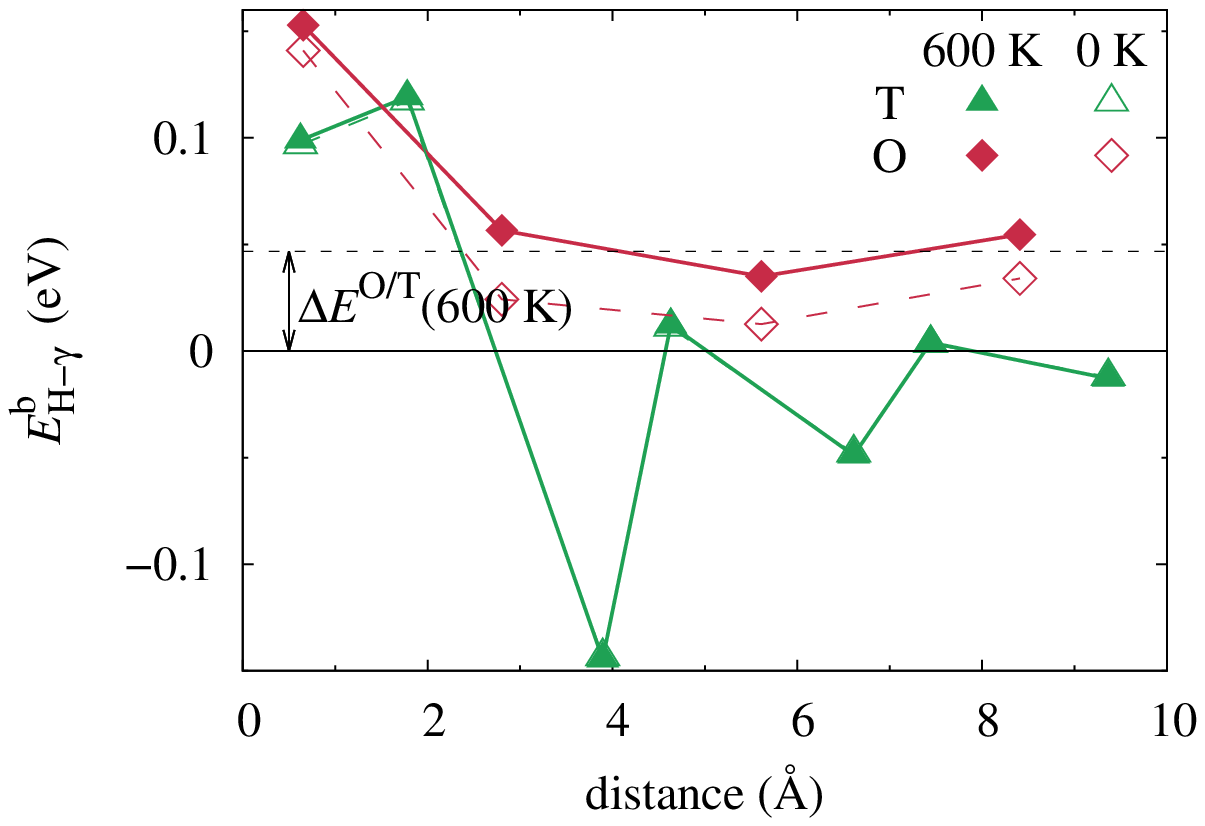}}
\end{tabular}
\caption{Interaction between H and a prismatic stacking fault.
(a) Projected positions in the $[1\bar{2}10]$ plane of the investigated T and O sites for the insertion of H in the vicinity of the stacking fault (vertical dashed line).
Zr atoms are shown by white and grey circles, depending on their position along the $[1\bar{2}10]$ direction in the faulted crystal.
(b) and (c) Binding energies $E^{\rm b}_{\textrm{H}-\gamma}$ between H and the prismatic stacking fault, versus their initial separation distance, 
calculated (b) without and (c) with H vibration free energy.}
\label{fig:Eb_prism_SF}
\end{figure}

Calculations are performed in a simulation box duplicated 4 times in the $[10\bar{1}0]$ direction, 2 times in the $[0001]$ direction, and 4 times in the $[1\bar{2}10]$ direction, \ie in the direction perpendicular to the fault plane. This corresponds to a total of 128 Zr atoms. The use of this simulation box to model a prismatic stacking fault has been already validated \cite{Clouet2012}. It is also large enough to model an isolated H atom: with such a simulation box, the energy difference between the O and T insertion sites is $\Delta E^{\rm O/T}=57$\,meV in the perfect crystal, in good agreement with the value obtained in conventional supercells (Tab. \ref{tab:Zr_H_prop}). 

Different insertion positions for the H atom, lying either in a T or an O interstitial site 
at varying distances from the stacking fault, are considered 
(Fig.~\ref{fig:Eb_prism_SF}a). 
\Abinitio calculations lead to a binding of the hydrogen with the prismatic stacking fault (Fig.~\ref{fig:Eb_prism_SF}b and c), 
with a maximal binding energy of $0.1$~eV. The most attractive interaction is obtained 
when the H atom is one plane apart from the stacking fault into a T site.
This attractive interaction is in agreement with previous \abinitio studies, 
although our maximal binding energy is slightly lower than the one obtained 
by Domain \etal \cite{Domain2004} (0.14\,eV)
or by Udagawa \etal \cite{Udagawa2010} (0.16\,eV).
The binding energy rapidly decays when the distance between H and the stacking fault increases:
it goes to zero for the T site and to $-\Delta E^{\rm O/T}$ for the O site. 
The only octahedral site leading to an attractive interaction with the stacking fault 
is the closest to the fault plane. 
This octahedral site did not exist before the introduction of the stacking fault
and is created by the shearing of the lattice \cite{Ghazisaedi2014}. 
The position corresponding exactly to the octahedron position in this site is unstable.
For each such octahedral site, two stable positions inside each half-octahedron exist, where the H atom is linked to five Zr first nearest neighbors (Fig.~\ref{fig:Eb_prism_SF}a). For all other octahedral sites, as well as all the tetrahedral sites, the relaxed position of the H atom is close to the ideal position. 

The same energy shift as in the bulk, corresponding to a stabilization by vibrations
of the O sites compared to the T sites, is observed in presence of the stacking fault
(Fig. \ref{fig:Eb_prism_SF}c).
Apart from this energy shift, the H vibrations decrease the H interaction with the stacking fault for the O site,
whereas it increases this interaction for the T sites.

\subsection{Basal stacking faults} 

Condensation of vacancies in a basal plane results in
the creation of a dislocation loop 
of Burgers vector $\vec{b}_1=1/2\,[0001]$. 
This corresponds to the removal of a platelet of one atomic layer 
in the perfect stacking $BABABA$ of basal planes 
and leads to the formation of a highly energetic stacking sequence, 
$BAB.BABA$.
The stacking then evolves so as to lower the energy of the vacancy loop
by creating one of the two following stacking faults \cite{Hull2011}: 
an extrinsic fault E, with a $BABCABA$ stacking sequence
preserving the same Burgers vector, 
or an intrinsic fault I$_1$, with a $BABCBCB$ stacking sequence 
leading to a dislocation loop with Burgers vector $\vec{b}_2 = 1/6\,\langle 20\bar{2}3 \rangle$.
We investigate here the interaction of H with these two basal stacking faults,
which gives insights on the influence of this impurity on the stability of $\left\langle c \right\rangle$ loops. 

\begin{figure*}[!bth]
\centering
\begin{tabular}{c c}
\subfigure[H insertion sites (I$_1$ fault)]{\label{fig:scheme_basal_SF-a}\includegraphics[scale=0.13]{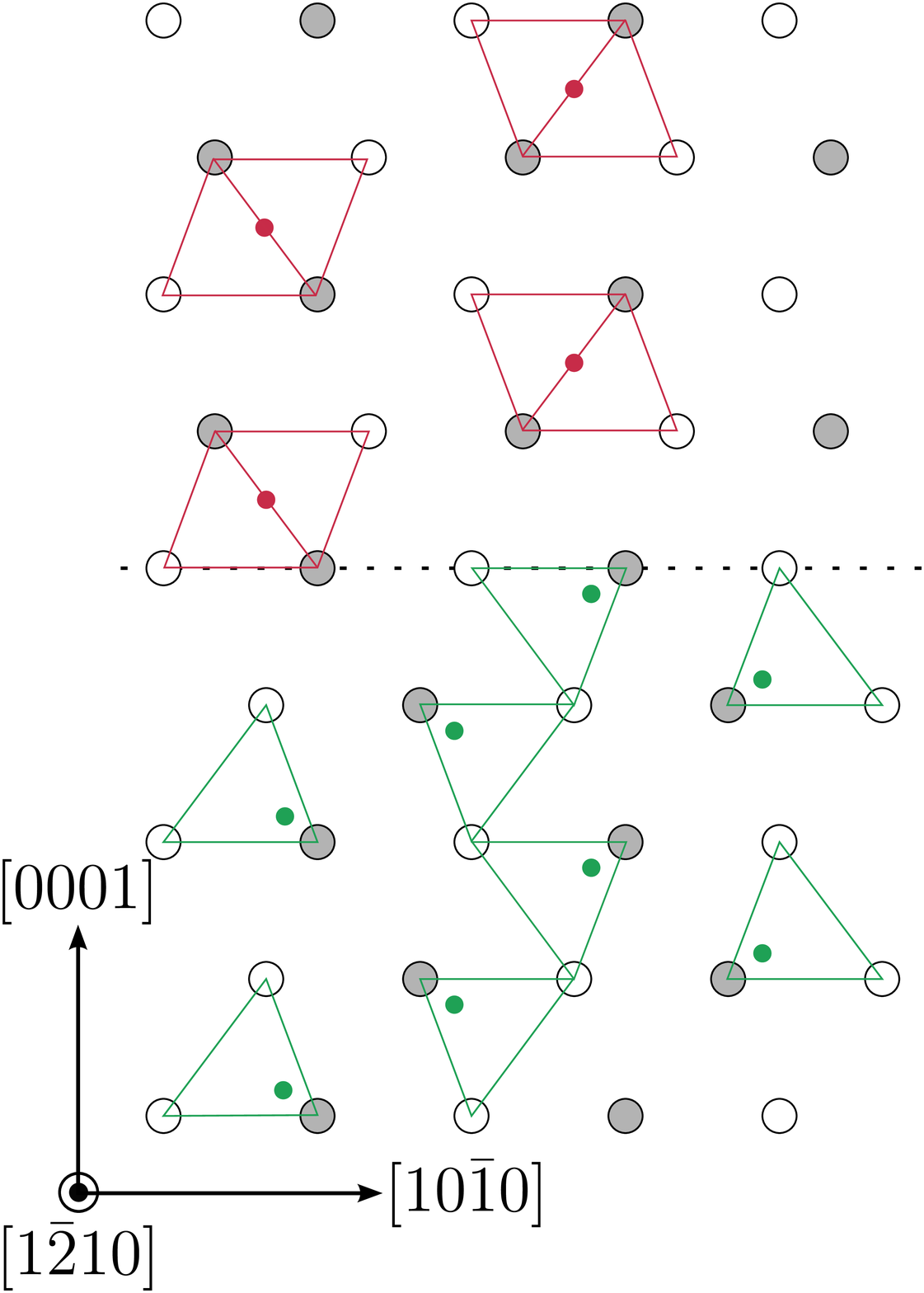}} &
\subfigure[H insertion sites (E fault)    ]{\label{fig:scheme_basal_SF-b}\includegraphics[scale=0.13]{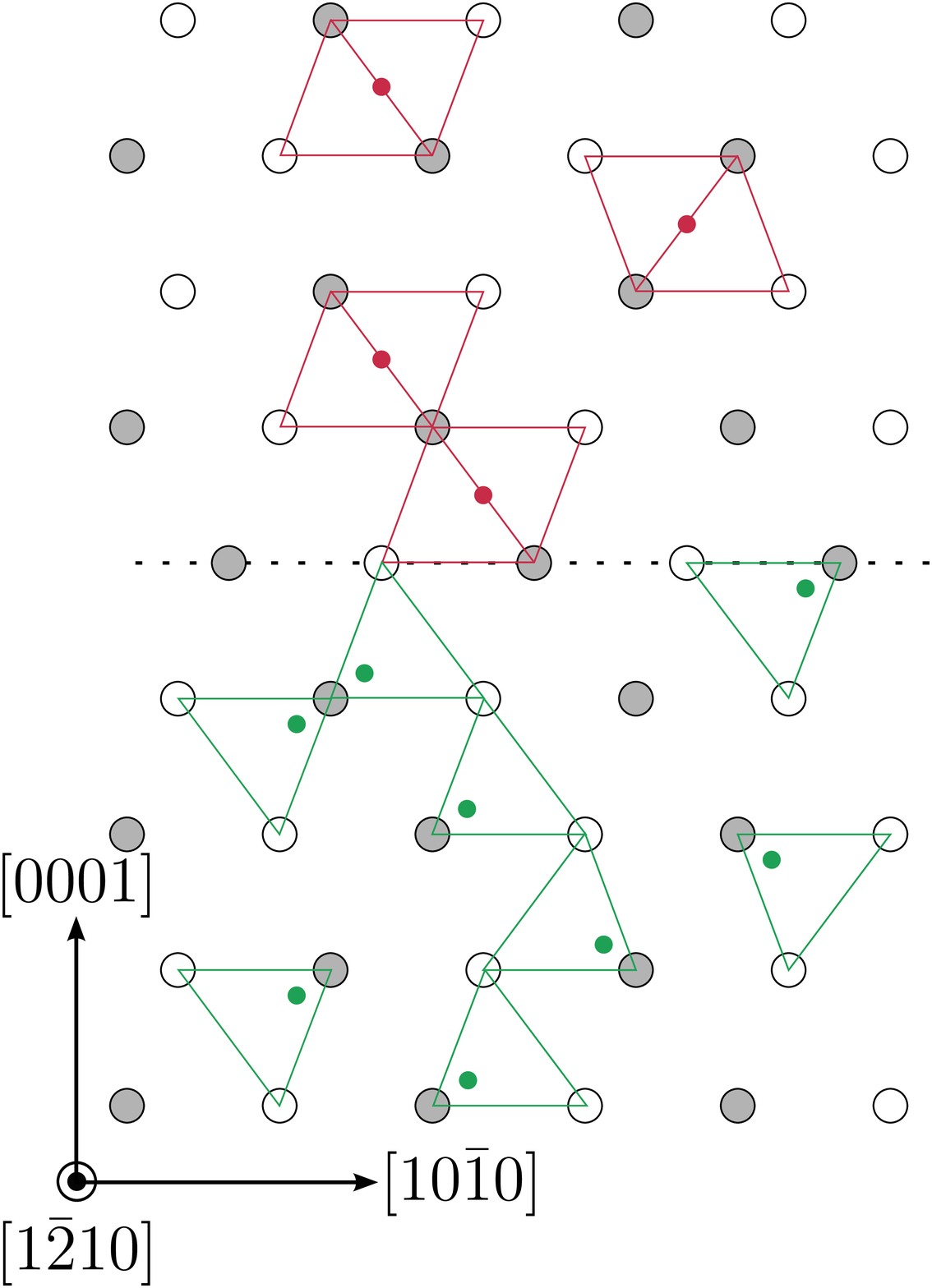}}\\
\subfigure[Binding energy without H vibration (I$_1$ fault)]{\includegraphics[scale=0.6]{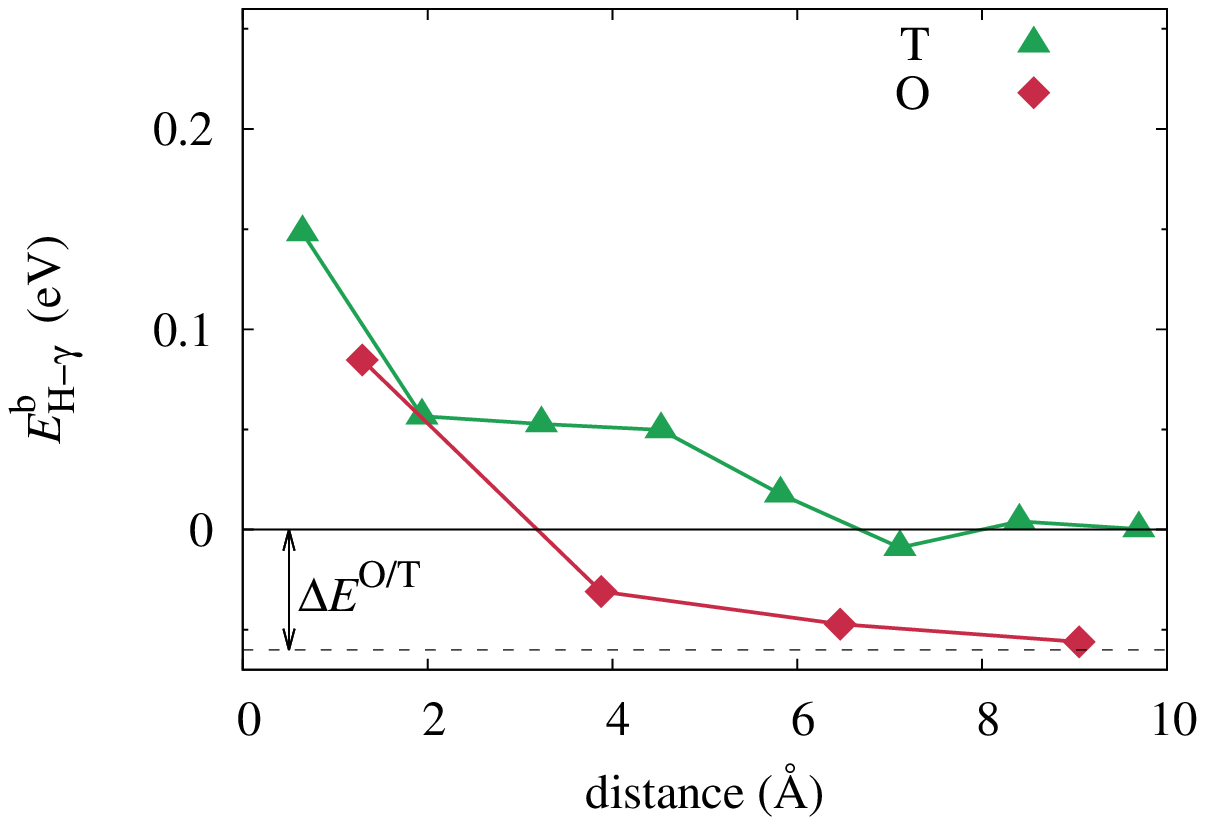}} &
\subfigure[Binding energy without H vibration (E fault)]{\includegraphics[scale=0.6]{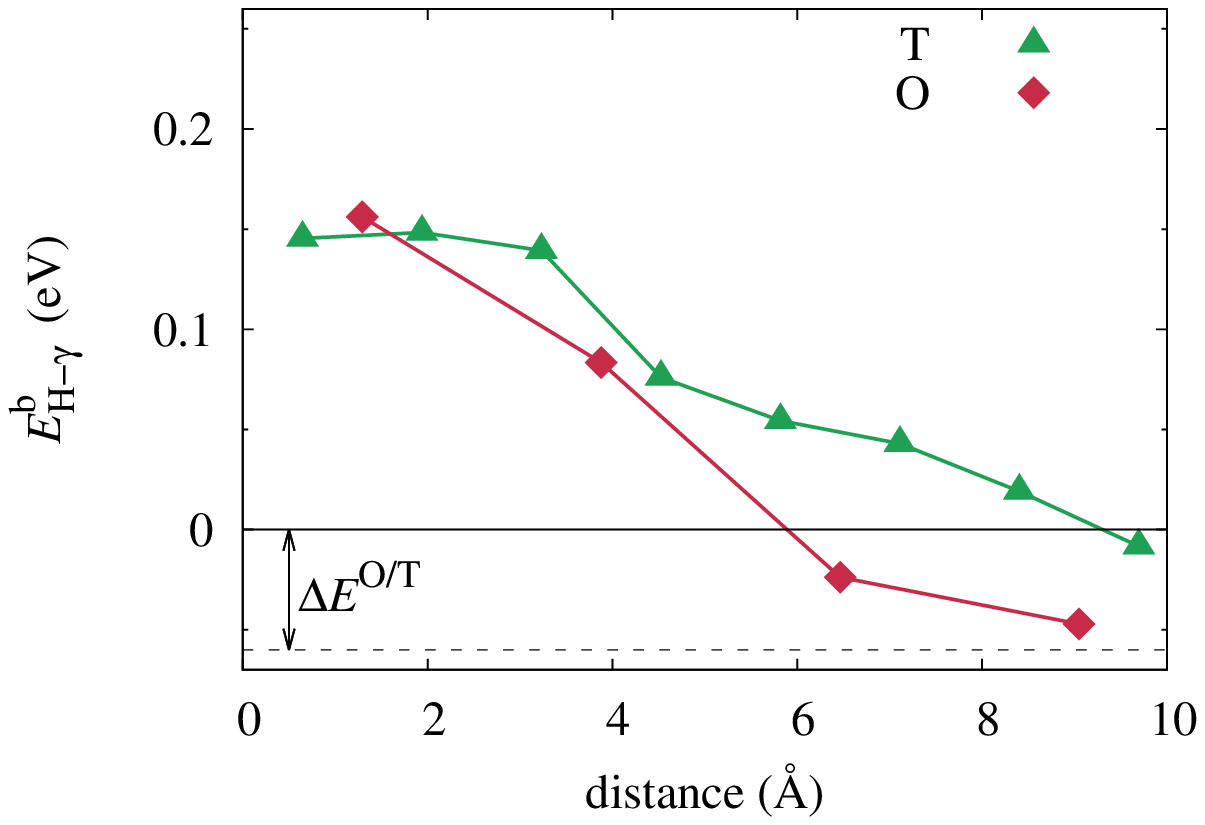}} \\
\subfigure[Binding energy with H vibration (I$_1$ fault)]{\includegraphics[scale=0.6]{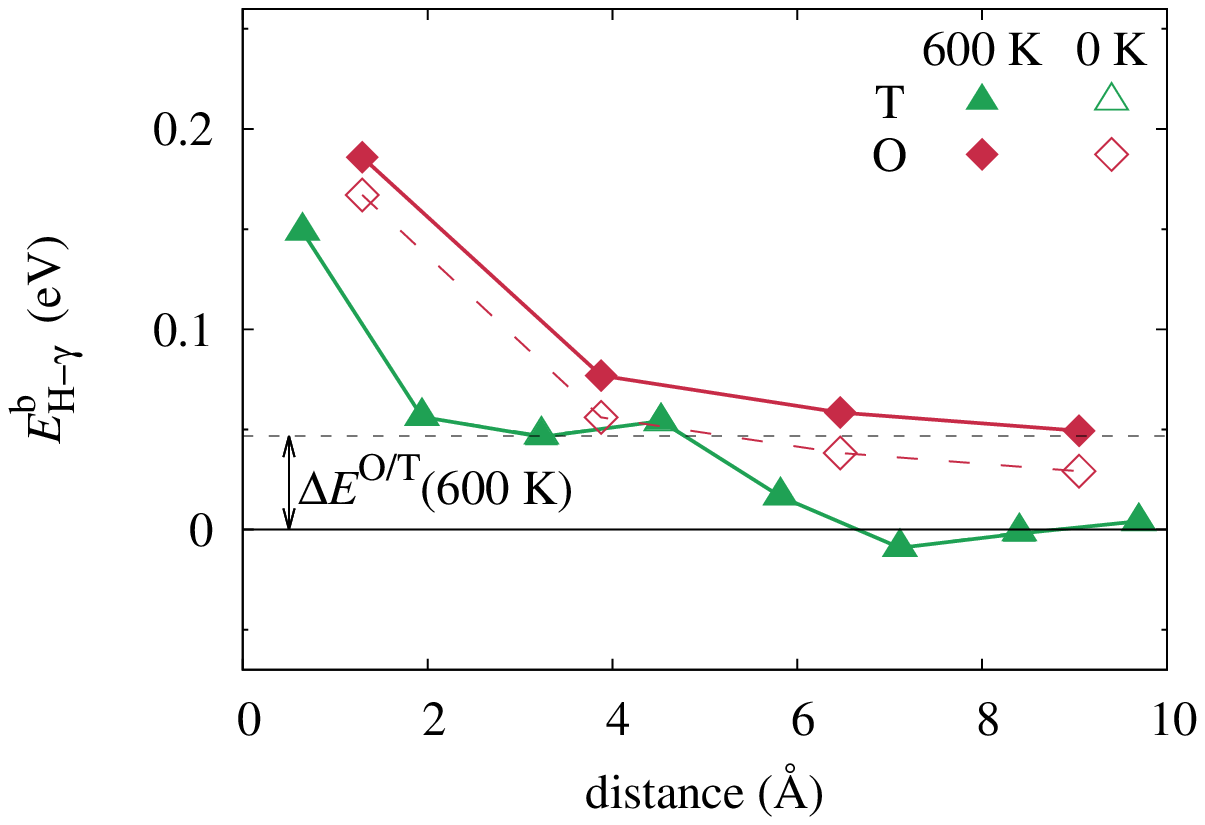}} &
\subfigure[Binding energy with H vibration (E fault)]{\includegraphics[scale=0.6]{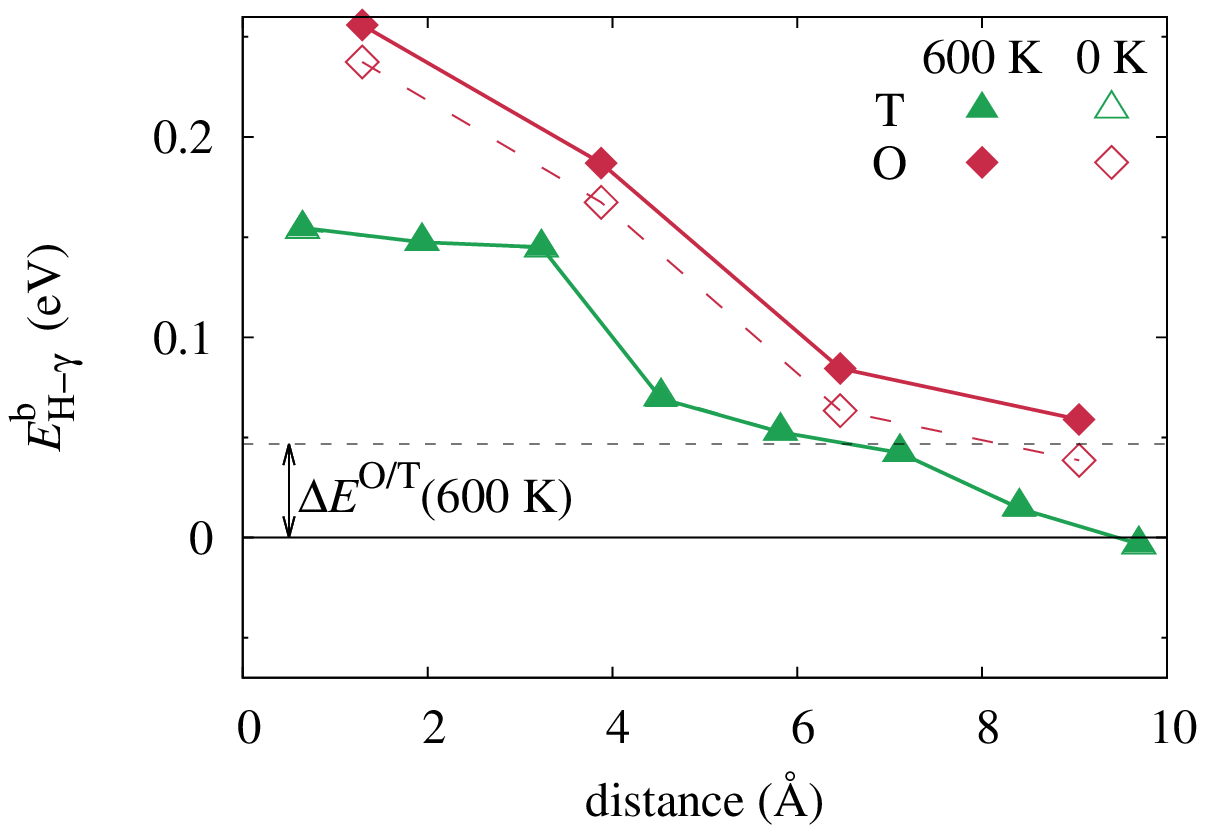}} \\
\end{tabular}
\caption{Interaction between an H atom and basal stacking faults.
The different T and O interstitial sites, where an H atom has been inserted, are sketched in (a) for the I$_1$
and in (b) for the E basal stacking fault.
The corresponding binding energies $E^{\rm b}_{\textrm{H}-\gamma}$ are respectively shown
in (c) and (d) without H vibration free energy,
and in (e) and (f) with H vibration free energy.}
\label{fig:Eb_basal_SF}
\end{figure*}

\Abinitio calculations are performed on simulation boxes duplicated 3 times in the $[2\bar{1}\bar{1}0]$ direction and 3 times in the $[\bar{1}2\bar{1}0]$ direction.
To build the basal stacking faults, one needs to remove one atomic layer normal to the $[0001]$ direction, thus leading to a total of $2n_z-1$ atomic layers, if $n_z$ is the number of duplicated unit cells in the $[0001]$ direction. Our simulations are carried out using $n_z=6$. 
This box size, which corresponds to a total of $99$ Zr atoms, has already been validated for both basal stacking faults in a previous study \cite{Varvenne2014}, and is also large enough to simulate an isolated H atom. 
The binding energy of H to the basal stacking fault then reads:
\begin{eqnarray}
	E^{\rm b}_{\rm H-\gamma} & = & E^{2n_z-1}_{\rm ZrH} + E^{2n_z-1}_{\rm Zr+\gamma} \nonumber \\
& & - E^{2n_z-1}_{\rm ZrH+\gamma}-E^{2n_z-1}_{\rm Zr}
\label{eq:Eb_H_SF}.
\end{eqnarray}
$E^{2n_z-1}_{\rm ZrH}$ is not accessible directly, as a periodic simulation box with $2n_z-1$ atomic layers automatically includes a stacking fault. We thus make the following approximation: 
\begin{eqnarray}
 E^{2n_z-1}_{\rm ZrH} - E^{2n_z-1}_{\rm Zr} & \simeq & E^{2n_z}_{\rm ZrH} - E^{2n_z}_{\rm Zr}, 
\end{eqnarray}
which allows us to obtain the binding energies of H with the basal stacking faults. 

As for the prismatic fault, our \abinitio calculations lead to a binding of H with basal faults, 
but with a higher maximal value of $0.16$~eV for the two basal faults. 
The strongest bindings are obtained when H is close to the stacking fault. This is true for both T and O sites, and for both I$_1$ and E faults. 
For the I$_1$ fault, the T sites are always more favorable than the O sites to insert the H atom. 
A higher number of attractive positions in the vicinity of the stacking fault are found for the insertion in T sites. 
The binding energies rapidly decrease with the distance, and become null or equal to $-\Delta E^{\rm O/T}$ for distances larger than $6$\,\AA. 
For the E fault, the most attractive interaction is found for the O site.
A largest number of attractive positions can be found in T sites than in O sites.  
Finally, interactions of H are longer ranged with the E fault than with the I$_1$ one,
as non-interacting behaviors are recovered only for separation distances larger than $9$\,\AA. 
This might be related to the fact that in the E stacking sequence, more atomic layers have an environment different from the hcp one than in the I$_1$ stacking sequence. 
The effect of the H vibration is the same as for the prismatic stacking fault,
\ie mainly a shift of the binding energies for the O sites.

\section{Hydrogen and vacancy loops}

To simulate the properties of extended defects such as dislocation loops, 
one cannot rely directly on \abinitio methods, 
because of the limited number of atoms, only a few hundreds, 
which can be modeled with such simulations.
In a previous study \cite{Varvenne2014}, we adopted a continuous description of large vacancy clusters.
For dislocation loops, the cluster energies were obtained through a line tension model, 
incorporating the stacking fault energy for faulted loops. 
After validating the approach with large scale atomistic simulations relying on an empirical potential, 
the main parameters of the model were calibrated on \abinitio data.

The influence of hydrogen on the stability of dislocation loops can be introduced in a very simple way
within this approach, by considering the modification of the stacking fault energies 
arising from hydrogen segregation \cite{Zinkle1987}. 
Here, we first model the segregation of hydrogen on the different stacking faults,
as well as the evolution of the stacking fault energies with the H content. 
We then discuss the effect of H on the stability of the different vacancy loops
formed under irradiation in hcp Zr. 

\subsection{Segregation profiles}

Following common approaches for solute segregation at a surface or an interface \cite{Treglia1999, Mutschele1987}, we make use of the previously calculated binding energies $E^{\rm b}_{\rm H-\gamma}$
to model the equilibrium concentration profiles of H in the vicinity of each stacking fault. 

We consider a bicrystal consisted of two semi infinite bulk spaces separated by an interface
corresponding to the stacking fault. The different insertion sites for H, either T or O sites,
are sitting in planes $p$ located at different distances $d_p$ from the stacking fault. 
We assume that the system is large enough so that each side of the stacking fault
acts as a reservoir for hydrogen.
The hydrogen chemical potential $\mu_H$ is then fixed at equilibrium.
We assume that the concentration $c_p \in [0,1]$ of each plane is homogeneous at equilibrium and that the configurational entropy is equal to the one of an ideal solid solution. 
We neglect the interaction between H atoms
and only the H interaction with the stacking fault is considered,
a valid assumption for not too concentrated solid solutions. 
Minimizing the grand potential of the system with respect to the average H occupation $c_p$ of each plane $p$, 
the  equilibrium distribution obey the equations
\begin{eqnarray}
	E^{\rm sol}_{\rm H} -E^{\rm b}_{\rm H-\gamma}(d_p) + kT \ln \frac{c_p}{1-c_p} & = & \mu_{\rm H},
\label{eq:minimisation_mu_H} 
\end{eqnarray}
where $E^{\rm sol}_{\rm H}$ is the hydrogen solution energy in the perfect zirconium matrix
for the T insertion sites.
These equations are valid for any plane of the bicrystal, 
as each plane, defined by its distance $d_p$ to the fault plane, has the same density of insertion sites.
This is true both for the prismatic (Fig. \ref{fig:Eb_prism_SF}a) and the basal faults (Fig. \ref{fig:Eb_basal_SF}a and b),
the insertion sites being either of the T or of the O type.
Far from the fault plane, concentrations and energies converge to their bulk values:
$c_p=c_{\rm T}^0$ and $E^{\rm b}_{\rm H-\gamma}(d_{\infty})=0$ for planes corresponding to T insertion sites,
$c_p=c_{\rm O}^0$ and $E^{\rm b}_{\rm H-\gamma}(d_{\infty})=-\Delta E^{\rm 0/T}$ for O planes.
This allows us to eliminate $\mu_{\rm H}$ in Eq.~\ref{eq:minimisation_mu_H} and to obtain for each plane $p$: 
\begin{equation}
	\frac{c_p}{1-c_p}  =  \frac{c_{\rm T}^0}{1-c_{\rm T}^0}\exp{\left( \frac{E^{\rm b}_{{\rm H}-\gamma}(d_p)}{kT} \right)}.
\label{eq:segregation_profile1} 
\end{equation}
There are two interstitial site of T type and one of O type per Zr atom. 
Their bulk concentrations is therefore linked to the nominal hydrogen concentration $x_{\rm H}$
trough the equations:
\begin{align}
	c_{\rm T}^0 &= \frac{1}{2 + \exp{\left( -\frac{\Delta E^{\rm O/T}}{kT} \right)}} x_{\rm H}, \\
	c_{\rm O}^0 &= \frac{\exp{\left( -\frac{\Delta E^{\rm O/T}}{kT} \right)}}{2 + \exp{\left( -\frac{\Delta E^{\rm O/T}}{kT} \right)}} x_{\rm H}, 
\end{align}
in the low concentration regime ($x_{\rm H}\ll1$).
We finally obtain the concentration profile as a function of the nominal concentration:
\begin{equation}
	c_p = \frac{x_{\rm H} \exp{\left( \frac{E^{\rm b}_{{\rm H}-\gamma}(d_p)}{kT} \right)}}
	{2 + \exp{\left( -\frac{\Delta E^{\rm O/T}}{kT} \right)}
	+ x_{\rm H} \exp{\left( \frac{E^{\rm b}_{{\rm H}-\gamma}(d_p)}{kT} \right)}}.
	\label{eq:segregation_profile2}
\end{equation}

\begin{figure}[!bth]
\centering
\subfigure[without H vibration]{\includegraphics[scale=0.6]{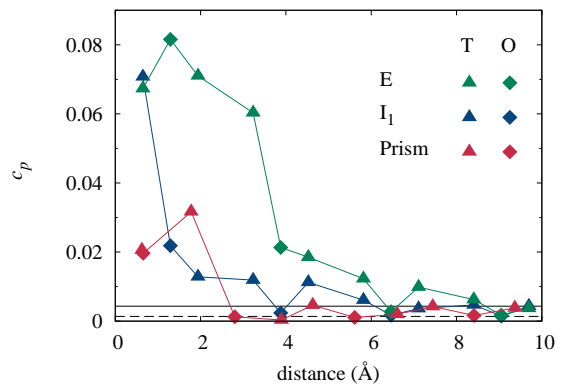}}
\subfigure[with H vibration]{\includegraphics[scale=0.6]{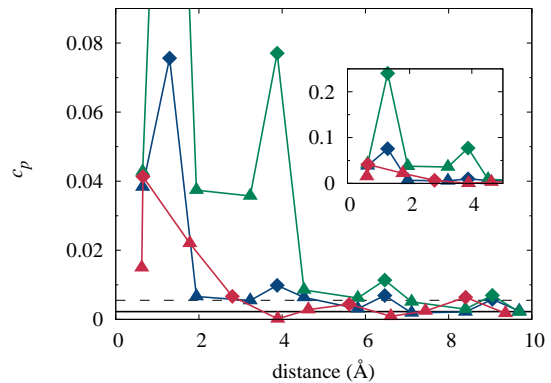}}
\caption{Concentration profiles of hydrogen at $T=600$\,K and for a nominal concentration $x_{\rm H}=0.01$,
	plotted as a function of the separation distance from the prismatic, basal I$_1$ and basal E stacking faults,
	(a) neglecting or (b) considering H vibration free energy. 
	Small triangles refer to the T sites and small diamonds to the O sites.
	The continuous and dashed horizontal lines correspond to the concentrations
	of the, respectively, T and O sites in the bulk.}
\label{fig:cp_profiles}
\end{figure}

Fig.~\ref{fig:cp_profiles} displays the concentrations $c_p$ for the prismatic, basal I$_1$ and basal E faults as a function of the separation distance from the stacking fault. 
The segregation profiles are given for $T=600$\,K, 
the service temperature of the pressurized water reactors,
and for a bulk nominal concentration $x_{\rm H}=0.01$. 
This corresponds to an upper limit of the H content that can be reached, as the solubility limit, at this temperature, is 
0.008 in pure Zr \cite{Zuzek1990}
and 0.009 in Zircaloy-4 \cite{Vizcaino2002}, a zirconium alloy currently used in the nuclear industry.
All faults lead to a strong increase of the H concentration in their vicinity.  
As these faults correspond to the ones met in vacancy loops,
this agrees with experiments suggesting a trapping of H by these loops \cite{Lewis1984,Vizcaino2002,Vizcaino2007}. 
We also notice that the basal faults concentrate more H atoms than the prismatic fault,
in particular the E fault which can capture hydrogen up to the fourth layer from the stacking faults.
The H trapping by basal faults is therefore stronger, explaining why annealing at higher temperatures 
is needed to recover the unirradiated solubility limits in Zr samples where vacancy basal loops have been formed 
after an irradiation at a high dose \cite{Vizcaino2007}.

H vibrations do not change these general trends between the different stacking faults, 
although they modify the segregation profiles. 
The inclusion of the vibration free energy in the binding energies leads to an increase of the enrichment for the O sites and a decrease for the T sites (Fig. \ref{fig:cp_profiles}).

\subsection{Variation of stacking fault energies}

As first pointed out by Suzuki \cite{Suzuki1962}, solute segregation modifies 
the stacking fault energies. 
We thus need to deduce from our segregation model,
the evolution of the basal and prismatic fault energies in the presence of hydrogen. 
As the system is open with respect to H, we define the stacking fault energy $\gamma(\{c_p\})$ directly from the grand potential:
\begin{equation}
	\gamma(\{c_p\}) = \frac{\Omega_f(\{c_p\})-\Omega_i(x_{\rm H})}{S_0},
\label{eq:gamma_grand_potentiel}
\end{equation}
where $S_0$ is the area of the dividing interface 
for one insertion site per plane: 
$S_0=a^2\sqrt{3}/2$ for the basal faults and $S_0=ac/2$ for the prismatic fault.
$\Omega_f$ and $\Omega_i$ are the grand potentials of respectively
the faulted crystal with the segregation profile
and the perfect crystal with the homogeneous solid solution.  
The thermodynamic model used in the previous section to obtain the segregation profiles
leads to the following expressions:
\begin{gather}
	\begin{split}
		\Omega_i(x_{\rm H})	
		= & \sum_{p, {\rm T}} \left\{
			c_{\rm T}^0 \left[E^{\rm sol}_{\rm H}-\mu_{\rm H}\right] \right.\\
			& \left. + kT \left[ c_{\rm T}^0 \ln{( c_{\rm T}^0 )} + (1 - c_{\rm T}^0 ) \ln{( 1 - c_{\rm T}^0 )} \right] \right\} \\
		& + \sum_{p, {\rm O}} \big\{
			c_{\rm O}^0 \left[E^{\rm sol}_{\rm H} + \Delta E^{\rm O/T} - \mu_{\rm H}\right] \\
			&  + kT \left[ c_{\rm O}^0 \ln{( c_{\rm O}^0 )} + ( 1 - c_{\rm O}^0 ) \ln{( 1 - c_{\rm O}^0 )} \right] \big\}, 
	\label{eq:omega_i}
	\end{split}
	\\
	\begin{split}
		\Omega_f(\{c_p\}) = & 
		S_0\gamma_0 +\sum_p c_p\left[E^{\rm sol}_{\rm H} - E^{\rm b}_{\rm H-\gamma}(d_p)-\mu_{\rm H}\right] \\
		&   + \sum_p kT\left[c_p\ln c_p +(1-c_p)\ln (1-c_p)\right],  
	\label{eq:omega_f}
	\end{split}
\end{gather}  
with $\gamma_0$ the stacking fault energy without hydrogen. 
The summation over the planes in Eq. \ref{eq:omega_i} has been partitioned between the planes
containing T and O interstitial sites.
Inserting Eqs.~\ref{eq:omega_i} and \ref{eq:omega_f} into Eq.~\ref{eq:gamma_grand_potentiel}
and eliminating the chemical potential $\mu_{\rm H}$ using Eq.~\ref{eq:minimisation_mu_H},
we obtain the stacking fault energy in presence of the segregation profile:
\begin{equation}
	\begin{split}
		\gamma(\{c_p\}) = & \gamma_0 + \frac{kT}{S_0}\sum_{p, {\rm T}} \ln{\left( \frac{1-c_p}{1-c_{\rm T}^0} \right)} \\
			& + \frac{kT}{S_0}\sum_{p, {\rm O}} \ln{\left( \frac{1-c_p}{1-c_{\rm O}^0} \right)}.
	\end{split}
	\label{eq:gamma_fct_cp_c0}
\end{equation}
The same result is obtained using the macroscopic thermodynamic approach 
summarized by Hirth \cite{Hirth1970}. 

\begin{figure}
\centering
\includegraphics[scale=0.6]{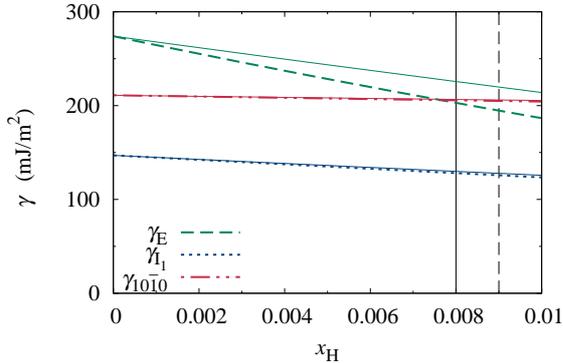}
\caption{Evolution of the basal and prismatic stacking fault energies
with the bulk concentration $x_{\rm H}$ of hydrogen
calculated at 600\,K.
The thick dashed lines and the thin continuous lines correspond to the data 
respectively with and without H vibration free energy.
The continuous and dashed vertical lines correspond to H solubility
at this temperature respectively in pure Zr and in Zircaloy-4.}
\label{fig:gamma_c0_600K}
\end{figure}

Using Eq.~\ref{eq:gamma_fct_cp_c0}, we show in Fig.~\ref{fig:gamma_c0_600K} 
the evolution at $T=600$\,K of the two basal and the prismatic stacking faults,
when the nominal hydrogen content varies in a concentration range
$x_{\rm H}<0.01$. 
All the stacking fault energies decrease when the hydrogen content increases.
The decay is small for the prism fault, slightly higher for the I$_1$ fault
and fast for the E fault. 
The consideration of H vibration only affects the E fault: 
it leads to a stronger decrease of the fault energy with the H content.
In the range of nominal concentrations allowed for hydrogen at this temperature, 
the relative order of the different fault energies is not modified,
except close to the solubility limit where the basal E and the prismatic stacking faults
have almost the same energy, with the basal E becoming slightly most stable
with the inclusion of vibration free energies in the thermodynamic modeling. 

\subsection{Stability of vacancy loops}

We now study how hydrogen may modify the stability of the different 
vacancy loops, using the continuous model we previously developed
to describe the energetics of vacancy clusters in pure zirconium \cite{Varvenne2014}.
Within this model, the formation energy of a loop containing $n$ vacancies
is given by:
\begin{equation}
E^{\rm f}_{\rm loop} (n)  =  \pi {R_1}^2 \gamma  n 
+ 2\pi f R_1 \bar{K} \sqrt{n} \ \ln{\left( \frac{R_1 \sqrt{n}}{r_{\rm c}} \right)}.
\label{eq:nrj_Ebcle}
\end{equation}
with $\gamma$ the stacking fault energy,
$f$ a shape factor close to unity,
$r_{\rm c}$ the dislocation core radius close to the Burgers vectors,
and $\bar{K}$ an average value of the prefactor appearing in the elastic energy.
The values of these parameters for the different vacancy loops in zirconium
can be found in Ref. \cite{Varvenne2014}. 
The quantity $R_1$ is a scaling distance with 
$R_1 = a(\sqrt{3}/2\pi)^{1/2}$ for the basal loops 
and $R_1=\sqrt{ac/2\pi}$ for the prismatic loops. 

\begin{figure}[!bth]
\centering
\subfigure[Without hydrogen]{\label{fig:Ef_bclesDFT1}\includegraphics[scale=0.6]{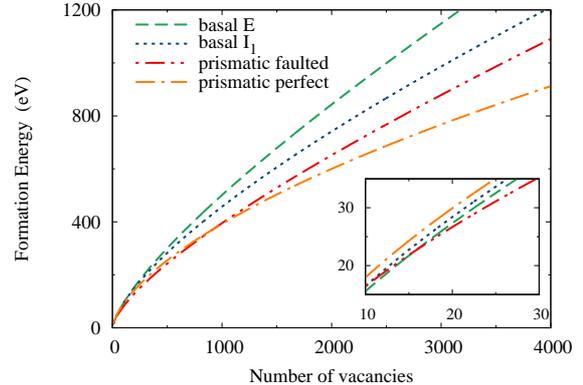}} \\
\subfigure[With $x_{\rm H}=0.01$ and $T=600$\,K]{\label{fig:Ef_bclesDFT2}\includegraphics[scale=0.6]{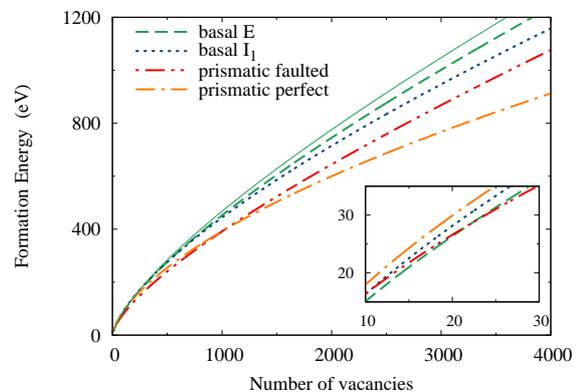}} 
\caption{Formation energies of large vacancy loops as predicted by continuous laws parameterized on DFT results for 
(a) pure hcp Zr 
and (b) Zr containing hydrogen at  $T=600$\,K for a nominal concentration $x_{\rm H}=0.01$.
The H vibration free energy is included in (b).
The result obtained without this vibration free energy
is also shown by a thin continuous line for the basal loops with an E fault.
The insets show the stability inversion at small sizes between basal and prismatic loops.
}
\label{fig:Ef_bclesDFT}
\end{figure}

As suggested in Refs.~\cite{Domain2004,Zinkle1987,Udagawa2011}, 
the effect of hydrogen on cluster stability 
can be included in a very simple way in the Eq.~\ref{eq:nrj_Ebcle}, 
through the modification of the stacking fault energy, which is the leading term for large vacancy clusters. 
This is done by replacing $\gamma$ in Eq.~\ref{eq:nrj_Ebcle} by the modified stacking fault energies $\gamma(\{c_p\})$, calculated using Eq.~\ref{eq:gamma_fct_cp_c0}. 

In Fig.~\ref{fig:Ef_bclesDFT}, the predicted formation energies of the loops lying in the basal and prismatic planes are displayed in the absence of hydrogen 
and for a hydrogen nominal concentration $x_{\rm H}=0.01$ at $T=600$\,K. 
The presence of hydrogen tends to increase the stability of $\left\langle c \right\rangle$ loops 
lying in the basal planes with respect to $\left\langle a \right\rangle$ loops lying in the prismatic planes.
The effect is stronger for $\left\langle c \right\rangle$ loops with an E fault,
because of the highest amount of hydrogen captured by this stacking fault.
These $\langle c \rangle$ loops with an E fault are the most stable vacancy clusters 
at very small size (loops containing less than 16 vacancies in pure Zr)
and H segregation shifts the stability crossover with prismatic loops to larger sizes,
an effect enhanced by the contribution of H vibrations.
The hydrogen segregation on stacking faults can thus explain the higher amount of $\left\langle c \right\rangle$ component loops
observed when Zr samples are pre-hydrided and then submitted to irradiation \cite{Tournadre2013,Tournadre2014},
and the correlated consequence on the breakaway growth at high irradiation dose \cite{McGrath2011}.

\section{Conclusion}

The influence of hydrogen in solid solution on vacancy loops energetics has been studied in hcp Zr using an \abinitio approach, including H vibrations. 
DFT calculations performed on small vacancy clusters showed that hydrogen is able to attract up to 4 vacancies. 
The most favorable insertion site for H in the vicinity of a vacancy cluster is not a T site
but an O site with 2 or 3 vacancies as first nearest neighbors.
Hydrogen therefore prefers to sit at the boundary of vacancy clusters, and not inside them.
No strong discriminating effect of hydrogen on the stability of the different cluster configurations
were found for small vacancy clusters.

Hydrogen binding to larger vacancy clusters has been modeled through its interaction with the stacking faults.
We computed, with \abinitio calculations, the interactions between H and the prismatic, basal E and basal I$_1$ faults.
These interactions are mainly attractive.
Hydrogen binding is stronger for the basal faults,
especially in the case of the E fault. 
Using these binding energies together with a thermodynamic modeling of H segregation on the faults,
we could predict the evolution of the stacking fault energies with the H nominal concentration. 
Including these variations into continuous laws describing the energetics of vacancy loops,
we could finally demonstrate that H stabilizes the vacancy loops, 
and that the stronger variation is observed for the basal loops, 
\ie the defects responsible for the breakaway growth observed at high irradiation dose.
This binding of H to faulted vacancy loops agrees with the trapping of hydrogen 
by vacancy clusters observed experimentally in irradiated zirconium alloys.

\appendix

\section{Convergence with supercell size\label{ap:CVg_SC_size}}

\begin{figure}[!bth]
\includegraphics[scale=0.63]{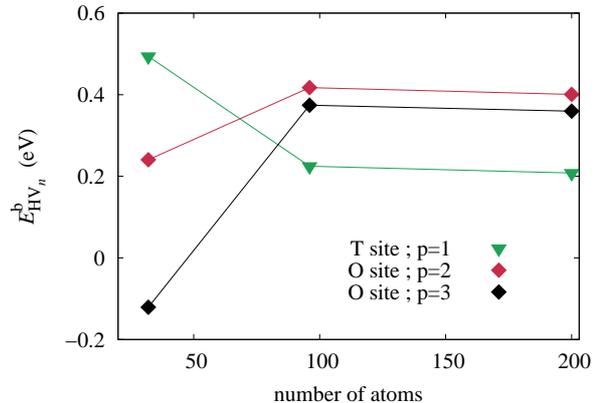}
\caption{Binding energy of H with an octahedral vacancy cluster ${\rm V}_6$ as a function of the supercell size. Three different insertion sites are shown for H: T site with one nearest neighbor HV interaction and O site with $p=2$ or $p=3$ nearest neighbor HV interactions.}
\label{fig:CVgce_DFT_supercell}
\end{figure}

In Fig.~\ref{fig:CVgce_DFT_supercell}, the binding energies of H with an octahedral V$_6$ cluster are displayed versus the number of atoms in the supercell. The three more favorable insertion sites are tested for the H atom: T site with one nearest neighbor HV interaction,  and O site with 2 or 3 nearest neighbor HV interactions ($p=2$ and $3$, respectively).  

The resulting binding energies only slightly vary between $100$ and $200$ atoms, assessing thus our choice of a $200$ atom supercell ($5\times 5\times 4$) to study the influence of H on small vacancy clusters.

\section{Validity of the harmonic approximation}
\label{sec:appendix_harmonic}

As pointed out by Christensen \etal \cite{Christensen2015b}, T insertion sites in the hcp lattice
form pairs of close sites and the migration barrier for H jumping from one T site to its neighbour site is small.
Using the simulation cell containing 200 lattice sites we compute 
the H migration energies between its different insertion sites (Tab. \ref{tab:Hmig})
and obtain the lowest energy for the T-T migration.
This migration energy is in the range of the zero point energy of H vibration. 
One can therefore wonder if the harmonic approximation used to calculate 
the contribution of H vibrations to the free energy is a good approximation.
We examine this point more closely in this appendix, 
first for a single H atom in a perfect crystal,
then for a H atom close to a vacancy.

\begin{table}[!hbt]
	\caption{H migration energies $E^{\rm mig}$ between neighbour T and O sites.}
	\label{tab:Hmig}
	\begin{center}
	\begin{tabular}{lcc}
		\hline
		& \multicolumn{2}{c}{$E^{\rm mig}$ (ev)} \\
		& unrelaxed & relaxed \\
		\hline
		T $\leftrightarrow$ T & 0.21 & 0.13 \\
		T $\rightarrow$ O     & 0.46 & 0.42 \\
		O $\leftrightarrow$ O & 0.54 & 0.50 \\
		\hline
	\end{tabular}
	\end{center}
\end{table}

\subsection{Single H atom}
\label{sec:appendix_harmonic_bulk}

\begin{figure}[!bht]
	\begin{center}
		\includegraphics[width=0.99\linewidth]{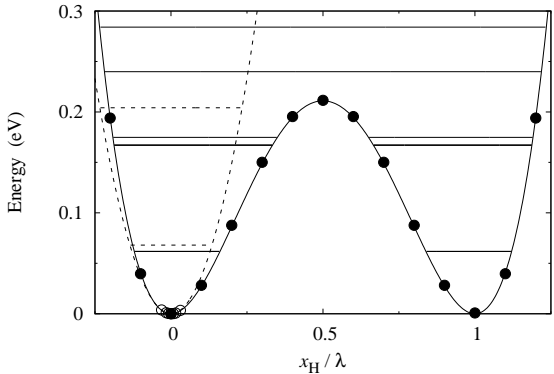}
	\end{center}
	\caption{Energy variation of a H atom migrating between two T sites in a hcp Zr lattice.
	The double well profile given by the solid line has been fitted to the \abinitio energies
	indicated by filled circles 
	and the corresponding quantized energy levels are given by horizontal solid lines. 
	The results obtained for a single well in the harmonic approximation are shown with dashed lines.}
	\label{fig:H_double_well}
\end{figure}

We show in Fig.~\ref{fig:H_double_well} the energy profile of an H atom between two neighbour insertion T sites. 
Only the equilibrium configurations, corresponding to $x_{\rm H}/\lambda=0$ and 1 with $\lambda=c/4$, have been relaxed. 
They are then used to obtain the other intermediate configurations by linear interpolation. 
The obtained energy profile is well fitted by an order 4 polynomial function.
We then compute the eigenstates of this double well energy profile using the numerical approach described in \cite{Proville2005a}.
The double well shape has two main effects on these eigenstates compared to the results obtained with the harmonic approximation for two single wells (Fig.~\ref{fig:H_double_well}):
it lowers the energy of the eigenstates, by 6\,meV for the fundamental level,
and it lifts the degeneracy of the higher levels.

\begin{figure}[!bht]
	\begin{center}
		\includegraphics[width=0.99\linewidth]{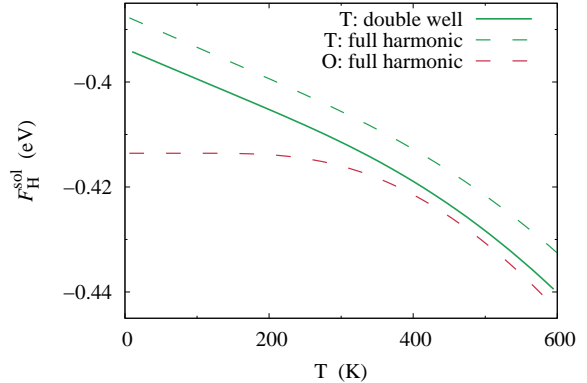}
	\end{center}
	\caption{Solution free energy $F^{\rm sol}_{\rm H}$ (Eq. \ref{eq:Esol}) of an H single atom in a Zr hcp matrix
	as a function of the temperature calculated for its T and O insertion sites.
	The free energy is computed either within the harmonic approximation 
	or using the eigenstates of the double well energy profile (Fig.~\ref{fig:H_double_well}).}
	\label{fig:FH_bulk}
\end{figure}

We then compute the H vibration free energy, 
following the approach proposed by Christensen \etal \cite{Christensen2015b}.
We factorize the partition function into vibration modes corresponding to
the H migration between two neighbouring T sites and vibration modes orthogonal to this migration direction.
The first contribution is deduced from the eigenstates of the double-well energy profile (Fig.~\ref{fig:H_double_well}),
whereas the harmonic approximation (Eq.~\ref{eq:Fvib}) is used for the two last vibration modes.
The obtained H solution free energy is shown in Fig. \ref{fig:FH_bulk}
where we compare it to the solution free energy obtained within harmonic approximation. 
The difference between the two plots is mainly a constant energy shift
corresponding to the shift of the fundamental eigenstate noticed before. 
The linear decrease with the temperature of the solution free energy close to 0\,K 
is well captured by the harmonic approximation, once the $-kT\ln{(2)}$ has been included 
in the free energy to account for the configurational entropy associated with the presence
of two energy wells.
The harmonic approximation looks therefore reasonable to account for H vibration in an hcp Zr matrix,
with a precision of the order of 10\,meV. 
This precision being larger than the energy difference between the T and the O configurations 
(Fig. \ref{fig:FH_bulk}), a more precise treatment of vibrations will nevertheless be needed
to be able to conclude on the preferential insertion site of H in Zr.

\subsection{H close to a vacancy}

\begin{figure}[!bht]
	\begin{center}
		\subfigure[Configurations b]{
			\includegraphics[width=0.99\linewidth]{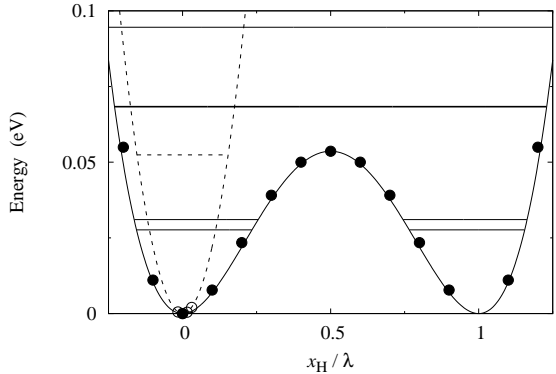}}
		\subfigure[Configurations a and c]{
			\includegraphics[width=0.99\linewidth]{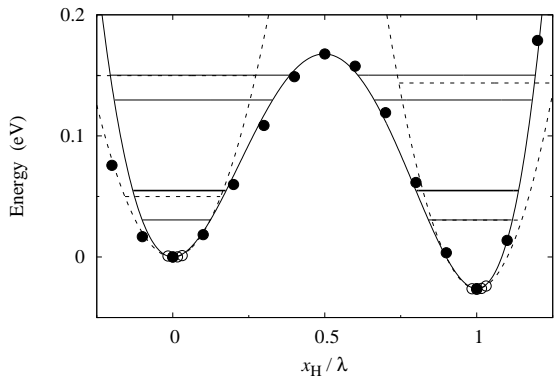}}
	\end{center}
	\caption{Energy variation of a H atom migrating between two T sites
	in the vicinity of a vacancy in a hcp Zr lattice. 
	In a), both equilibrium positions correspond to b configurations (Fig. \ref{fig:conf_HV_1nn}).
	In b), $x_{\rm H}=0$ and 1 correspond respectively to the a and c configurations.}
	\label{fig:HV_double_well}
\end{figure}

We now examine the case where the H atom lies in a T site close to a vacancy. 
We examine two different configurations, both corresponding to the H atom and the vacancy
being first nearest neighbours.  The HV pair can either form a configuration b (see Fig. \ref{fig:conf_HV_1nn}
for a sketch of the configuration), with both wells of the T positions corresponding to this b configuration,
or the HV complex can form a configuration a, with the other well corresponding to configuration c. 
The corresponding energy profiles are shown in Fig. \ref{fig:HV_double_well}. 
The presence of the vacancy leads to a decrease of the migration barrier between the two H equilibrium positions
compared to the migration barrier in the bulk crystal (Fig. \ref{fig:H_double_well}). 
As a consequence, the effect of the potential anharmonicity are more pronounced 
in presence of a vacancy: the lowering of the eigenstates is more important 
and the degeneracy is lifted even for the fundamental eigenstate. 

\begin{figure}[!bhtp]
	\begin{center}
		\includegraphics[width=0.99\linewidth]{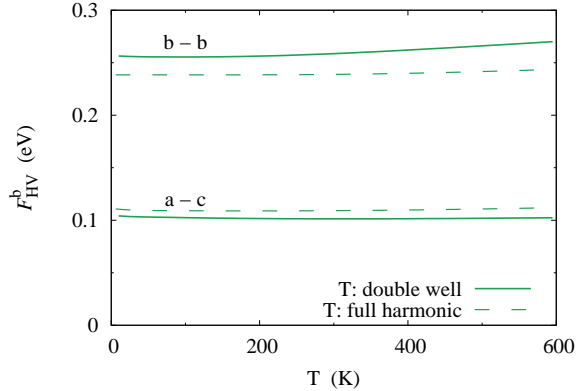}
	\end{center}
	\caption{Binding free energy $F_{\rm HV}^{\rm b}$ between a vacancy and an H atom lying in a double-well 
	composed of two T sites with a configuration either b-b or a-c.
	The free energy is computed either within the harmonic approximation
	or using the eigenstates of the double well energy profiles (Figs. \ref{fig:H_double_well}
	and \ref{fig:HV_double_well}).}
	\label{fig:FHV_binding}
\end{figure}

We then compute the binding energy between H and a vacancy for these two configurations, 
including in Eq. \ref{Eb_HVn} the vibration contribution in the free energy of the ZrHV and the ZrH configurations.
Like in the previous case, we consider either a fully harmonic approximation, 
or we include the eigenstates corresponding to the exact energy profile along the migration direction.
For the asymmetric double-well energy profile corresponding to the transition between the configurations a and c 
of the HV complex (Fig. \ref{fig:FHV_binding}b), one cannot separate the vibration mode in the migration direction
from the modes orthogonal to this direction.  An approximate treatment is thus needed. 
For such a double-well, the free energy is defined by
\begin{multline}
	F(T) = - kT \ln{\left[ 
		  \exp{\left(-\frac{           F^{\rm vib}_{\rm a}}{kT} \right)} 
		  \right.} \\ {\left.
		+ \exp{\left(-\frac{\Delta E + F^{\rm vib}_{\rm c}}{kT} \right)}
		\right]},
	\label{eq:Fvib_double_well}
\end{multline}
where the vibration contributions $F^{\rm vib}_{\rm a}$ and $F^{\rm vib}_{\rm c}$
are given by Eq.~\ref{eq:Fvib} for both configurations a and c,
and $\Delta E = E_{\rm c} - E_{\rm a}$ is the energy difference between both configurations.
The free energy deduced from the exact eigenstates of the 1D double-well (Fig. \ref{fig:HV_double_well})
can be added to this free energy, after having withdrawn the corresponding contribution in the 
harmonic approximation:
\begin{multline}
	F^{\rm vib}_{\rm 1D}(T) = -kT\ln{\left[ 
	\frac{1}{2\sinh{(\hbar\omega^3_{\rm a}/kT)}} 
		\right.} \\ {\left.
	+ \frac{\exp{\left( -\Delta E/kT \right)}}
			{2\sinh{\left(\hbar\omega^3_{\rm c}/kT\right)}}\right]},
	\label{eq:Fvib_double_well_1D}
\end{multline}
with $\omega^3_{\rm a}$ and $\omega^3_{\rm c}$ the pulsations of the vibration modes
in the migration direction.

The binding free energies are shown in Fig. \ref{fig:FHV_binding}. 
We recover a difference of the order of 10\,meV between the result of the harmonic approximation
and the more exact treatment of H vibrations.
This difference is much smaller than the absolute values of the binding energies
and does not modify the relative stability between different configurations. 
The harmonic approximation appears clearly to be well suited to study H interaction with vacancies
and a more precise treatment of H vibrations is thus unnecessary.

\section*{Acknowledgments}
  This work was performed using HPC resources from GENCI-[CINES/CCRT/IDRIS]
  (Grant 2014-096847). 
  AREVA is acknowledged for financial support.

\section*{References}
\bibliographystyle{elsarticle-num} 
\bibliography{varvenne2015}

\begin{thebibliography}{10}
\expandafter\ifx\csname url\endcsname\relax
  \def\url#1{\texttt{#1}}\fi
\expandafter\ifx\csname urlprefix\endcsname\relax\def\urlprefix{URL }\fi
\expandafter\ifx\csname href\endcsname\relax
  \def\href#1#2{#2} \def\path#1{#1}\fi

\bibitem{Sandrock1999}
G.~Sandrock, A panoramic overview of hydrogen storage alloys from a gas
  reaction point of view, J. Alloys Compd. 293--295 (1999) 877--888.
\newblock \href {http://dx.doi.org/10.1016/s0925-8388(99)00384-9}
  {\path{doi:10.1016/s0925-8388(99)00384-9}}.

\bibitem{Taketomi2008}
S.~Taketomi, R.~Matsumoto, N.~Miyazaki, Atomistic study of hydrogen
  distribution and diffusion around a $\{112\}\langle 111 \rangle$ edge
  dislocation in alpha iron, Acta Mater. 56 (2008) 3761--3769.
\newblock \href {http://dx.doi.org/10.1016/j.actamat.2008.04.011}
  {\path{doi:10.1016/j.actamat.2008.04.011}}.

\bibitem{Garner2006}
F.~Garner, E.~Simonen, B.~Oliver, L.~Greenwood, M.~Grossbeck, W.~Wolfer,
  P.~Scott, Retention of hydrogen in fcc metals irradiated at temperatures
  leading to high densities of bubbles or voids, J. Nucl. Mater. 356 (2006)
  122--135.
\newblock \href {http://dx.doi.org/10.1016/j.jnucmat.2006.05.023}
  {\path{doi:10.1016/j.jnucmat.2006.05.023}}.

\bibitem{Oriani1977}
R.~A. Oriani, P.~H. Josephic, Equilibrium and kinetic studies of the
  hydrogen-assisted cracking of steel, Acta Metall. 25 (1977) 979--988.
\newblock \href {http://dx.doi.org/10.1016/0001-6160(77)90126-2}
  {\path{doi:10.1016/0001-6160(77)90126-2}}.

\bibitem{Birnbaum1994}
H.~K. Birnbaum, P.~Sofronis, Hydrogen-enhanced localized plasticity --- a
  mechanism for hydrogen-related fracture, Mater. Sci. Eng. A 176 (1994)
  191--202.
\newblock \href {http://dx.doi.org/10.1016/0921-5093(94)90975-x}
  {\path{doi:10.1016/0921-5093(94)90975-x}}.

\bibitem{Song2013}
J.~Song, W.~A. Curtin, Atomic mechanism and prediction of hydrogen
  embrittlement in iron, Nat. Mater. 12 (2013) 145--151.
\newblock \href {http://dx.doi.org/10.1038/nmat3479}
  {\path{doi:10.1038/nmat3479}}.

\bibitem{Lewis1984}
M.~B. Lewis, Deuterium-defect trapping in ion-irradiated zirconium, J. Nucl.
  Mater. 125 (1984) 152--159.
\newblock \href {http://dx.doi.org/10.1016/0022-3115(84)90542-7}
  {\path{doi:10.1016/0022-3115(84)90542-7}}.

\bibitem{McMinn2000}
A.~McMinn, E.~C. Darby, J.~S. Schofield, The terminal solid solubility of
  hydrogen in zirconium alloys, in: Zirconium in the Nuclear Industry: Twelfth
  International Symposium, ASTM International, 2000, p. 173.
\newblock \href {http://dx.doi.org/10.1520/stp14300s}
  {\path{doi:10.1520/stp14300s}}.

\bibitem{Vizcaino2002}
P.~Vizcaíno, A.~Banchik, J.~Abriata, Solubility of hydrogen in {Z}ircaloy-4:
  irradiation induced increase and thermal recovery, J. Nucl. Mater. 304 (2002)
  96--106.
\newblock \href {http://dx.doi.org/10.1016/s0022-3115(02)00883-8}
  {\path{doi:10.1016/s0022-3115(02)00883-8}}.

\bibitem{Vizcaino2007}
P.~Vizcaíno, A.~D. Banchik, J.~P. Abriata, Hydrogen in {Z}ircaloy-4: effects
  of the neutron irradiation on the hydride formation, J. Mater. Sci. 42 (2007)
  6633--6637.
\newblock \href {http://dx.doi.org/10.1007/s10853-007-1525-x}
  {\path{doi:10.1007/s10853-007-1525-x}}.

\bibitem{Domain2004}
C.~Domain, R.~Besson, A.~Legris, Atomic-scale ab initio study of the {Z}r-{H}
  system: {II}. interaction of {H} with plane defects and mechanical
  properties, Acta Mater. 52 (2004) 1495--1502.
\newblock \href {http://dx.doi.org/10.1016/j.actamat.2003.11.031}
  {\path{doi:10.1016/j.actamat.2003.11.031}}.

\bibitem{Udagawa2010}
Y.~Udagawa, M.~Yamaguchi, H.~Abe, N.~Sekimura, T.~Fuketa, Ab initio study on
  plane defects in zirconium-hydrogen solid solution and zirconium hydride,
  Acta Mater. 58 (2010) 3927--3938.
\newblock \href {http://dx.doi.org/10.1016/j.actamat.2010.03.034}
  {\path{doi:10.1016/j.actamat.2010.03.034}}.

\bibitem{McGrath2011}
M.~A. McGrath, S.~Yagnik, Experimental investigation of irradiation creep and
  growth of recrystallized zircaloy-4 guide tubes pre-irradiated in {PWR}, J.
  ASTM Int. 8 (2011) 103770.
\newblock \href {http://dx.doi.org/10.1520/jai103770}
  {\path{doi:10.1520/jai103770}}.

\bibitem{Carpenter1988}
G.~J.~C. Carpenter, R.~H. Zee, A.~Rogerson, Irradiation growth of zirconium
  single crystals: A review, J. Nucl. Mater. 159 (1988) 86--100.
\newblock \href {http://dx.doi.org/10.1016/0022-3115(88)90087-6}
  {\path{doi:10.1016/0022-3115(88)90087-6}}.

\bibitem{Fidleris1988}
V.~Fidleris, The irradiation creep and growth phenomena, J. Nucl. Mater. 159
  (1988) 22--42.
\newblock \href {http://dx.doi.org/10.1016/0022-3115(88)90083-9}
  {\path{doi:10.1016/0022-3115(88)90083-9}}.

\bibitem{Tournadre2013}
L.~Tournadre, F.~Onimus, J.-L. Béchade, D.~Gilbon, J.-M. Cloué, J.-P. Mardon,
  X.~Feaugas, Toward a better understanding of the hydrogen impact on the
  radiation induced growth of zirconium alloys, J. Nucl. Mater. 441 (2013)
  222--231.
\newblock \href {http://dx.doi.org/10.1016/j.jnucmat.2013.05.045}
  {\path{doi:10.1016/j.jnucmat.2013.05.045}}.

\bibitem{Tournadre2014}
L.~Tournadre, F.~Onimus, J.-L. Béchade, D.~Gilbon, J.-M. Cloué, J.-P. Mardon,
  X.~Feaugas, Impact of hydrogen pick-up and applied stress on c-component
  loops: toward a better understanding of the radiation induced growth of
  recrystallized zirconium alloys, J. ASTM Int. 17 (2014) 853--894.
\newblock \href {http://dx.doi.org/10.1520/STP154320120200}
  {\path{doi:10.1520/STP154320120200}}.

\bibitem{Legrand1984}
B.~Legrand, Relations entre la structure {\'e}lectronique et la facilit{\'e} de
  glissement dans les m{\'e}taux hexagonaux compacts, Philos. Mag. B 49 (1984)
  171--184.
\newblock \href {http://dx.doi.org/10.1080/13642818408227636}
  {\path{doi:10.1080/13642818408227636}}.

\bibitem{Clouet2012}
E.~Clouet, Screw dislocation in zirconium: An \textit{ab initio} study, Phys.
  Rev. B 86 (2012) 144104.
\newblock \href {http://dx.doi.org/10.1103/PhysRevB.86.144104}
  {\path{doi:10.1103/PhysRevB.86.144104}}.

\bibitem{Varvenne2014}
C.~Varvenne, O.~Mackain, E.~Clouet, Vacancy clustering in zirconium: An
  atomic-scale study, Acta Mater. 78 (2014) 65--77.
\newblock \href {http://dx.doi.org/10.1016/j.actamat.2014.06.012}
  {\path{doi:10.1016/j.actamat.2014.06.012}}.

\bibitem{Christensen2015a}
M.~Christensen, W.~Wolf, C.~Freeman, E.~Wimmer, R.~Adamson, L.~Hallstadius,
  P.~Cantonwine, E.~Mader, Diffusion of point defects, nucleation of
  dislocation loops, and effect of hydrogen in hcp-{Z}r: Ab initio and
  classical simulations, J. Nucl. Mater. 460 (2015) 82--96.
\newblock \href {http://dx.doi.org/10.1016/j.jnucmat.2015.02.013}
  {\path{doi:10.1016/j.jnucmat.2015.02.013}}.

\bibitem{Christensen2015b}
M.~Christensen, W.~Wolf, C.~Freeman, E.~Wimmer, R.~B. Adamson, L.~Hallstadius,
  P.~E. Cantonwine, E.~V. Mader, H in $\alpha$-{Z}r and in zirconium hydrides:
  solubility, effect on dimensional changes, and the role of defects, J. Phys.:
  Condens. Matter 27 (2015) 025402.
\newblock \href {http://dx.doi.org/10.1088/0953-8984/27/2/025402}
  {\path{doi:10.1088/0953-8984/27/2/025402}}.

\bibitem{Giannozzi2009}
P.~Gianozzi, {et al.}, Quantum espresso: a modular and open-source software
  project for quantum simulations of materials, J. Phys.: Condens. Matter 21
  (2009) 395502.
\newblock \href {http://dx.doi.org/10.1088/0953-8984/21/39/395502}
  {\path{doi:10.1088/0953-8984/21/39/395502}}.

\bibitem{Perdew1996}
J.~P. Perdew, K.~Burke, M.~Ernzerhof, Generalized gradient approximation made
  simple, Phys. Rev. Lett. 77 (1996) 3865--3868.
\newblock \href {http://dx.doi.org/10.1103/PhysRevLett.77.3865}
  {\path{doi:10.1103/PhysRevLett.77.3865}}.

\bibitem{Varvenne2013}
C.~Varvenne, F.~Bruneval, M.-C. Marinica, E.~Clouet, Point defect modelling in
  materials: coupling ab initio and elasticity approaches, Phys. Rev. B 88
  (2013) 134102.
\newblock \href {http://dx.doi.org/10.1103/PhysRevB.88.134102}
  {\path{doi:10.1103/PhysRevB.88.134102}}.

\bibitem{Ashcroft1976}
N.~W. Ashcroft, N.~D. Mermin, Solid State Physics, Saunders College,
  Philadelphia, 1976.

\bibitem{Fukai1993}
Y.~Fukai, The metal-hydrogen system, Springer Verlag, 1993.

\bibitem{Domain2002}
C.~Domain, R.~Besson, A.~Legris, Atomic-scale ab-initio study of the {Z}r-{H}
  system: {I}. bulk properties, Acta Mater. 50 (2002) 3513--3526.
\newblock \href {http://dx.doi.org/10.1016/S1359-6454(02)00173-8}
  {\path{doi:10.1016/S1359-6454(02)00173-8}}.

\bibitem{Lumley2014}
S.~C. Lumley, R.~W. Grimes, S.~T. Murphy, P.~A. Burr, A.~Chroneos, P.~R.
  Chard-Tuckey, M.~R. Wenman, The thermodynamics of hydride precipitation: The
  importance of entropy, enthalpy and disorder, Acta Mater. 79 (2014) 351--362.
\newblock \href {http://dx.doi.org/10.1016/j.actamat.2014.07.019}
  {\path{doi:10.1016/j.actamat.2014.07.019}}.

\bibitem{Burr2013a}
P.~A. Burr, S.~T. Murphy, S.~C. Lumley, M.~R. Wenman, R.~W. Grimes, Hydrogen
  accommodation in {Z}r second phase particles: Implications for {H} pick-up
  and hydriding of {Z}ircaloy-2 and {Z}ircaloy-4, Corrosion Science 69 (2013)
  1--4.
\newblock \href {http://dx.doi.org/10.1016/j.corsci.2012.11.036}
  {\path{doi:10.1016/j.corsci.2012.11.036}}.

\bibitem{Narang1977}
P.~Narang, G.~Paul, K.~Taylor, Location of hydrogen in $\alpha$-zirconium, J.
  Less Common Met. 56 (1977) 125--128.
\newblock \href {http://dx.doi.org/10.1016/0022-5088(77)90225-9}
  {\path{doi:10.1016/0022-5088(77)90225-9}}.

\bibitem{Hayward2013}
E.~Hayward, C.-C. Fu, Interplay between hydrogen and vacancies in
  $\alpha$-{F}e, Phys. Rev. B 87 (2013) 174103.
\newblock \href {http://dx.doi.org/10.1103/PhysRevB.87.174103}
  {\path{doi:10.1103/PhysRevB.87.174103}}.

\bibitem{Ghazisaedi2014}
M.~Ghazisaedi, D.~R. Trinkle, Interaction of oxygen interstitials with lattice
  faults in ti, Acta Materialia 76 (2014) 82--86.

\bibitem{Hull2011}
D.~Hull, D.~J. Bacon, Introduction to Dislocations, 5th Edition,
  Butterworth-Heinemann, Oxford, UK, 2011.

\bibitem{Zinkle1987}
S.~J. Zinkle, W.~G. Wolfer, G.~L. Kulcinski, L.~E. Seitzman, Stability of
  vacancy clusters in metals {II}. {E}ffect of oxygen and helium on void
  formation in metals, Philos. Mag. A 55 (1987) 127--140.
\newblock \href {http://dx.doi.org/10.1080/01418618708209804}
  {\path{doi:10.1080/01418618708209804}}.

\bibitem{Treglia1999}
G.~Tréglia, B.~Legrand, F.~Ducastelle, A.~Saúl, C.~Gallis, I.~Meunier,
  C.~Mottet, A.~Senhaji, Alloy surfaces: segregation, reconstruction and phase
  transitions, Comp. Mater. Sci. 15 (1999) 196--235.
\newblock \href {http://dx.doi.org/10.1016/s0927-0256(99)00004-x}
  {\path{doi:10.1016/s0927-0256(99)00004-x}}.

\bibitem{Mutschele1987}
T.~Mütschele, R.~Kirchheim, Segregation and diffusion of hydrogen in grain
  boundaries of palladium, Scripta Metall. 21 (1987) 135--140.
\newblock \href {http://dx.doi.org/10.1016/0036-9748(87)90423-6}
  {\path{doi:10.1016/0036-9748(87)90423-6}}.

\bibitem{Zuzek1990}
E.~Zuzek, J.~P. Abriata, A.~San-Martin, F.~D. Manchester, The {H}-{Z}r
  (hydrogen-zirconium) system, Bulletin of Alloy Phase Diagrams 11 (1990)
  385--395.
\newblock \href {http://dx.doi.org/10.1007/bf02843318}
  {\path{doi:10.1007/bf02843318}}.

\bibitem{Suzuki1962}
H.~Suzuki, Segregation of solute atoms to stacking faults, J. Phys. Soc. Jpn.
  17 (1962) 322--325.
\newblock \href {http://dx.doi.org/10.1143/jpsj.17.322}
  {\path{doi:10.1143/jpsj.17.322}}.

\bibitem{Hirth1970}
J.~P. Hirth, Thermodynamics of stacking faults, Metall. Trans. 1 (1970)
  2367--2374.
\newblock \href {http://dx.doi.org/10.1007/BF03038365}
  {\path{doi:10.1007/BF03038365}}.

\bibitem{Udagawa2011}
Y.~Udagawa, M.~Yamaguchi, T.~Tsuru, H.~Abe, N.~Sekimura, Effect of {S}n and
  {N}b on generalized stacking fault energy surfaces in zirconium and gamma
  hydride habit planes, Philos. Mag. 91~(12) (2011) 1665--1678.
\newblock \href {http://dx.doi.org/10.1080/14786435.2010.543651}
  {\path{doi:10.1080/14786435.2010.543651}}.

\bibitem{Proville2005a}
L.~Proville, Biphonons in the {K}lein-{G}ordon lattice, Phys. Rev. B 71 (2005)
  104306.
\newblock \href {http://dx.doi.org/10.1103/physrevb.71.104306}
  {\path{doi:10.1103/physrevb.71.104306}}.

\end{thebibliography}
\end{document}